\documentclass[journal]{IEEEtran}

\usepackage[utf8]{inputenc}
\usepackage{physics}
\usepackage{amsmath}
\usepackage{amssymb}
\usepackage{subfigure}
\usepackage{graphicx}
\usepackage{graphics}
\usepackage{color}
\usepackage{psfrag}
\usepackage{cite}
\usepackage{balance}
\usepackage{algorithm}
\usepackage{accents}
\usepackage{amsthm}
\usepackage{bm}
\usepackage{url}
\usepackage{algorithmic}
\usepackage[english]{babel}
\usepackage{multirow}
\usepackage{enumerate}
\usepackage{cases}
\usepackage{stfloats}
\usepackage{dsfont}
\usepackage{soul}
\usepackage{amsfonts}
\usepackage{fancyhdr}
\usepackage{hhline}
\usepackage{array}
\usepackage{booktabs}
\usepackage{framed}
\usepackage{float}
\usepackage{soul}
\usepackage{url}
\usepackage{hyperref}
\usepackage{mathtools}
\usepackage{threeparttable}
\usepackage{booktabs}
\usepackage{tabularx}
\usepackage{array}
\usepackage{makecell}
\usepackage{tabularray}
\usepackage{array,tabularx,multirow,makecell}
\newcolumntype{C}{>{\centering\arraybackslash}X}

\begin{document}
\bstctlcite{BSTcontrol-main-v7}

\title{{Rydberg Atomic Receivers for Wireless Physical-Layer}}



\author{Minze Chen, Mingyao Cui,~\IEEEmembership{Graduate Student Member, IEEE}, Zhiao Zhu, Jingyi Feng, Haodong Zhang, \\Yuanbin Chen,~\IEEEmembership{Member, IEEE}, Qunsong Zeng,~\IEEEmembership{Member, IEEE}, Tianqi Mao,~\IEEEmembership{Member, IEEE}, Weibin Li,\\ Linglong Dai,~\IEEEmembership{Fellow, IEEE}, Chau Yuen,~\IEEEmembership{Fellow, IEEE}, Merouane Debbah,~\IEEEmembership{Fellow, IEEE}, \\Dezhi Zheng, Jun Zhang, and Kaibin Huang,~\IEEEmembership{Fellow, IEEE}
\thanks{M. Chen, Z. Zhu, J. Feng, T. Mao, D. Zheng, J. Zhang are with State Key Laboratory of Environment Characteristics and Effects for Near-space, Beijing Institute of Technology, Beijing 100081, China (emails: \{chenmz22, zza14606, fengjy, maotq, zhengdezhi\}@bit.edu.cn, buaazhangjun@vip.sina.com). }
\thanks{M. Cui, H. Zhang, Q. Zeng, and K. Huang are with the Department of Electrical and Electronic Engineering, The University of Hong Kong, Hong Kong (emails: \{mycui, qszeng\}@eee.hku.hk, huangkb@hku.hk).}
\thanks{Y. Chen, C. Yuen are with the School of Electrical and Electronics Engineering, Nanyang Technological University, Singapore (email: \{yuanbin.chen, chau.yuen\}@ntu.edu.sg).}
\thanks{W. Li is with the School of Physics and Astronomy,   University of Nottingham, United Kingdom (email: weibin.li@nottingham.ac.uk).}
\thanks{L. Dai is with Department of Electronic Engineering, Tsinghua University, China (email: daill@tsinghua.edu.cn).}
\thanks{M. Debbah is with KU 6G Research Center, Khalifa University of Science and Technology, Abu Dhabi, UAE (email: merouane.debbah@ku.ac.ae).}
\thanks{Minze Chen and Mingyao Cui contributed equally to this work.
(\emph{Corresponding authors: Tianqi Mao, Qunsong Zeng.})}}


\IEEEaftertitletext{\vspace{-1em}}
\maketitle

\begin{abstract} 
With the escalating development of the quantum information technologies, the Rydberg atomic receiver has been envisioned as a promising alternative to its classical electronic counterparts, attracting extensive attention from both physics and electronic engineering disciplines. By exploiting the electric-field sensitivity of alkali atoms with diverse Rydberg energy states for electromagnetic field measurement, a Rydberg atomic receiver is capable of surmounting the inherent limitations on the operating bandwidth and detection sensitivity of the classical electronic counterparts. Such technology is envisioned to establish an unprecedented quantum-based paradigm for next-generation communication and sensing, which has motivated a thorough revolution on the transceiving mechanisms to fully unleash the potential of Rydberg atomic receivers. Against this background, this article provides an extensive overview of Rydberg atomic communication and sensing from fundamental theory, critical technologies to practical implementations. We first review key milestones in the development of Rydberg atomic receivers. Then we provide a clear understanding of Rydberg atomic receiver mechanisms from the perspective of classical communication and sensing, and highlight the unprecedented advantages over traditional electronic receivers as well as the technical challenges. Afterwards, necessary fundamentals are presented including the Rydberg atoms, typical non-coherent/coherent structures, and explicit experimental configurations, followed by discussions on state-of-the-art development of the Rydberg atomic receiver. Specifically, advancements on electric-field detection methods for accurate measurement of amplitude, phase, frequency and polarization, are extensively investigated. We then review recent advances in Rydberg atomic communication and sensing, from waveform design and array processing to localization, imaging, and remote sensing, and discuss the reported experimental results. Finally, we summarize the promising research trends for Rydberg atomic receivers, providing visionary insights to facilitate future breakthrough on this new paradigm of Rydberg atomic communication and sensing.
\end{abstract} 

\vspace{-2mm}
\section{Introduction}

Sixth-generation (6G) wireless networks are expected to support ubiquitous connectivity and sensing of the physical environment. Space-air-ground integrated networks (SAGIN) are an important part of this vision, extending terrestrial networks through cooperation with aerial platforms and satellites. Such integration brings together links operating over different propagation distances and widely separated carrier frequencies. Reliable communication over these links requires receivers to recover transmitted waveforms despite propagation loss. Sensing further requires accurate measurements of signal characteristics such as delay and phase to infer the location or motion of objects. These requirements place particular demands on receiver sensitivity and frequency coverage, as well as the ability to preserve the information carried by the received field. Therefore, the physical mechanism used to detect this field and convert it into a measurable signal is an  important consideration in receiver design.

Conventional wireless receivers use antennas to convert incident electromagnetic fields into electrical signals. Electronic RF circuits then perform filtering, amplification, and frequency conversion before digitization. Weak-signal reception depends on efficient antenna coupling and low-noise RF circuitry. However, maintaining these conditions across widely separated frequency bands often requires multiple RF paths or reconfigurable components. The resulting hardware complexity complicates receiver designs that seek broad frequency coverage and sensitive reception in a compact system. These constraints have motivated interest in alternative reception mechanisms based on different physical principles.

Rydberg atomic receivers (RAREs) provide an alternative approach to RF detection based on the interaction of incident fields with highly excited atoms. Transitions between Rydberg states have large electric dipole moments, allowing weak RF fields to produce measurable changes in the atomic response. In a typical RARE, lasers prepare and probe Rydberg states in an alkali vapor cell, and the RF-induced response is observed through changes in probe-beam transmission. Different Rydberg transitions correspond to different RF frequencies, so the receiver can be configured for widely separated carrier bands by selecting the states addressed by the lasers. An applied RF reference field can make the optical readout sensitive to the phase of the incident signal through coherent atomic mixing. The overlap of the optical beams defines a localized interrogation region, allowing the RF field to be sampled at different positions within the vapor cell. These properties have supported the development of RAREs for modulated-signal reception and spatially resolved sensing.

Therefore, RAREs have progressed from laboratory measurements of RF electric fields to receiver-level demonstrations of wireless communication and sensing functions. This progression has shifted the research focus toward how information carried by an incident RF signal is represented in the atomic response and recovered through optical readout. The capabilities realized in a particular application depend on receiver sensitivity, bandwidth, and other performance measures, but more decisively on the detection scheme selected for the task and the engineering resources available to implement it. These developments motivate a systematic survey that connects atomic detection principles with receiver capabilities and examines how those capabilities are used and constrained in wireless systems.

\subsection{Evolution of Rydberg Atomic Receivers}

\begin{figure*}
	\centering
	\includegraphics[width=\textwidth]{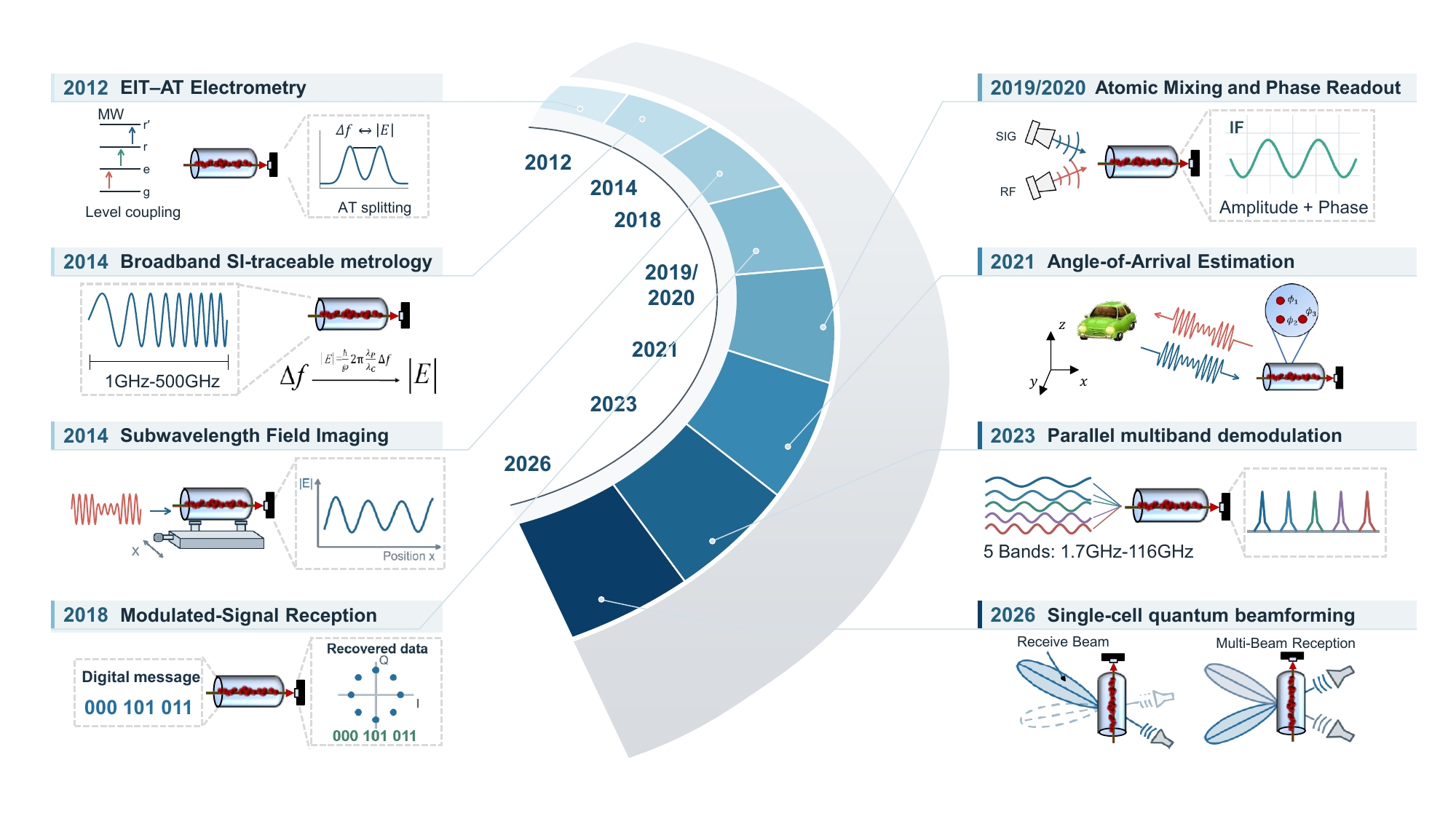}
	\vspace*{-3em}
	\caption{Milestones in the development of Rydberg atomic receivers.
    } 
	\vspace*{-1em}
\label{fig:milestones}
\end{figure*}

The development of Rydberg atomic receivers, summarized in Fig.~\ref{fig:milestones}, began with studies of cold Rydberg atoms. These studies examined their response to electric fields through Stark shifts and microwave spectroscopy~\cite{BohlouliZanjani2007}. In 2007, Charles Adams's group in Durham University demonstrated ladder-type Rydberg electromagnetically induced transparency (EIT) in room-temperature rubidium vapor, providing a nondestructive optical probe of highly excited Rydberg states~\cite{Intro-Mohapatra2007}. In 2011, the same group observed Autler--Townes (AT) splitting of the EIT resonance under resonant microwave coupling in a cold rubidium gas~\cite{Tanasittikosol2011}. Building on these advances, the experimental paradigm of room-temperature Rydberg microwave electrometry was established by the landmark 2012 work of J. P. Shaffer's group at the University of Oklahoma~\cite{Intro-Sedlacek2012}. Their experiment used changes in EIT probe transmission to measure microwave electric fields in a vapor cell.

Subsequent work examined the metrological basis of the EIT–AT method, particularly the accuracy of electric-field strengths inferred from the measured spectral splitting. In 2014, Christopher L. Holloway and colleagues at the U.S. National Institute of Standards and Technology (NIST) developed a self-calibrating, SI-traceable vapor-cell probe based on the relation between the AT splitting, the Rydberg transition dipole moment, and the microwave electric field~\cite{Intro-Holloway2014Broadband}. They measured fields at several widely separated frequencies using a single vapor cell and identified Rydberg transitions suitable for measurements across the 1--500~GHz range. The spatially localized optical probe also made it possible to examine how the field varied within the cell. In a separate study that year, Holloway and colleagues used this capability to map microwave electric fields with subwavelength spatial resolution~\cite{Intro-Holloway2014Sub}.

The optical transduction of time-varying atomic responses subsequently opened a direct path toward wireless signal reception. In 2018, Deb and Kj{\ae}rgaard demonstrated reception of amplitude-modulated microwave signals through a Rydberg-atom-based radio-over-fiber link, while Meyer et al. demonstrated digital communication at megabit-per-second rates~\cite{Comm-RoF-Deb2018,Comm-AM/FM-Meyer2018}. These studies marked a transition from electric-field metrology to wireless signal reception. In 2019, the introduction of a reference RF field enabled coherent mixing inside the atomic medium. The frequency difference between the signal and reference fields appeared as a low-frequency optical beat note, while its phase retained the relative RF phase~\cite{Comm-PM-Simons2019}. This capability supported phase-sensitive field measurement and the reception of phase-modulated signals~\cite{Comm-PM-Holloway2019}. In 2020, the atomic superheterodyne receiver further incorporated a strong microwave local oscillator into the Rydberg energy-level system, enabling high-sensitivity recovery of the signal amplitude, frequency, and phase through an optically detected intermediate-frequency response~\cite{Comm-PM-Jing2020}.

Following the introduction of atomic superheterodyne receivers, researchers have continued to investigate alternative measurement and readout schemes, including microwave-to-optical conversion through four- and six-wave mixing~\cite{FourWaveMixing-Ji2024,SixWaveMixing-Borowka2024} and microwave electrometry enhanced by many-body criticality~\cite{Critical-Ding2022}. Nevertheless, the superheterodyne scheme remains a standard architecture for Rydberg reception. It directly produces a downconverted signal analogous to the output of a conventional receiver, retaining the amplitude and relative phase information of the incident RF field. This relatively simple excitation scheme and compatibility with conventional intermediate-frequency and baseband processing have encouraged further investigation of atomic receivers in wireless sensing and communication systems. In 2021, Robinson et al. demonstrated RF angle-of-arrival estimation by measuring the phase difference between two spatially separated sensing regions within a single vapor cell~\cite{Sensing-LOC-Robinson2021}. For communication, Meyer et al. demonstrated simultaneous reception of digital data in four widely separated frequency bands through a single atomic sensing medium~\cite{Comm-MB-Meyer2023}. In 2026, Cui et al. introduced the continuous quantum aperture (CQA) concept, enabling reconfigurable beamforming within a single vapor cell~\cite{Comm-CA-Cui2026a,Comm-CA-Cui2026b}. Thus, Rydberg atomic reception has expanded beyond electric-field metrology and is emerging as a physical-layer technology for wireless communication and sensing.

\subsection{Comparison with Classical Electronic Receivers}

\begin{figure*}[hbt]
	\centering
	\includegraphics[width=\textwidth]{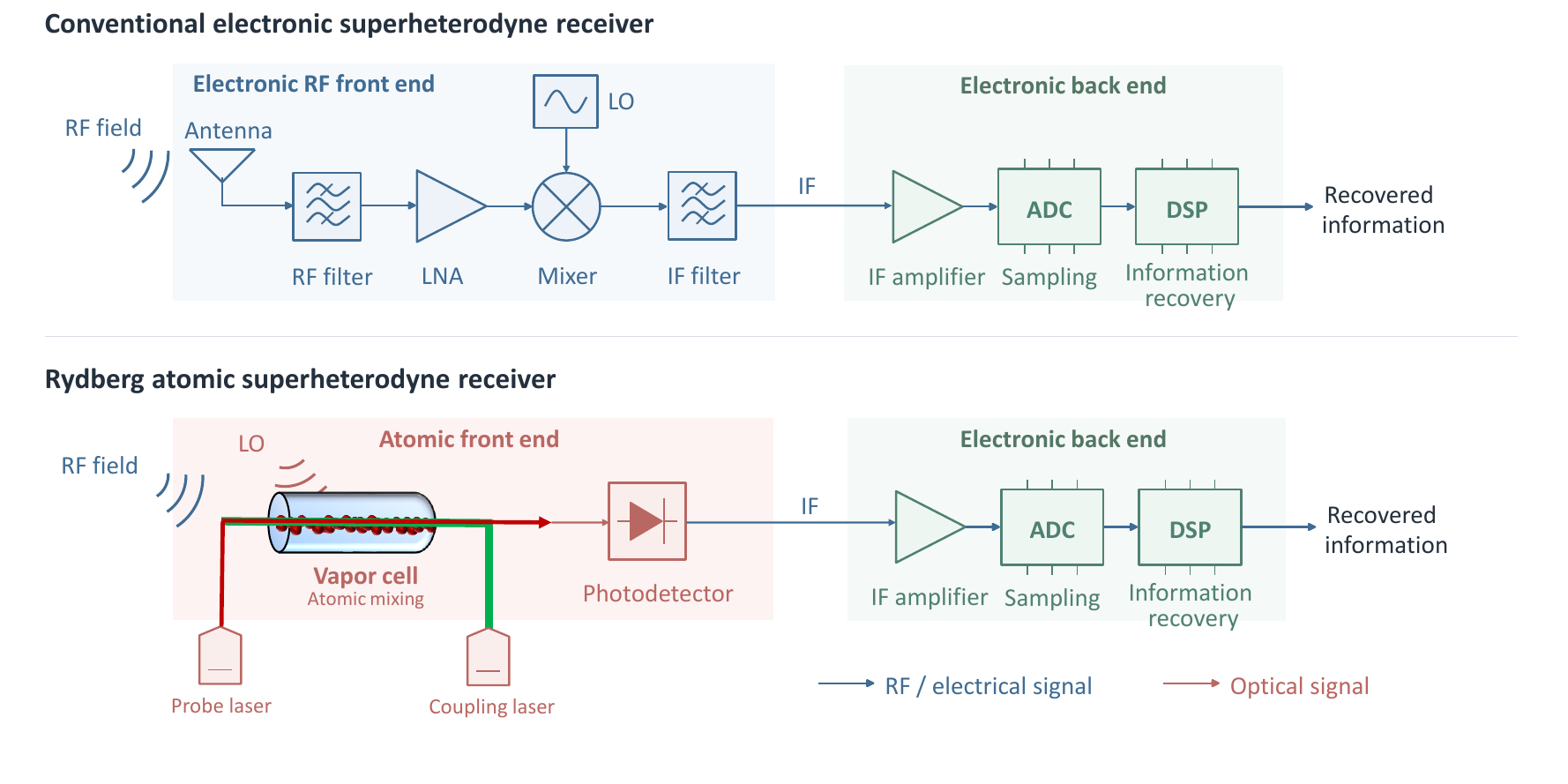}
	\vspace*{-4em}
	\caption{Architectural comparison of a conventional electronic superheterodyne receiver (top) and a Rydberg atomic superheterodyne receiver (bottom).
    } 
	\label{fig:receiver_comparison}
\end{figure*}

In the conventional superheterodyne receiver shown in Fig.~\ref{fig:receiver_comparison} (top), the incident RF field first induces an electrical signal at the antenna terminals. Before frequency conversion, RF filtering suppresses unwanted spectral components, and low-noise amplification reduces the relative contribution of noise from subsequent stages. The amplified signal is then mixed with a local oscillator (LO) to produce an intermediate-frequency (IF) component that retains the original modulation. IF filtering selects this component from the mixer output for further amplification and digitization. Subsequent digital processing recovers the transmitted information or estimates the signal parameters required for sensing.

In contrast, the atomic superheterodyne receiver in Fig.~\ref{fig:receiver_comparison} (bottom) performs field detection and frequency downconversion in the vapor cell before electrical readout. The incident RF signal and the LO jointly drive the optically prepared Rydberg atoms, and their frequency difference appears as a modulation of the probe transmission. Within the linear operating range, this optical modulation preserves the signal amplitude and relative phase information~\cite{Comm-PM-Simons2019,Comm-PM-Jing2020}. Photodetection therefore delivers an electrical IF signal directly to the electronic back end for filtering, amplification, and digitization. The electronic processing chain thus begins at the IF, because the atomic interaction and optical readout have already converted the incident RF waveform into a form suitable for conventional signal processing.

In conventional receivers, frequency coverage and spatial sampling are largely determined by the RF hardware and antenna layout. Rydberg receivers introduce atomic and optical control over these aspects of the front end. The operating band can be changed by selecting and tuning Rydberg transitions. Within the vapor cell, the interrogation beams define the regions that sample the RF field. Adjusting the optical arrangement therefore changes the sampling geometry, and multiple regions can provide spatially resolved measurements within a single cell. These frequency and spatial configurations can share parts of the atomic and optical hardware. Such sharing offers opportunities to support multiband reception and array processing within a common receiver platform.

The architectural flexibility comes with a shift in the factors that govern receiver performance and implementation. Atomic reception reduces the need for electronic stages at the carrier frequency, but the atomic response and optical readout determine the bandwidth available for recovering the signal. Broad carrier-frequency coverage therefore does not imply a correspondingly broad instantaneous bandwidth. The optical subsystem also contributes to the receiver noise and requires stable laser operation. Its lasers and control electronics must consequently be included when assessing receiver size, power consumption, and complexity. Conventional electronic receivers benefit from mature integration and established implementations for wideband reception. The comparison in Table~\ref{tab:receiver_comparison} thus concerns the balance between configurable reception functions, usable performance, and the hardware needed to realize them.

\begin{table*}[!t]
\centering
\caption{Representative Comparison of Classical Electronic and Rydberg Atomic Receivers}
\label{tab:receiver_comparison}

\footnotesize
\renewcommand{\arraystretch}{1.25}
\setlength{\tabcolsep}{3.5pt}
\renewcommand{\tabularxcolumn}[1]{m{#1}}

\begin{tabularx}{\linewidth}{
>{\centering\arraybackslash}m{1.55cm}
C
C
C
C
C
C
}
\toprule

\textbf{Receiver Type}
&
\makecell{\textbf{Sensitivity}\\\textbf{(NEF)}}
&
\makecell{\textbf{Carrier-Frequency}\\\textbf{Coverage}}
&
\makecell{\textbf{Instantaneous}\\\textbf{Bandwidth}}
&
\makecell{\textbf{Calibration and}\\\textbf{Traceability}}
&
\makecell{\textbf{System}\\\textbf{Integration}}
&
\makecell{\textbf{Interference}\\\textbf{Resilience}}
\\

\midrule

\textbf{\makecell{Classical\\electronic\\receiver}}
&
\makecell{$\approx 0.10$\textsuperscript{a}\\Antenna + LNA}
&
\makecell{Band-limited\\LO tuning\\RF switching}
&
\makecell{MHz--GHz~\cite{Eng-USRPX4xx}\\Hardware-dependent}
&
Requires antenna-factor,
cable-loss, gain/noise,
and impedance calibration
&
\makecell{Mature RF ICs\\Compact modules}
&
Mature filtering,
AGC, and spatial
interference suppression
\\[3pt]

\textbf{\makecell{Rydberg\\atomic\\receiver}}
&
\makecell{$1.25$\textsuperscript{b}\\Warm vapor}
&
\makecell{MHz--THz~\cite{Review-Schlossberger2024}\\Across states\\and setups}
&
\makecell{kHz--MHz (typ.)}
&
AT splitting enables
SI-traceable field
measurement~\cite{Intro-Holloway2014Broadband}
&
\makecell{Prototype stage\\Laser/control\\overhead}~\cite{Sensing-Amarloo2025}
&
Strong out-of-band selectivity;
in-band blockers may
perturb operation~\cite{Review-Xiang2026}
\\

\bottomrule
\end{tabularx}
\par\vspace{3pt}
\begin{minipage}{\linewidth}
\footnotesize\raggedright
\textit{Notes:} NEF: noise-equivalent field, in $\mu\mathrm{V\,m^{-1}\,Hz^{-1/2}}$ (lower is better). Sensitivity values are representative X-band benchmarks, not performance limits.
\textsuperscript{a}Calculated for a matched, lossless antenna ($G=1.5$) and an LNA ($T_{\mathrm{LNA}}\approx100~\mathrm{K}$), neglecting thermal background~\cite{Sensitivity-Botello2022}.
\textsuperscript{b}Measured at 10.68~GHz in warm vapor~\cite{Sensitivity-Cai2022Photonics}.
\end{minipage}
\vspace{-6pt}
\end{table*}

\subsection{Technical Advantages of Rydberg Atomic Receivers for Communication and Sensing}

The strong coupling between Rydberg states and RF electric fields supports sensitive reception of weak signals. Their large transition dipole moments enable weak resonant fields to produce measurable changes in the optical response. With an RF local oscillator, atomic superheterodyne detection also recovers the signal amplitude and phase relative to the reference~\cite{Comm-PM-Jing2020}. This coherent readout connects atomic field sensitivity to waveform reception. It supports phase-sensitive demodulation in communications and coherent processing of received echoes in sensing.

Atomic spectroscopy also provides a direct relation between the measured response and the RF electric-field amplitude. In the resolved Autler--Townes regime, the field amplitude can be inferred from the spectral splitting and the transition dipole moment~\cite{Intro-Holloway2014Broadband}. This relation enables SI-traceable field measurement and gives atomic receivers a metrological role alongside signal reception. Such measurements are useful for characterizing RF fields and calibrating electromagnetic measurement systems.

The rich atomic level structure further supports reception across widely separated frequency bands. Different Rydberg transitions address different carrier frequencies, and suitable excitation schemes can support multiband operation. Parts of the atomic medium and optical hardware can therefore serve several reception bands~\cite{Comm-MB-Cui2026}. This sharing offers a route to multiband communication and sensing without duplicating the entire receiver for each band.

Optical interrogation extends this hardware sharing to the spatial domain. The interrogation volume defines a local sensing region that can be much smaller than the RF wavelength. Multiple probe regions can therefore sample the field at different positions within a vapor cell~\cite{Eng-Multi-Otto2021}. With coherent readout, these samples provide the spatial phase information needed for beamforming and direction estimation. Optical control of the sampling geometry also offers opportunities for reconfigurable arrays. A shared set of spatial measurements can consequently support communication reception and electromagnetic sensing within the same atomic platform.

\subsection{Core Challenges and Open Problems }
RAREs remain largely at the laboratory stage, with several limitations that must be addressed before practical deployment. The main challenges can be summarized as follows.

\subsubsection{\textit{Limited instantaneous bandwidth and response speed}}
Although RAREs can access widely separated carrier frequencies, the instantaneous bandwidth around a given operating point remains limited. Typical values range from tens to hundreds of kilohertz, while some systems reach the MHz range \cite{Sens-wideband-meyer2020}. Multiphoton and dressed-state schemes have extended the response range to approximately ±10.2 MHz \cite{Comm-BW-Yang2024}, but this remains narrow for broadband communication and high-rate sensing. Increasing the bandwidth may also reduce spectral resolution or field sensitivity.

\subsubsection{\textit{Gap between intrinsic atomic sensitivity and system sensitivity}}
The high intrinsic sensitivity of Rydberg atoms does not directly translate into receiver-level performance. In warm-vapor systems, atomic motion and collisions, laser and photodetection noise, finite atom number, and long-term drift can all limit sensitivity\cite{noise-yang2024,noise-tang2025}. For centimeter-scale receivers, the measured sensitivity can even be lower than that of electrically small antennas with active amplification \cite{Sensing-Backes2024}.

\subsubsection{\textit{Nonlinearity, dynamic range, and interference}}
The atomic response is approximately linear only over a limited range of field strengths. Strong signals can cause saturation, broadening, and gain compression, while multitone excitation can generate harmonics and intermodulation products\cite{Comm-LinearDynamicRange-Wu2023}. These nonlinearities complicate multicarrier reception and linear signal processing. Strong interference can also shift the receiver away from its optimum operating point or distort the desired atomic response\cite{Comm-LinearDynamicRange-gonccalves2026}.

\subsubsection{\textit{Full-system integration and deployment}}
Although the atomic sensing region can be only millimeters or centimeters in size, a complete receiver still requires lasers, optical readout, temperature control, data acquisition, and electronic control. These peripheral subsystems dominate much of the size, power consumption, and cost, and introduce additional sensitivity to environmental and technical perturbations\cite{Sensing-Amarloo2025}.

\subsection{Motivation and Contributions}

\begin{table*}[!t]
\caption{Comparison of representative reviews, perspectives, and modeling studies on Rydberg-atom RF sensing and reception from an experimental physical-layer viewpoint.}
\label{tab:related_surveys}
\centering
\scriptsize
\renewcommand{\arraystretch}{1.10}
\setlength{\tabcolsep}{3.2pt}
\newcommand{\reviewyes}{\ensuremath{\checkmark}}
\newcommand{\reviewpartial}{\ensuremath{\triangle}}
\begin{tabular}{@{}lccccccp{1.90in}@{}}
\toprule
\textbf{Reference} &
\textbf{Exp. hardware} &
\textbf{Measured perf.} &
\textbf{PHY reception} &
\textbf{Comm.} &
\textbf{Sensing} &
\textbf{Classical map.} &
\textbf{Main emphasis} \\
\midrule
Fan \textit{et al.}~\cite{Review-Fan2015}
& \reviewyes & \reviewyes & -- & -- & \reviewyes & --
& RF electric-field electrometry \\

Fancher \textit{et al.}~\cite{Review-Fancher2021}
& \reviewyes & \reviewyes & \reviewpartial & \reviewyes & \reviewyes & \reviewpartial
& Communications and sensing applications \\

Simons \textit{et al.}~\cite{Review-Simons2021}
& \reviewyes & \reviewyes & \reviewpartial & \reviewpartial & \reviewyes & --
& RF quantum electric-field sensors \\

Artusio-Glimpse \textit{et al.}~\cite{Review-Artusio2022}
& \reviewyes & \reviewyes & \reviewpartial & \reviewpartial & \reviewyes & \reviewpartial
& Hot-atom RF measurements and metrology \\

Liu \textit{et al.}~\cite{Review-Liu2023}
& \reviewyes & \reviewyes & -- & \reviewpartial & \reviewyes & --
& DC-to-THz field measurement \\

Schlossberger \textit{et al.}~\cite{Review-Schlossberger2024}
& \reviewyes & \reviewyes & \reviewpartial & \reviewpartial & \reviewyes & \reviewpartial
& SI-traceable probes and communication receivers \\

Zhang \textit{et al.}~\cite{Review-Zhang2024}
& \reviewyes & \reviewyes & \reviewpartial & \reviewpartial & \reviewyes & \reviewpartial
& Metrology, communications, and hybrid systems \\

Somaweera \textit{et al.}~\cite{Review-Somaweera2025}
& \reviewyes & \reviewyes & \reviewpartial & \reviewpartial & \reviewyes & --
& Sensor principles, metrics, and applications \\

Gong \textit{et al.}~\cite{Review-Gong2025}
& \reviewpartial & \reviewpartial & \reviewyes & \reviewyes & \reviewyes & \reviewpartial
& RAQ receiver models, SISO, and MIMO \\

Chen \textit{et al.}~\cite{Comm-ISAC-Chen2025}
& \reviewpartial & \reviewpartial & \reviewyes & \reviewyes & \reviewyes & \reviewpartial
& Rydberg-enabled ISAC architecture \\

Gong \textit{et al.}~\cite{Comm-Theory-Gong2026}
& \reviewpartial & \reviewpartial & \reviewyes & \reviewyes & \reviewyes & \reviewpartial
& Communication and sensing models and performance \\

Chen \textit{et al.}~\cite{Comm-Theory-Chen2025}
& \reviewpartial & \reviewpartial & \reviewyes & \reviewyes & -- & \reviewyes
& End-to-end modeling from quantum physics to wireless communications \\

Cui \textit{et al.}~\cite{Comm-ISAC-Cui2026}
& \reviewpartial & \reviewpartial & \reviewyes & \reviewyes & \reviewpartial & \reviewyes
& Receiver perspective for wireless communications \\

Allinson \textit{et al.}~\cite{allinson2026rydbergreceiversspaceapplications}
& \reviewpartial & \reviewpartial & \reviewpartial & \reviewpartial & \reviewyes & \reviewpartial
& Spaceborne receiver opportunities \\

Babusenan \textit{et al.}~\cite{Review-Babusenan2026}
& \reviewyes & \reviewyes & \reviewpartial & \reviewpartial & \reviewyes & \reviewpartial
& Principles, metrics, and technology roadmaps \\

\midrule
\textbf{This survey}
& \reviewyes & \reviewyes & \reviewyes & \reviewyes & \reviewyes & \reviewyes
& \textbf{Experiment-oriented physical-layer analysis from atom--light--RF hardware to communication and sensing functions} \\
\bottomrule
\end{tabular}

\vspace{1mm}
\parbox{0.98\textwidth}{\scriptsize
\reviewyes: primary focus;\quad
\reviewpartial: partial coverage;\quad
--: not a primary focus. ``Exp. hardware'' includes atomic energy-level schemes, laser excitation and readout, vapor-cell implementations, RF coupling, and electronic interfaces. ``Measured perf.'' indicates analysis based on reported experimental conditions and results.
}
\end{table*}


To the best of the authors' knowledge, several reviews, perspectives, and modeling studies on Rydberg-atom RF sensing and reception have been published, as summarized in Table~\ref{tab:related_surveys}. Early studies mainly review RF electrometry, experimental sensing architectures, SI-traceable measurements, and their applications~\cite{Review-Fan2015,Review-Artusio2022,Review-Schlossberger2024,Review-Zhang2024}. More recent works extend the discussion to atomic receiver models, wireless communications, MIMO, ISAC, and space applications~\cite{Review-Gong2025,Comm-Theory-Gong2026,Comm-Theory-Chen2025,Comm-ISAC-Cui2026,allinson2026rydbergreceiversspaceapplications}. Detailed connections among experimental hardware, measured atomic responses, and physical-layer receiver functions receive limited treatment in these studies. The remainder of this survey is organized around these hardware, communication, and sensing dimensions. Compared with the existing reviews, our main contributions are summarized below.

\begin{enumerate}

\item We establish a hardware-to-performance framework for Rydberg atomic receivers. The framework connects atomic excitation schemes, vapor-cell implementations, RF coupling, optical readout, and electronic interfaces with the experimentally observed sensitivity, bandwidth, dynamic range, and frequency-coverage limits.

\item We develop a physical-layer taxonomy for Rydberg atomic communications according to the degrees of freedom exploited in reception, including waveform modulation, spatial reception and array processing, spectral-domain operation, and other receiver-enabled communication functions.

\item We formulate an application-oriented taxonomy for Rydberg atomic sensing that distinguishes field imaging, source localization, radar sensing, and other sensing tasks. For each category, the survey relates the available electromagnetic information, receiver configuration, and experimentally demonstrated sensing capability.

\item We identify the receiver-to-system transition requirements for Rydberg atomic wireless platforms. The discussion highlights how hardware constraints and atomic-response characteristics affect channel estimation, synchronization, and the practical integration of atomic receivers into wireless communication and sensing systems.

\end{enumerate}

Therefore, this survey provides an experiment-oriented physical-layer account of Rydberg atomic receivers, complementing existing studies centered on atomic metrology or communication models. Section~II introduces the physical fundamentals and receiver principles. Section~III reviews detection technologies and performance metrics. Sections~IV and V discuss communication and sensing applications, respectively. Section~VI presents future directions, followed by the conclusion in Section~VII.

\section{Fundamentals}
\subsection{Fundamentals of Rydberg Atoms}

\subsubsection{Atomic and Energy-Level Structure}
An atom consists of a positively charged nucleus surrounded by negatively charged electrons occupying discrete energy levels. 
For alkali atoms such as rubidium {(Rb)} or cesium {(Cs)}, which are widely used in Rydberg-atom receivers, the outermost electron dominates their optical and microwave response.
An {energy level} is commonly characterized by a set of quantum numbers:  $n$, $\ell$, $s$, $j$, and $m_{j}$.  
The principal quantum number $n$ {describes the electron shell of an atom} and the associated binding energy. 
The orbital angular momentum quantum number $\ell$ specifies the orbital shape, which corresponds to the spectroscopic labels $S,P,D,F,\cdots$ for $\ell = 0,1,2,3,\cdots$, respectively.
For a single valence electron, the spin quantum number is $s=1/2$.
The total angular momentum quantum number $j$ results from the coupling between the orbital and spin angular momenta, and therefore takes the values {$j=1/2, 3/2, 5/2, \cdots$}. The magnetic quantum number $m_j$ gives the projection of the total angular momentum along a chosen quantization axis. 
In the Rydberg-atom literature, a state is commonly written in compact spectroscopic notation {$n\ell_j$}, such as $66D_{5/2}$, where $n=66$ {and} the spectroscopic label $D$ corresponds to $\ell=2$ and $j=5/2$. The magnetic quantum number $m_j$ is often omitted unless Zeeman splitting or polarization-dependent selection rules are explicitly considered.

A Rydberg atom is an atom in which at least one electron is excited to a state with a large principal quantum number, typically $n \gtrsim 20$.  The binding energy of an alkali Rydberg state is approximately described by the quantum-defect-modified Rydberg formula 
\begin{align}
    E_{n\ell j} = -\frac{hcR_y}{(n-\delta_{\ell j})^2}.
\end{align}
In this equation, $h$ is the Planck constant, $c$ {is} the speed of light, $R_y$ is the Rydberg constant, and $\delta_{\ell j}$ is the quantum defect. The quantum defect accounts for the deviations from the hydrogen-like energy level by the finite ionic core. Typical values of $\delta_{\ell j}$ can be found in~\cite{QD1,QD2}. 

\subsubsection{Energy-Level Transitions}
Atoms can exchange energy with electromagnetic fields through transitions between distinct energy levels. 
When an atom absorbs a photon from an external field, the valence electron is promoted from an initial state $\ket{i}$ to a higher-energy final state $\ket{f}$.  Conversely, photon emission transfers the atom to a lower-energy state.

A transition is resonantly driven when the photon energy closely matches the energy gap between the two atomic states:
\begin{align}
    h\nu \approx E_f-E_i ,
\end{align}
where \(\nu\) is the photon frequency. In practice, “closely matches” means that the frequency detuning $\left|\nu - \frac{E_f-E_i}{h}\right|$ is within the linewidth of the transition.
In addition to resonance condition, a transition must also satisfy the corresponding selection rules.
These rules specify whether the transition matrix element between two states is nonzero, and they arise from angular-momentum conservation, parity, and the form of the atom-field interaction. For electric-dipole transitions between fine-structure states, the most important selection rules are 
\begin{align} \notag 
\Delta\ell &= \pm 1, \:\:
\Delta j = 0,\pm 1\:(j=0 \nleftrightarrow j=0), \:\: 
\Delta m_j = 0,\pm 1.
\end{align}
The allowed value of \(\Delta m_j\) depends on the polarization of the driving field. A linearly polarized light along the quantization axis drives \(\Delta m_j=0\) transitions, while circularly polarized components drive \(\Delta m_j=\pm1\). These selection rules determine whether a photon with the current frequency can actually couple two atomic states.

\subsubsection{Properties of Rydberg atoms}
Owing to the large principal quantum number \(n\), Rydberg atoms have valence electrons located far from the ionic core. As a result, they can be approximately described by a hydrogen-like model and exhibit several distinctive properties that are highly favorable for electric-field sensing.
    

    
    
    

\begin{table}[!t]
\renewcommand{\arraystretch}{1.15}
\caption{Properties of Rydberg atoms.}
\label{tab:rydberg_properties}
\centering
\setlength{\tabcolsep}{5pt}
\begin{tabular}{lcc}
\hline
\textbf{Property} & \textbf{$n$ Dependence} & \textbf{Cs $(50D_{5/2})$ estimate} \\
\hline
Binding energy
& $n^{-2}$
& $6.0\times10^{-3}\:\mathrm{eV}$ \\
    Energy spacing 
    & $n^{-3}$ 
    & $2.5\times10^{-4}\:\mathrm{eV}$ \\
    
    Orbital radius 
    & $n^{2}$ 
    & $3.4\times10^{3}\:a_0$ \\

    Transition dipole moment
    & $n^{2}$ 
    & $3.4\times10^{3}\:ea_0$ \\
    
    Polarizability 
    & $n^{7}$ 
    & $1.2\times10^{4}\mathrm{MHz\cdot cm^2/V^2}$ \\
    
    Radiative lifetime& $n^{3}$ 
     & $1.0\times10^{2}\:\mu\mathrm{s}$\textsuperscript{*}\\

    Fine-structure interval 
    & $n^{-3}$ 
    & $50$--$100\:\mathrm{MHz}$ \\
    \hline
    \multicolumn{3}{@{}l}{\footnotesize
  \textsuperscript{*}\,Evaluated at $T=0~\mathrm{K}$.} \\
\end{tabular}

\end{table}

First, the spatial size of a Rydberg atom increases rapidly with \(n\). In a hydrogen-like picture, the average orbital radius scales as
\(\langle r\rangle \propto a_0n^2\), where \(a_0 \approx 0.053~\mathrm{nm}\) is the Bohr radius. Clearly, high-\(n\) Rydberg atoms are orders of magnitude larger than ground-state atoms. 
The transition dipole moment between neighboring Rydberg states also grows approximately as
\(
d \propto e a_0 n^2
\), with $e$ {being} the elementary charge. For example, the dipole moment is roughly \(6.4\times 10^3\) {D} for \(n=50\), and about \(2.5\times 10^4\) {D} for \(n=100\).
Such large dipole moments allow Rydberg atoms to couple strongly to external electromagnetic fields. 

Moreover, the energy spacing between adjacent Rydberg states decreases rapidly with \(n\). Since the binding energy scales approximately as
\(E_n \propto -\frac{1}{n^2},\) the adjacent level spacing follows
\(|E_n - E_{n+1}|\propto \frac{1}{n^3}.\) 
Because of this narrowed energy-level spacing, transitions between Rydberg states can be tuned into resonance with many frequency bands used in modern wireless systems, spanning from MHz to THz frequencies. For example, the transition $\ket{60D_{5/2}}\rightarrow \ket{63P_{3/2}}$ couples to 60.8 GHz{, while the} transition $\ket{59D_{5/2}}\rightarrow \ket{60P_{3/2}}$ couples to 3.39 GHz. 
This broad tunability is one {of the} major {reasons} why Rydberg atoms are attractive for wireless communication and sensing. 

Other properties also show favorable scaling. 
 The radiative lifetime of a Rydberg state increases approximately as
 \(
 \tau \propto n^3,
 \) allowing highly excited atoms to persist long enough for coherent atom-light interaction. 
The polarizability grows even faster, following
\(
\alpha \propto n^7,
\) which makes Rydberg energy levels highly responsive to low-frequency and off-resonant electric fields through Stark shifts. A summary of these Rydberg-atom properties is presented in Table~\ref{tab:rydberg_properties}. 
Together, these features make Rydberg atoms a powerful platform for quantum-based electric-field receivers.

\subsection{Principles of Rydberg atomic receivers}
\begin{figure*}
	\centering
	\includegraphics[width=6.5in]{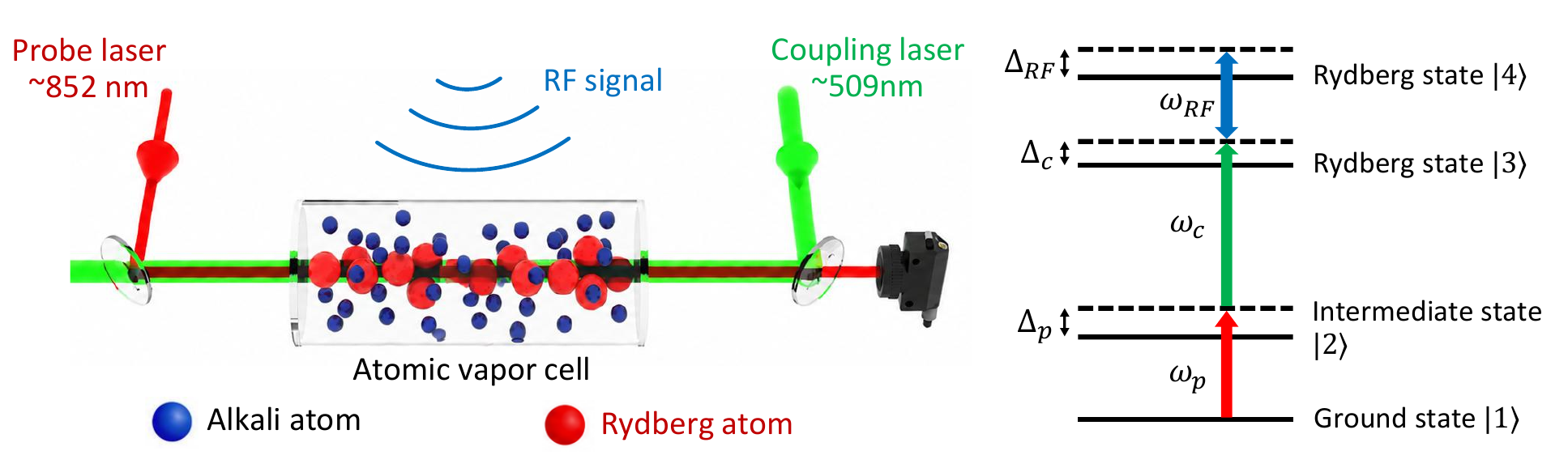}
	\vspace*{-1em}
	\caption{Architecture of a typical Rydberg atomic receiver and the associated four-level quantum system.
    } 
	\vspace*{-1em}
	\label{fig:receiver_architecture}
\end{figure*}
\subsubsection{Preparation of Rydberg atoms}
For electromagnetic (EM) signal detection with Rydberg atomic receivers (RAREs), the first step is to excite alkali atoms into a selected Rydberg state that couples strongly to the incident {EM} field.
Direct one-photon excitation from the ground state to a Rydberg state is generally inefficient and may require high laser power because of the small wavefunction overlap between the ground and Rydberg states.
In practice, Rydberg states are commonly prepared via a two-photon ladder-type excitation process, as illustrated in Fig.~\ref{fig:receiver_architecture}. 
In this configuration, a weak probe laser and a strong coupling laser are overlapped inside an alkali-atom vapor cell. 
The probe laser drives the transition from the ground state $\ket{1}$ to an intermediate state $\ket{2}$, while the coupling {laser} further excites the atom from $\ket{2}$ to a high-lying Rydberg state $\ket{3}$.
For rubidium atoms, typical excitation pathways are $5S_{1/2}\rightarrow 5P_{3/2}\rightarrow nS_{1/2},nD_{5/2}$, with the probe laser near 780~nm and the coupling laser near 480~nm. 
For cesium atoms, the transitions are $6S_{1/2}\rightarrow 6P_{3/2}\rightarrow nS_{1/2},nD_{5/2}$, with probe and coupling wavelengths of approximately 852 nm and 509 nm, respectively. 
The exact coupling wavelength depends on the selected Rydberg level.

Moreover, the probe and coupling lasers are commonly arranged to propagate in opposite directions. 
This configuration can partially suppress the Doppler broadening because the two-photon Doppler shift depends on the residual wave-vector mismatch, $\Delta = (k_p-k_c)v$, where $k_p$ and $k_c$ denote the wavenumbers of counter-propagating probe and coupling lasers, and $v$ is the atomic velocity {of thermal motion}. 
Several recent studies have also employed three-photon excitation schemes to further reduce Doppler broadening.  For example, cesium atoms are excited via $6S_{1/2}\rightarrow6P_{3/2}\rightarrow9S_{1/2} \rightarrow nP_{3/2}$ using probe, dressing, and coupling lasers at 895 nm, 636 nm, and 2245~nm, respectively  \cite{Three-photon-Prajapati2023}. 
Once the Rydberg-state population and coherence are established, the atoms become highly sensitive to incident EM fields and can therefore serve as wireless sensors.

\subsubsection{Electromagnetically induced transparency (EIT)}
EIT is a coherent quantum effect that occurs in three-level or multi-level atomic systems, making the atomic medium transparent to the probe laser within a narrow spectral window~\cite{RevModPhys.77.633}. In RAREs, the EIT effect provides a real-time and nondestructive optical method for reading out the response of Rydberg atoms {by measuring} the transmitted probe laser.

\begin{figure*}
	\centering
	\includegraphics[width=6.5in]{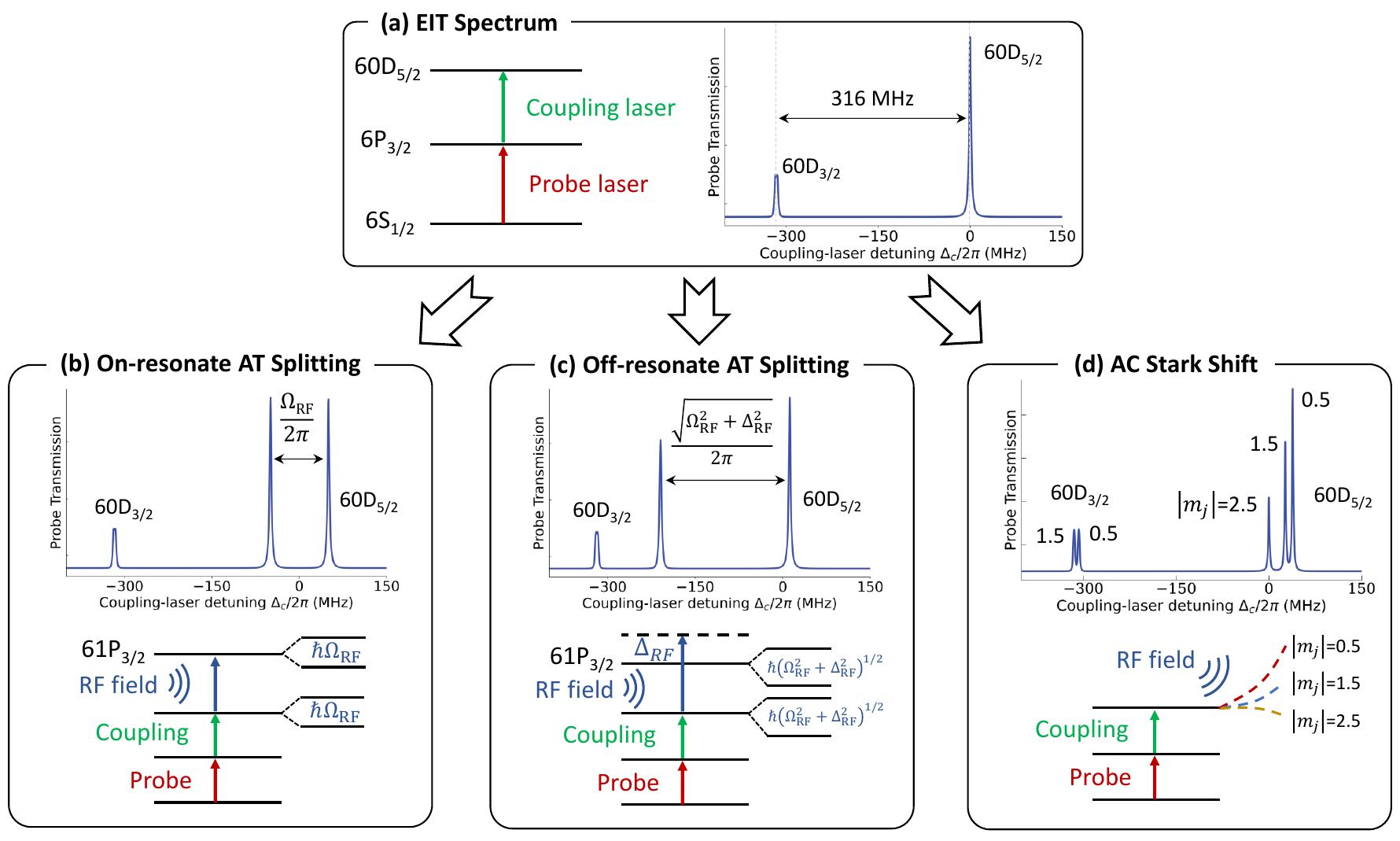}
	\vspace*{-1em}
	\caption{Key mechanisms of a RARE: (a) the EIT effect, (b) on-resonance AT splitting, (c) off-resonance AT splitting, and (d) AC Stark shift. 
    } 
	\vspace*{-1em}
	\label{img:mechanism}
\end{figure*}

Specifically, in an ordinary two-level system driven only by the probe laser, atoms absorb probe photons and are excited from $\ket{1}$ to $\ket{2}$. Thus, the probe light undergoes strong attenuation. 
By contrast, in a three-level quantum system, the additional coupling laser induces two excitation pathways: $\ket{1}\rightarrow\ket{2}$ and $\ket{2}\rightarrow \ket{3}$. When the two lasers satisfy the two-photon resonance condition and the coupling laser is sufficiently strong, the atoms are pumped into a coherent superposition of the ground state $\ket{1}$ and the Rydberg state $\ket{3}$, known as a \emph{dark state}~\cite{RevModPhys.77.633}.
In the ideal dark state, the intermediate state $\ket{2}$ has zero population, and probe absorption through $\ket{2}$ is strongly suppressed.
Consequently, the probe light can pass through the atomic vapor cell with greatly reduced absorption, giving rise to EIT. 
When the frequency of the probe or coupling laser is scanned, a narrow transmission window with a sharp peak thus appears near the resonance. 
An example {of} EIT spectrum obtained by scanning the coupling laser is shown in Fig.~\ref{img:mechanism}(a), where the excitation pathways ${6S_{1/2}}\rightarrow{6P_{3/2}}\rightarrow{60D_{5/2}}$ and ${6S_{1/2}}\rightarrow{6P_{3/2}}\rightarrow{60D_{3/2}}$ are considered. 
An EIT peak appears whenever the lasers are resonant with one of these pathways. The peak associated with ${60D_{5/2}}$ is much stronger than that associated with ${60D_{3/2}}$, because the ${6P_{3/2}}\rightarrow{60D_{5/2}}$ path has a larger transition dipole moment than the ${6P_{3/2}}\rightarrow{60D_{3/2}}$ path.

EIT provides the fundamental optical approach for reading out the Rydberg state. In particular, when external fields perturb Rydberg states, these changes are mapped onto the measured EIT spectrum, as discussed in the next subsection. 
In addition to the two-photon EIT process described above, several studies have exploited quantum interference in three-photon-excitation systems. 
In such systems, both EIT and electromagnetically induced absorption (EIA) can occur, with EIA corresponding to an enhancement of probe absorption. 
Three-photon schemes therefore provide additional readout mechanisms in which changes in the Rydberg state can be detected through either transparency or absorption features.
In this survey, we mainly focus on EIT/EIA-based readout because most RARE-based communication and sensing applications are built on them. Other readout approaches have also been explored recently, including readout of coherent Rabi oscillations~\cite{romalis2024vaporcellrydbergatom}, six-wave mixing~\cite{Sixwavemixing_Han2018}, and fluorescence detection~\cite{Fluorescence-Prajapati2024}. Interested readers may refer to the review article~\cite{allinson2026rydbergreceiversspaceapplications}.

\subsubsection{Autler-Townes (AT) splitting}

When an EM field interacts with Rydberg atoms, it can shift, mix, or split the Rydberg energy levels. With Rydberg-EIT readout, these field-induced changes are mapped onto the probe transmission spectrum, allowing the external field to be inferred optically. {The resonant response, namely Autler-Townes (AT) splitting, is explained as follows}.

AT splitting occurs when an incident EM field is resonant, or nearly resonant, with a transition between two atomic energy levels~\cite{Tanasittikosol2011}. In RAREs, the RF field usually couples the prepared Rydberg state $\ket{3}$ to a nearby Rydberg state $\ket{4}$. Under resonant coupling, the two states form microwave-dressed states, which cause the original Rydberg level to split into two dressed sub-levels, as illustrated in Fig.~\ref{img:mechanism}(b). The energy separation {$\Delta U_{\rm AT}$} between the two sub-levels is proportional to the field amplitude $E_{\rm RF}$: 
\begin{align}
    \Delta U_{\rm AT} =\hbar \Omega_{\rm RF} = \mu_{34} E_{\rm RF},
\end{align}
where {$\hbar=\frac{h}{2\pi}$ is the reduced Planck constant,} $\Omega_{\rm RF} = \frac{\mu_{34}}{\hbar}E_{\rm RF}$ is the Rabi frequency induced by the incident field, {and} $\mu_{34}$ {is} the transition dipole moment associated with $\ket{3}\rightarrow \ket{4}$.  
{Therefore, measuring the AT splitting yields a direct, calibration‑free determination of the RF field amplitude, relying only on fundamental physical constants.}

The AT splitting can be optically measured using the Rydberg-EIT approach. When the state $\ket{3}$ is dressed by the EM field and splits into two dressed sub-levels, the original EIT pathway $\ket{1}\rightarrow\ket{2}\rightarrow\ket{3}$ is divided into two possible excitation pathways. Consequently, a single EIT transmission peak splits into two peaks when the probe or coupling laser frequency is scanned, as shown in Fig.~\ref{img:mechanism}(b). The peak separation, expressed in frequency units, is related to the RF Rabi frequency by
\begin{align}
    \Delta f_{\rm AT} = \left\{
    \begin{array}{cl}
       \frac{\lambda_c}{\lambda_p}\frac{\Omega_{\rm RF}}{2\pi}  &  (\rm probe\:laser\:scanned)\\
      \frac{\Omega_{\rm RF}}{2\pi} & (\rm coupling\:laser\:scanned)
    \end{array}
    \right..
\end{align}
Here, the wavelength ratio $(\lambda_c/\lambda_p)$ {compensates} the Doppler mismatch between the probe and coupling fields when the probe laser is scanned.
After measuring the peak separation, the RF field amplitude {is} recovered as $E_{\rm RF} = \frac{2\pi\hbar}{\mu_{34}} \Delta f_{\rm AT}$ {when the} coupling laser is scanned{, and as} $E_{\rm RF} = \frac{2\pi\hbar\lambda_p}{\mu_{34}\lambda_c} \Delta f_{\rm AT}$ {when the} probe laser is scanned. 
A key advantage of EIT-AT electrometry is that it directly relates the field amplitude to atomic parameters, such as $\hbar$ and $\mu_{34}$. This makes the measurement SI-traceable and, in principle, calibration-free. Owing to this property, EIT-AT-based Rydberg sensors have been widely used in antenna calibration and subwavelength near-field imaging~\cite{NF2}.

When the incident field is detuned from the Rydberg transition, the AT splitting is governed by the generalized Rabi frequency. 
{Denoting} the angular-frequency detuning {as} $\Delta_{\rm RF}$, the dressed-state {energy} separation becomes $\Delta U_{\rm AT} =\hbar\sqrt{\Omega_{\rm RF}^2 + \Delta_{\rm RF}^2}$, which {captures the combined dependence on both the field amplitude and the detuning.} 
Therefore, in principle, EIT-AT spectra can be used to extract both the frequency and strength of the incident field. 
In practice, however, the AT spectrum becomes increasingly asymmetric as the detuning increases. 
As shown in Fig.~\ref{img:mechanism}(c), one of the two peaks can rapidly {diminish}, {causing} the resolved AT doublet {to} eventually {collapse} into a single shifted peak. 
In this {regime}, the system is {more accurately} described {as} an off-resonant AC Stark shift rather than {as} AT splitting. 
For this reason, EIT-AT measurements are mainly suitable for resonant or near-resonant fields, and the detuning is typically required to be within about 10 MHz of the selected Rydberg transition.

The sensitivity of EIT-AT electrometry is limited by the EIT linewidth $\sigma_{\rm EIT}$. 
The two peaks can be resolved only when the induced splitting exceeds this linewidth; otherwise, the spectrum {collapses into} a single, weakened, and broadened peak. 
In warm vapor cells, the EIT linewidth is {dominated} by Doppler broadening. 
Counter-propagating probe and coupling lasers partially suppress this broadening, thereby {narrowing} the EIT linewidth. 
Three-photon excitation schemes can further reduce the linewidth. For example, a linewidth of $2\pi\times 39$~kHz has been demonstrated in a three-photon scheme, compared with $2\pi\times 3.5$~MHz in a conventional two-photon scheme~\cite{Three-photon-Prajapati2023}.
In addition, balanced detection is {commonly} employed to further improve signal quality~\cite{Comm-PM-Jing2020}. 
{In this method, subtracting a reference probe-absorption signal from the two‑photon signal removes the Doppler‑broadened background that appears during probe scanning and suppresses common‑mode laser power noise.}

\subsubsection{Stark shift} 
EIT-AT splitting is mainly applicable to discrete frequencies that are resonant or near-resonant with selected Rydberg transitions.
For {detecting fields with} continuous-{spectra} or ultra-low-frequency, off-resonant Stark shifts are more commonly exploited. 
In this regime, the external field {shifts} the entire Rydberg energy levels through field-induced polarization.  
Depending on the field frequency, this response is {typically classified} as either an AC Stark shift or a DC Stark shift.

Consider an off-resonant oscillating electric field with angular frequency $\omega$ and strength $E_{\rm AC}$.
{Applying} second-order {in time-dependent} perturbation theory, the {AC Stark} shift of a Rydberg state {$\ket{n\ell jm_j}$} has a quadratic dependence on the electric-field amplitude:
\begin{align}
    \Delta U_{\rm AC} = -\frac{1}{4} \alpha(\omega) E_{\rm AC}^2. 
\end{align}
Here, the polarizability $\alpha(\omega)$ quantifies the sensitivity of {the energy} shift to {the external field} and depends strongly on the selected Rydberg state {and the field frequency}. 
{For a linearly polarized field aligned with the quantization axis, it} can be decomposed into scalar and tensor components as
\begin{align}
    \alpha(\omega) = \alpha_{n\ell j}^{(0)}(\omega) + \alpha_{n\ell j}^{(2)}(\omega)\frac{3m_j^2-j(j+1)}{j(2j-1)}, {~{j\geq1,}}
\end{align}
where $\alpha_{n\ell j}^{(0)}(\omega)$ and $\alpha_{n\ell j}^{(2)}(\omega)$ {are the scalar and tensor polarizabilities, respectively.}
The $m_j$-dependence of $\alpha(\omega)$ can lead to the splitting of magnetic sub-levels. 
Specifically, an $nS_{1/2}$ Rydberg state mainly experiences an overall energy shift because the $S_{1/2}$ state has only $|m_j|=1/2$ magnetic sub-levels. 
In contrast, an $nD_{5/2}$ manifold contains {three groups of} magnetic sub-levels, $|m_j| = 1/2, 3/2, 5/2$. When an off-resonant field is applied, this state can exhibit both level shifts and splittings because different $|m_j|$ components have different polarizabilities. 
As shown {for} the $n = 60$ manifold in Fig.~\ref{img:mechanism}(d), the EIT peak {associated with} $\ket{60D_{5/2}}$  splits into three resolvable peaks corresponding to Stark-shifted $\ket{60D_{5/2}, |m_j| = 1/2, 3/2, 5/2}$  {sub-levels}, while {the peak associated with} $\ket{60D_{3/2}}$ splits into {two peaks corresponding to} {$\ket{60D_{3/2}, |m_j| = 1/2}$ and $\ket{60D_{3/2}, |m_j| = 3/2}$}. {Given the field frequency and the corresponding polarizabilities, the} field strength can then be recovered from these spectral shifts according to their quadratic dependence on the electric-field {amplitude}. 
Detection based on AC Stark shifts does not require exact resonance with a particular Rydberg transition. It {can thus} enable {continuous-frequency measurements over a broad spectral range, though the sensitivity remains frequency dependent}. 

For static or slowly varying electric fields, the response is described as the DC Stark shift. In the second-order perturbative regime, the energy shift of a Rydberg state is also quadratic in the field amplitude:
\begin{align}
\Delta U_{\rm DC} = -\frac{1}{2}\alpha(0)E_{\rm DC}^2,
\end{align}
where $\alpha(0)$ is the static polarizability. 
The DC Stark shift therefore provides a straightforward mechanism for measuring {the magnitude of} static electric fields, as well as low-frequency signals typically below 1 MHz. 
A major challenge for low-frequency field detection is the electric-field screening effect of the vapor cell. Alkali atoms can adsorb onto the inner surface of the cell and form a weakly conducting layer, which behaves like a Faraday cage and prevents DC or low-frequency fields from penetrating into the cell. 
To address this issue, {one approach is to apply} the field directly using a pair of electrodes placed inside the vapor cell. Alternatively, sapphire vapor cells can be used because they substantially {reduce the screen rate}. 
 {This extends the time window during which an externally applied low-frequency field can be measured to approximately one second under reported experimental conditions.}

\subsubsection{Superheterodyne approach}
Direct EIT-AT and Stark-shift readout methods face two major limitations. First, they primarily provide access to the amplitude and frequency of the incident RF field, while the phase of the RF signal is not directly measured. 
Second,  their sensitivity is limited by the EIT linewidth. The signal field  {must} be strong enough to produce a resolvable {splitting} or shift {of} the EIT peak, making weak-field detection challenging. To overcome these limitations, the Rydberg superheterodyne receiver, also known as {a} Rydberg atomic mixer, has been developed.

\begin{figure}
	\centering
	\includegraphics[width=\columnwidth]{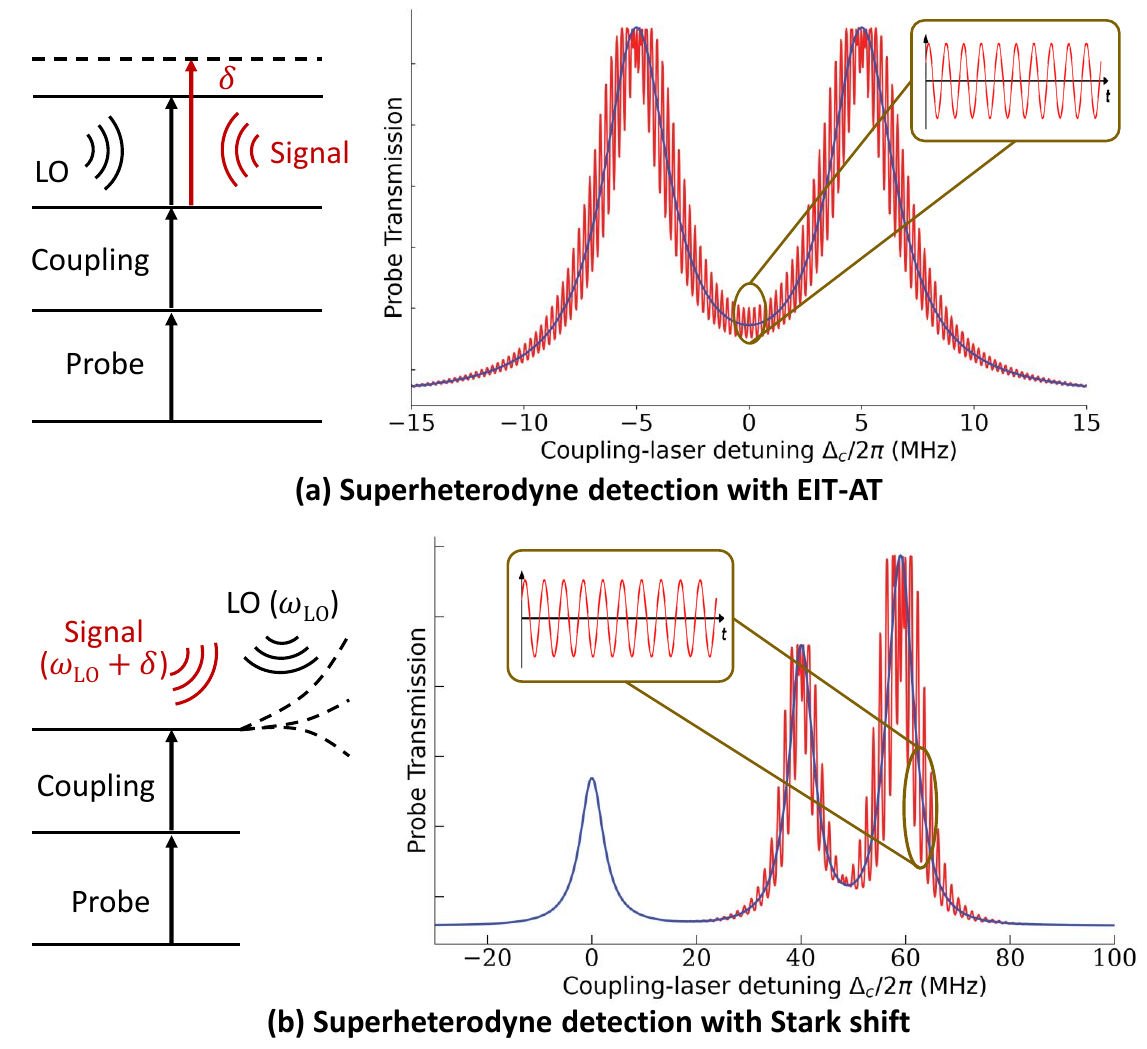}
	\vspace*{-1em}
	\caption{Superheterodyne detection under (a) the EIT-AT effect and (b) AC Stark shift effect. 
    } 
	\vspace*{-1em}
	\label{img:superheterodyne}
\end{figure}

The original Rydberg superheterodyne scheme was inspired by optical holography and was first applied to EIT-AT spectroscopy~\cite{Comm-PM-Simons2019,Comm-PM-Jing2020}.
As shown in Fig.~\ref{img:superheterodyne}, an additional local-oscillator (LO) microwave field with frequency $\omega_{l}$ is introduced to {interact with the atoms in the} vapor cell. 
The LO field is tuned to be resonant or near-resonant with the target Rydberg transition $\ket{3}\rightarrow\ket{4}$, thereby creating microwave-dressed Rydberg states. 
The signal {(SIG)} field to be detected has a frequency $\omega_{s}$ slightly different from the LO field and is much weaker {than the LO field}. The {coherent superposition of} the LO and SIG fields inside the vapor cell therefore produces a slowly varying Rabi frequency
\begin{align}
    \Omega_{\rm RF}(t) = |\Omega_l + \Omega_se^{i(\omega_\delta t +\varphi_\delta)}|
    \approx \Omega_l + \Omega_s\cos{(\omega_\delta t + \varphi_\delta)},
\end{align}
where $\omega_{\delta} = \omega_s - \omega_l$ is the beat {angular} frequency, or intermediate frequency (IF), and $\varphi_\delta = \varphi_s - \varphi_l$ denotes the LO-SIG phase difference. 
The approximation holds in the weak-signal regime, $\Omega_s \ll \Omega_l$. 
Consequently, the Rydberg state experiences a time-varying AT splitting $\hbar\Omega_{\rm RF}(t)$, which appears as a modulation of the transmitted probe power. 
As presented in Fig.~\ref{img:superheterodyne}(a), 
{probe transmission can be viewed as the response of the static EIT-AT spectrum produced by the strong LO field $\Omega_l$, together with}
a small oscillating perturbation proportional to the signal field $\Omega_s\cos(\omega_\delta t + \varphi_\delta)$. 
For signal detection, the laser frequency is usually locked to a fixed point on the EIT-AT spectrum, such as the {resonant transmission point, while the LO strength is optimized to maximize the response to the weak signal}.
The optical output can then be approximated as:
\begin{align}\label{eq:superh}
    P(t) = \bar{P}_0 + \kappa \Omega_s\cos(\omega_\delta t + \varphi_\delta),
\end{align}
where $\bar{P}_0$ denotes the static probe transmission {at the operating point}, and $\kappa = {\partial \bar{P}_0}/{\partial \Omega_l}$ is the {linear response coefficient evaluated at that point.}
A key observation from \eqref{eq:superh} is that the AC component of the optical readout directly carries the relative phase between the signal and LO fields. 
Since the LO phase is known and controllable, the phase of the incident signal can be recovered from the optical beat note.
{Moreover, the weak signal is converted linearly into an optical modulation with a conversion coefficient $\kappa$. Optimizing $\kappa$ and shifting the readout to a nonzero IF enable high-sensitivity detection.}
In the original work~\cite{Comm-PM-Jing2020}, a sensitivity down to 55~$\rm nV/cm/\sqrt{Hz}$ was achieved. 
Owing to its phase-sensitive and high-sensitivity capabilities, the Rydberg superheterodyne receiver has become a widely used technique in subsequent sensing and communication applications.

Another essential point is that the LO field strength is usually chosen to maximize the linear coefficient $\kappa$ and hence the detection sensitivity. 
{Within the simplified line-shape model used in the original work}~\cite{Comm-PM-Jing2020}, the LO Rabi frequency was set as
\begin{align}
    \Omega_l^\star = \Gamma_{\rm EIT}/\sqrt{3},
\end{align}
where $\Gamma_{\rm EIT}$ {denotes an effective spectral-width parameter associated with the EIT linewidth. The precise relation between $\Gamma_{\rm EIT}$ and the experimentally measured full width at half maximum depends on the linewidth convention and line-shape model.}
The underlying intuition is that the optimal LO field should split the EIT peak such that the locked laser frequency lies {in the steep-response region of the two dressed EIT peaks, thereby maximizing the change in probe transmission with respect to the LO Rabi frequency.}
Subsequent study in \cite{Comm-MB-Cui2026} further refined the optimal LO strength by maximizing the overall signal-to-noise ratio (SNR) in the presence of photon shot noise (PSN) and black-body radiation (BBR) noise. The optimal Rabi frequency is modified as 
\begin{align}
    \Omega_l^\star = \sqrt{\frac{\chi_0+4-\sqrt{\chi_0^2 +4\chi_0+16}}{\chi_0-4+\sqrt{\chi_0^2+4\chi_0+16}}}\Gamma_{\rm EIT},
\end{align}
where $\chi_0$ is the laser attenuation coefficient. In addition, \cite{Comm-Theory-Gong2026} showed that the LO detuning, laser powers, and laser detunings also affect the linear coefficient $\kappa$. These parameters are therefore jointly optimized through numerical search to achieve the best sensitivity.

Although the superheterodyne technique {was originally demonstrated using} resonant EIT-AT detection, the same principle has been extended to AC- and DC-Stark-shift-based sensing.
In the presence of a strong LO {or bias} field, the Stark shift becomes time-dependent:
\begin{align}
    \Delta U &= \eta[E_l + E_s\cos(\omega_\delta t +\varphi_\delta)]^2 \notag \\
    &\approx \eta E_l^2 + 2\eta E_lE_s\cos(\omega_\delta t + \varphi_\delta), 
\end{align}
where $\eta = -\frac{1}{4}\alpha(\omega)$ for {two closely spaced high-frequency fields, with $E_l$ and $E_s$ defined as their peak amplitudes}, and $\eta=-\alpha(0)/2$ for a {static or quasi-static bias field}.
The cross term $2\eta E_lE_s\cos(\omega_\delta t+\varphi_\delta)$ converts the weak signal into a time-varying Stark shift at the beat frequency $\omega_\delta$ and {enhances the linear response by a factor proportional to the LO-field strength,}
as illustrated in Fig.~\ref{img:superheterodyne}(b). 
By locking the laser to the 
{steepest-slope point of the EIT spectrum,}
the weak signal can be extracted from the oscillating probe transmission with high sensitivity. 
For example, 
{continuous carrier-frequency coverage from DC} to 20 GHz was demonstrated using an AC-Stark-based superheterodyne receiver in the milestone work of \cite{PhysRevApplied.15.014053}.  
A similar {DC-bias-assisted Stark-detection approach} was also applied to ultra-low-frequency field detection at $1$~kHz, which achieved a sensitivity of $5.7\:{\rm \mu V/cm/\sqrt{Hz}}$~\cite{lei2024high}.

\subsection{Signal Modeling}
\label{sec:signal_modeling}

\begin{table*}[!t]
    \centering
    \hypersetup{hidelinks}
    \caption{Representative Signal Models for Rydberg Atomic Receivers}
    \label{tab:signal_models}
    \scriptsize
    \renewcommand{\arraystretch}{1.0}
    \setlength{\tabcolsep}{2pt}
    \setlength{\extrarowheight}{0pt}
    \setlength{\arrayrulewidth}{0.35pt}
    \begin{tabularx}{\textwidth}{|>{\centering\arraybackslash}p{3.0cm}|>{\centering\arraybackslash}p{1.35cm}|>{\raggedright\arraybackslash}X|}
        \hline
        \textbf{Category} & \textbf{Ref.} & \textbf{Main contribution} \\
        \hline
        \multirow{7}{3.0cm}{\centering\makecell{\textbf{Superheterodyne}\\\textbf{model extensions}}}
        & \cite{Comm-Theory-Gong2026} & Joint modeling of probe, coupling, and RF LO detunings \\
        \cline{2-3}
        & \cite{Comm-Theory-Zhu2026} & Closed-form dynamic RF-to-photocurrent transfer function \\
        \cline{2-3}
        & \cite{Comm-BW-Yang2024} & IF bandwidth enhancement via optical sideband amplification \\
        \cline{2-3}
        & \cite{Comm-LinearDynamicRange-Wu2023} & Harmonic generation and linear dynamic range \\
        \cline{2-3}
        & \cite{SignalModel-Qu2026} & Finite-IF response and harmonic distortion in AC Stark readout \\
        \cline{2-3}
        & \cite{Comm-MB-Cui2026} & LO-dependent gain allocation across RF bands \\
        \cline{2-3}
        & \cite{SignalModel-Wang2023} & Noise spectra versus atom number and readout frequency \\
        \hline
        \multirow{5}{3.0cm}{\centering\makecell{\textbf{Alternative transduction}\\\textbf{and readout}}}
        & \cite{FourWaveMixing-Ji2024} & Resonant and off-resonant four-wave mixing models \\
        \cline{2-3}
        & \cite{SixWaveMixing-Borowka2024} & Continuous six-wave mixing in room-temperature vapor \\
        \cline{2-3}
        & \cite{Comm-CB-Chen2026} & Two-pole baseband response for six-wave mixing \\
        \cline{2-3}
        & \cite{Comm-PM-Borowka2025} & RF quadrature readout with optical bias and phase-noise correction \\
        \cline{2-3}
        & \cite{Sensing-Radar-CuiSelfHet2026} & Self-heterodyne autocorrelation using transmit leakage as the reference \\
        \hline
    \end{tabularx}
\end{table*}

Jing et al.~\cite{Comm-PM-Jing2020} demonstrated that a microwave-dressed
Rydberg transition converts a weak RF tone into an intermediate-frequency
modulation of the transmitted probe power, as described by
\eqref{eq:superh}. We use this small-signal response to model the
reception of a modulated waveform in a single RF band. Consider the
resonant four-level EIT-AT receiver in Fig.~\ref{img:superheterodyne}(a).
The probe, coupling, and RF LO fields are tuned to their respective
transitions. For negligible Rydberg-state dephasing, the steady-state
probe transmission follows the line shape used in~\cite{Comm-MB-Cui2026}.
The probe power and its slope at the LO operating point are
\begin{subequations}\label{eq:superhet_static_response}
\begin{align}
    \bar P_0 &= P_{\rm in}
    \exp\!\left(-\frac{\chi_0\Omega_l^2}{\Omega_l^2+\Gamma_0^2}\right),
    \label{eq:superhet_static_power}\\
    \kappa &= -\bar P_0
    \frac{2\chi_0\Gamma_0^2\Omega_l}{(\Omega_l^2+\Gamma_0^2)^2}.
    \label{eq:superhet_slope}
\end{align}
\end{subequations}
Here, $P_{\rm in}$ is the incident probe power. The line-shape parameter
is $\Gamma_0^2=2\Omega_p^2(\Omega_c^2+\Omega_p^2)/
(\gamma_2^2+2\Omega_p^2)$, where $\Omega_p$ and $\Omega_c$ are the probe
and coupling Rabi frequencies and $\gamma_2$ is the population decay
rate of $\ket{2}$. The dimensionless attenuation parameter is
$\chi_0=2N_0\mu_{12}^2 k_p L\gamma_2/
[\epsilon_0\hbar(\gamma_2^2+2\Omega_p^2)]$, where $N_0$ is the atomic
number density, $\mu_{12}$ is the probe transition dipole moment, $k_p$
is the probe wavenumber, $L$ is the vapor-cell length, and $\epsilon_0$
is the vacuum permittivity.

Let $x(t)$ denote the complex envelope of a waveform normalized to unit
average power. For an isotropic transmitter with average power $P_{\rm t}$
and carrier angular frequency $\omega_s$, a co-polarized free-space far-field
link of length $r$ introduces a delay $\tau=r/c$ and a peak-field
scaling of $\sqrt{Z_0P_{\rm t}/(2\pi r^2)}$. Here, $c$ and $Z_0$ are
the speed of light and the wave impedance in free space, respectively.
The RF signal produces a nonzero IF $\omega_\delta=\omega_s-\omega_l$.
The slope in \eqref{eq:superhet_slope} applies when the peak signal Rabi
frequency is much smaller than $\Omega_l$ and the atomic response is
quasi-static over the occupied IF band. A photodiode of responsivity
$\mathcal R$ and a flat transimpedance gain $R_{\rm T}$ convert the
probe-power modulation into the IF voltage. Coherent I/Q demodulation
referenced to the RF LO phase yields
\begin{align}
    y(t) &= \mathcal R R_{\rm T}\frac{\kappa\mu_{34}}{\hbar}
        \sqrt{\frac{Z_0 P_{\rm t}}{2\pi r^2}}\,
        e^{-i\omega_s\tau}x(t-\tau) + z(t),
    \label{eq:superhet_baseband}
\end{align}
where $z(t)$ is the equivalent baseband voltage noise. The I/Q
demodulator is normalized to recover the complex envelope of the real IF
voltage and reject its image. When photon shot noise dominates, the
mean photocurrent $\mathcal R\bar P_0$ sets its noise spectral density.
For an ideal baseband filter of total two-sided bandwidth $B$, the
corresponding complex noise variance is
\begin{equation}\label{eq:superhet_shot_noise}
    \mathbb E[|z(t)|^2] = 4q\mathcal R\bar P_0 R_{\rm T}^2 B,
\end{equation}
where $q$ is the elementary charge.

The conversion gain in \eqref{eq:superhet_baseband} depends on the
atomic operating point. Gong et al.~\cite{Comm-Theory-Gong2026}
incorporated probe, coupling, and RF LO detunings into the four-level
response and derived the corresponding baseband models for direct and
balanced coherent optical detection. At a fixed operating point,
the gain can also vary across the IF band. Zhu and
Dai~\cite{Comm-Theory-Zhu2026} linearized the time-dependent Lindblad
equation to obtain a frequency-dependent transfer function from the RF
field to the photocurrent. The dynamic response also explains how optical excitation can change
the usable IF band. Yang et al.~\cite{Comm-BW-Yang2024} showed that a
strong coupling field can amplify an optical sideband and increase the
instantaneous IF bandwidth.

The dynamic small-signal model still requires the signal amplitude to
lie within the linear range. Wu et
al.~\cite{Comm-LinearDynamicRange-Wu2023} analyzed how stronger inputs
generate harmonics and how the probe Rabi frequency affects the linear
dynamic range. For a continuously tunable AC Stark receiver, Qu and
Zhang~\cite{SignalModel-Qu2026} derived the finite-IF response and used
Floquet theory to quantify harmonic distortion. Cui et
al.~\cite{Comm-MB-Cui2026} developed a multiband superheterodyne model
in which several RF bands share one optical readout. The conversion gain
of each band depends on the joint LO operating point. At the
weak-signal end, Wang et al.~\cite{SignalModel-Wang2023} measured how
the noise spectrum varies with the number of probed atoms and the
readout frequency.

The models above track modulation of the transmitted probe. In
microwave-to-optical conversion, the RF signal can instead be encoded
in a generated optical field. Ji et al.~\cite{FourWaveMixing-Ji2024}
used the Maxwell--Bloch equations to describe resonant and off-resonant
four-wave mixing, and Bor\'owka et al.~\cite{SixWaveMixing-Borowka2024}
demonstrated continuous six-wave mixing in a thermal vapor. Chen et
al.~\cite{Comm-CB-Chen2026} modeled conversion and detection together to
obtain a two-pole approximation for the six-wave-mixing baseband
response. The choice of reference also changes the measured signal.
Bor\'owka et al.~\cite{Comm-PM-Borowka2025} used an optical bias to recover
the RF quadratures and corrected laser phase noise. For self-heterodyne
sensing, Cui et al.~\cite{Sensing-Radar-CuiSelfHet2026} used transmit
leakage as the reference, so the atomic response depends on the
waveform autocorrelation at the target delay. Table~\ref{tab:signal_models}
summarizes the additional effects modeled and the alternative readout
schemes.

\section{Detection Technologies}

\subsection{Performance Metrics}

Consider a Rydberg atomic receiver tuned to the carrier of a modulated RF signal. The incident electric field changes the transmission of the probe light, and the receiver recovers the waveform from this optical response. Two aspects of the response determine the range of usable signal amplitudes. At low field strengths, the modulation must be distinguishable from fluctuations in the optical readout. At higher field strengths, the output must remain proportional to the input if the receiver is to preserve the waveform. These considerations lead to the first two metrics in Table~\ref{tab:receiver_metrics}: electric-field sensitivity and linear dynamic range.

\begin{table}[t]
\centering
\caption{Key performance metrics of Rydberg atomic receivers.}
\label{tab:receiver_metrics}
\footnotesize
\begin{tabularx}{\columnwidth}{@{}p{0.32\columnwidth}X@{}}
\toprule
\textbf{Metric} & \textbf{Definition} \\
\midrule
Electric-field sensitivity & Input field noise amplitude spectral density, expressed as a noise-equivalent field (NEF) in $\mathrm{V\,m^{-1}\,Hz^{-1/2}}$. \\
Linear dynamic range & Ratio of the largest field within a specified linearity tolerance to the smallest detectable field in a given measurement bandwidth. \\
Instantaneous bandwidth & Width of the signal band that a single channel can receive at a fixed operating point within a specified response tolerance. \\
Carrier-frequency tuning range & Continuous interval over which the receiver can tune its operating frequency and detect a signal at a specified field strength and signal-to-noise ratio. \\
\bottomrule
\end{tabularx}
\end{table}

\begin{figure*}[t]
\centering
\includegraphics[width=\textwidth]{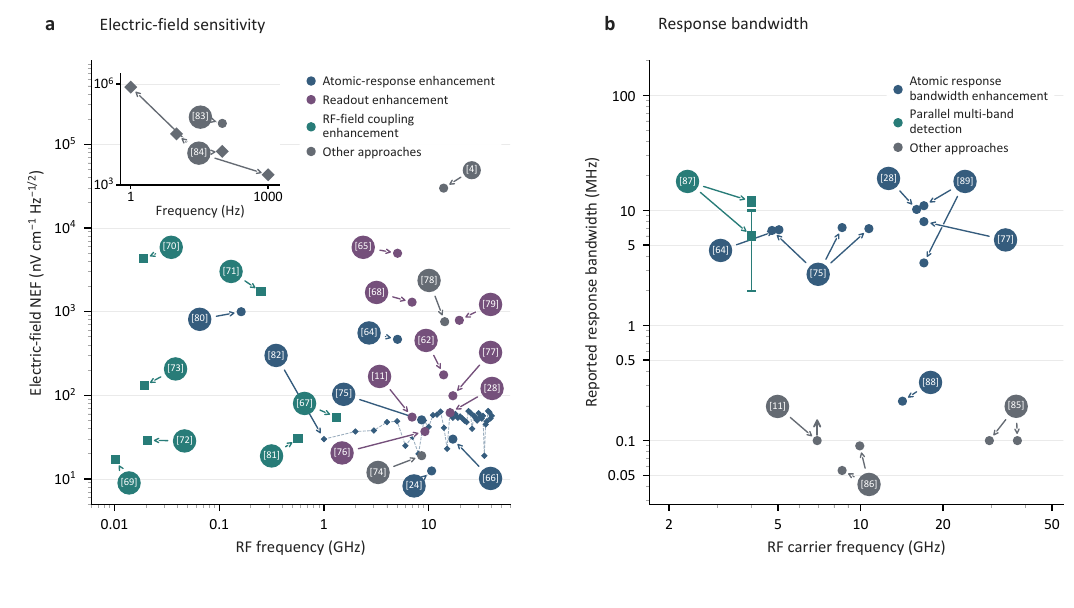}
\vspace{-13mm}
\caption{Experimental performance of Rydberg atomic receivers across RF frequencies.
(a) Electric-field noise-equivalent field (NEF)~\cite{Intro-Sedlacek2012,Perf-Homodyne-Kumar2017,Comm-PM-Jing2020,Perf-Repumping-Prajapati2021,Sensitivity-Cai2022Photonics,Eng-RF-SRR-Holloway2022,Sensing-Bandwidth-Hu2023,Perf-MZI-Yang2023,Perf-SW-Mao2024,Perf-HF-Mao2024,Eng-RF-Yang2024,Comm-BW-Yang2024,Perf-MSRR-Wan2026,Perf-TMR-Wan2025,Eng-Multi-Wu2025,Comm-PM-Borowka2025,Perf-Bandwidth-Wang2026,Perf-Readout-Shylla2026,Sensing-Bandwidth-Manchaiah2026,Perf-Heterodyne-Su2026,Perf-Mixer-Gordon2019,Perf-HAMMER-Prajapati2024,Perf-BFRR-Yang2026,Comm-CB-Yao2026}; inset: low-frequency electrometry~\cite{Perf-LF-Li2023,Perf-LF-Chandra2026}.
(b) Response bandwidth~\cite{Comm-PM-Jing2020,Comm-AM/FM-Anderson2021,Sensing-Bandwidth-Hu2023,Comm-AM/FM-Yang2023,Perf-Comb-ArtusioGlimpse2024,Comm-AM/FM-Yuan2024,Comm-BW-Yang2024,Sensing-Sideband-Shylla2025,Perf-Bandwidth-Wang2026,Sensing-Bandwidth-Manchaiah2026}.
Colors indicate enhancement methods. Connected diamonds represent separately tuned measurements. Error bars show reported uncertainties; the upward arrow indicates a bandwidth lower bound.
The panels use independently optimized configurations. Field-enhanced NEFs follow external-field calibration. Bandwidths use source-specific criteria~\cite{Perf-Comb-ArtusioGlimpse2024,Sensing-Bandwidth-Manchaiah2026}; the 10.2-MHz value is detector-corrected~\cite{Comm-BW-Yang2024}. The spectrum-derived equivalent NEF~\cite{Perf-HAMMER-Prajapati2024} and preprint values~\cite{Sensing-Bandwidth-Manchaiah2026,Perf-Heterodyne-Su2026,Perf-LF-Chandra2026,Sensing-Sideband-Shylla2025} are included.}
\label{fig:receiver_performance_overview}
\end{figure*}

Electric-field sensitivity is commonly reported as a noise-equivalent field (NEF). A calibrated weak RF signal first establishes the change in output per unit electric field. Dividing the output noise amplitude spectral density by this response slope gives the NEF. The field used for calibration determines what the reported value describes. For example, a resonator can increase the field inside the vapor cell for a given incident field. Comparing the complete receiver with another receiver therefore requires the NEF referred to the incident field~\cite{Sensing-Backes2024}. Figure~\ref{fig:receiver_performance_overview}(a) compares reported NEFs across RF carrier frequencies. The smallest field detectable in a particular measurement also depends on the measurement bandwidth and the required signal-to-noise ratio. This field sets the lower end of the linear dynamic range. To find the upper end, the applied field is increased until the output departs from its small-signal slope by a specified amount, such as 1~dB of gain compression~\cite{Comm-LinearDynamicRange-Wu2023}.

The frequency performance of the receiver has two distinct aspects. The first is its \emph{instantaneous bandwidth}: how wide a signal it can receive while tuned to one carrier. Consider an RF carrier whose amplitude varies sinusoidally. At low modulation frequencies, the probe transmission follows the changes in field amplitude. As the modulation becomes faster, the finite response times of the atoms and the readout electronics can reduce the output modulation. This response can be measured by holding the carrier frequency and modulation depth fixed and sweeping the modulation frequency. For a low-pass response, the one-sided 3-dB bandwidth is the modulation frequency at which the output amplitude falls to $1/\sqrt{2}$ of its low-frequency value~\cite{Sensing-Bandwidth-Hu2023}. Figure~\ref{fig:receiver_performance_overview}(b) compares reported response bandwidths across RF carrier frequencies.

The second aspect is the \emph{carrier-frequency tuning range}: how far the carrier itself can be moved while the receiver continues to recover the modulation. Different pairs of Rydberg states provide resonant reception at different carrier frequencies. Selecting among these transitions gives access to widely separated bands, but the sensitivity can decrease between resonances. To measure a continuous tuning range, the carrier is swept with the modulation held fixed, and the receiver settings are adjusted as needed. The interval over which the signal meets the chosen detection threshold at a given incident field defines the usable range.

\subsection{Sensitivity Enhancement}

Sensitivity characterizes the minimum RF electric field that can be resolved above the measurement noise and is commonly expressed as a noise-equivalent electric field in units of $\mathrm{V\,m^{-1}\,Hz^{-1/2}}$. In a Rydberg atomic receiver, the RF field couples Rydberg states and is transduced into a change in the optical transmission or phase through the atomic response. The achievable sensitivity is therefore governed by the RF-to-optical response slope and the noise of the optical readout. Although Rydberg transitions provide large electric-dipole moments and strong RF coupling, practical sensitivity is still limited by the finite atomic response, readout noise, and the efficiency with which the incident RF field is coupled into the atomic sensing volume.

Accordingly, sensitivity can be improved at three points along the detection chain: by increasing the atomic response to a given local RF field, reducing the noise introduced during optical readout, or increasing the local RF field produced by a given incident signal. The following subsections review these approaches as atomic-response enhancement, readout-noise reduction, and RF-field coupling enhancement.

\subsubsection{Atomic-Response Enhancement}

Atomic-response enhancement aims to increase the change in the optical signal produced by a given RF field before the readout noise is considered. Early Rydberg electrometry relied primarily on spectroscopic changes induced by RF coupling. Rydberg-EIT spectroscopy established how the optical powers affect the contrast and linewidth of the Rydberg resonance~\cite{HighSensitivity-Zhao2009}. Sedlacek et al.~\cite{Intro-Sedlacek2012} subsequently showed that an RF field coupling two Rydberg states produces Autler--Townes (AT) features from which the field strength can be determined through the corresponding transition dipole moment. This direct spectral readout is effective when the RF-induced splitting is well resolved, but becomes less effective as the splitting approaches the EIT linewidth.

One way to improve weak-field detection is to modify the spectral response. Three-photon excitation can suppress Doppler and recoil broadening and produce narrower Rydberg resonances with longer coherence times~\cite{ThreePhoton-Ryabtsev2011}. For two-photon Rydberg EIT, Schlossberger et al.~\cite{Linewidth-Schlossberger2026} derived the residual Doppler contribution and measured a linewidth close to the predicted limit in a warm vapor cell. Comparisons of two- and three-photon electrometry show that narrower resonances do not necessarily improve field sensitivity~\cite{Three-photon-Prajapati2023}. The optical response to a weak RF field also depends on laser intensities, optical depth, detuning, and absorption. Simons et al.~\cite{Sensing-ContTune-Simons2016} used off-resonant AT spectra to infer weak fields that were difficult to resolve on resonance. Meyer et al.~\cite{OptimalEIT-Meyer2021} identified optimum EIT operating conditions and a sensitivity limit associated with absorption loss.

The RF-dependent optical response can also be modified through the choice of atomic states and population distribution. Chopinaud and Pritchard~\cite{OptimalState-Chopinaud2021} compared several microwave-coupled Rydberg-state pairs and showed that neighboring transitions can distort the symmetry and linearity of the AT spectrum. Appropriate state selection can therefore provide a more robust field-dependent response. Prajapati et al.~\cite{Repumping-Prajapati2021} increased the number of atoms participating in the EIT process using a ground-state repumping laser. The EIT amplitude was nearly doubled without a comparable increase in linewidth, and the measured field sensitivity improved by approximately a factor of two. Related schemes have combined optical pumping, multilevel excitation, and detuning control to redistribute atomic population and strengthen the microwave-induced optical response~\cite{Sensitivity-Cai2022,Xue-Multilevel2025}.

Interactions between Rydberg atoms provide a different mechanism for increasing the response. Near the critical point of a driven many-body system, a small microwave perturbation can produce a collective change in the Rydberg population and optical transmission. Ding et al.~\cite{Critical-Ding2022} quantified this enhancement using the Fisher information and reported an increase of about three orders of magnitude relative to an independent-particle system. Here, the amplification arises from the collective susceptibility of the interacting ensemble rather than from the response of an individual atomic transition.

\subsubsection{Readout Enhancement}

Even when a measurable atomic response is produced, the receiver sensitivity still depends on how efficiently that response is extracted from the optical readout. In conventional EIT--AT electrometry, the RF field is inferred from changes in the transmission, phase, or spectral structure of the probe light. For weak fields, these changes can be small relative to the optical background and its associated noise. Readout enhancement therefore seeks either to extract more information from the RF-induced atomic response or to convert that information into an optical observable with a higher signal-to-noise ratio.

One approach is to make better use of the information carried by the optical field. Conventional transmission measurements mainly probe the absorptive response, although the RF field modifies both the absorptive and dispersive components of the atomic susceptibility. Yang et al.~\cite{Comm-AM/FM-Yang2023} used a Mach--Zehnder interferometer to measure both quadratures of the probe field and optimized their combination for microwave electrometry. The full complex susceptibility provided a sensitivity improvement of approximately $12$~dB over conventional transmission readout. The optical background can also be changed by choosing a different observable. Prajapati et al.~\cite{Fluorescence-Prajapati2024} compared transmission and fluorescence readout in a three-photon Rydberg system. Fluorescence isolates the Rydberg-dependent signal from the much stronger transmitted probe field and improved the electric-field sensitivity by nearly a factor of four over the corresponding EIT/EIA measurement.

A different strategy is to convert the weak RF response into a coherent intermediate-frequency signal. Simons et al.~\cite{Comm-PM-Simons2019} demonstrated that Rydberg atoms can act as an RF mixer: a signal field and a nearby reference field produce a low-frequency optical beat whose phase follows their relative RF phase. Jing et al.~\cite{Comm-PM-Jing2020} developed this principle into an atomic superheterodyne receiver. A strong microwave local oscillator dresses the Rydberg transition, and the weak signal field produces a coherent optical beat at their difference frequency. The weak RF signal can therefore be detected at a selected intermediate frequency rather than through a small static change in the optical spectrum. The experiment achieved a sensitivity of $55~\mathrm{nV\,cm^{-1}\,Hz^{-1/2}}$ and a minimum detectable field of $780~\mathrm{pV\,cm^{-1}}$, while preserving the amplitude, frequency, and phase information of the RF signal.

Once the signal is transferred to an intermediate frequency, the noise spectrum at that frequency becomes a direct determinant of receiver sensitivity. The original atomic superheterodyne experiment was limited primarily by laser-frequency noise. Tang et al.~\cite{noise-tang2025} modeled quantum projection noise as a theoretical sensitivity limit and examined the noise contributions that affect practical receiver sensitivity. Yang et al.~\cite{noise-yang2024} further quantified the influence of laser frequency and phase noise on the atomic coherence and the resulting electrometric sensitivity.

\subsubsection{RF-Field Coupling Enhancement}

The sensitivity discussed in atomic electrometry is usually referenced to the RF field experienced by the atoms, whereas a practical receiver is characterized by the incident field that can be detected outside the vapor cell. These two fields are not necessarily the same. The cell geometry, dielectric boundaries, polarization, surrounding structures, and source configuration can all modify the field inside the sensing volume. Receiver-level sensitivity therefore depends on the RF coupling between the external field and the atoms in addition to the atomic response and optical readout. This distinction becomes especially important when the local field is deliberately enhanced by an RF structure~\cite{Sensing-Backes2024}.

Resonant field concentration provides the most direct way to improve this coupling. A resonator stores electromagnetic energy and increases the electric field over the atomic sensing region, so the same incident signal produces a larger Rabi frequency and a stronger optical response. Holloway et al.~\cite{Eng-RF-SRR-Holloway2022} combined a subwavelength split-ring resonator with a Rydberg vapor cell at approximately $1.3$~GHz and obtained nearly two orders of magnitude improvement in incident-field sensitivity. The same field enhancement was retained when the atomic signal was detected through heterodyne mixing. Loading resonators have also been used to increase the local field in the vapor-cell region~\cite{Eng-RF-Hao2022}. The enhancement, however, is inherently linked to the electromagnetic mode of the resonator and is therefore accompanied by constraints on frequency range, spatial uniformity, polarization, and sensor placement.

As resonant structures became part of the receiver front end, the relevant question shifted from the local enhancement factor alone to its effect on the sensitivity referred to the external field. Sandidge et al.~\cite{Eng-RF-Sandidge2024} introduced a receiver-level framework that relates resonant field enhancement to the noise-equivalent field and equivalent receiver noise. In an X-band implementation, a resonant cavity improved the measured sensitivity from approximately $23~\mu\mathrm{V\,m^{-1}\,Hz^{-1/2}}$ to $1.58~\mu\mathrm{V\,m^{-1}\,Hz^{-1/2}}$. Such calibration is necessary because the atoms measure the enhanced local field, whereas the performance of the receiver must ultimately be referenced to the field incident on the complete sensing structure.

The narrow frequency response of a resonator introduces a different limitation. Large enhancement is usually obtained only near resonance, which can conflict with the wide frequency agility available from Rydberg transitions. Reconfigurable structures have therefore been explored to move the enhancement band together with the operating frequency. Yang et al.~\cite{Eng-RF-Yang2024}, for example, developed an electrically tunable PCB resonator with continuous tuning from approximately $187$ to $293.6$~MHz. The resulting design extends field enhancement over a wider carrier range, but also introduces a tradeoff among enhancement factor, tuning range, RF bandwidth, and field uniformity. These quantities must be considered together when a resonant structure is used to improve the incident-field sensitivity of a Rydberg receiver.

\subsection{Bandwidth Extension}

Instantaneous bandwidth refers to the frequency range over which a Rydberg-atom receiver can respond to RF signals without retuning its operating point. When the signal varies faster than the atomic system can establish a new steady state, the atomic response can no longer fully follow the input. The instantaneous bandwidth is therefore commonly characterized by the frequency range over which the output response remains within 3 dB of its low-frequency or center-frequency value.

Existing approaches can be grouped into three categories. The first focuses on increasing the response bandwidth of individual atomic channels. The second uses transient responses to extract RF information before steady state is reached. The third employs multiple atomic channels in parallel to broaden the simultaneously accessible frequency range.

\subsubsection{Atomic Response Bandwidth Enhancement}
Common approaches include reducing the beam waist and adjusting the optical readout. These methods may, however, reduce the effective atom number or add noise, leading to a trade-off between bandwidth and sensitivity. Therefore, how to maintain high-sensitivity reception is a key issue to consider in enhancing the bandwidth of atomic response.
Early studies first examined the dynamic capability of Rydberg-atom receivers through high-speed modulation experiments. Meyer et al. used thermal-atom Rydberg EIT to detect amplitude modulation on a $17~\mathrm{GHz}$ carrier and demonstrated a photon-shot-noise-limited channel capacity of up to $8.2~\mathrm{Mbit/s}$ \cite{Comm-AM/FM-Meyer2018}. Cox et al. subsequently demonstrated quantum-limited atomic reception over a measurement bandwidth from $10~\mathrm{kHz}$ to $30~\mathrm{MHz}$ \cite{Comm-Theory-Cox2018}. These experiments showed that Rydberg atoms can respond to signal variations on the megahertz scale and provided an experimental basis for later work on instantaneous-bandwidth enhancement.

Building on this, Hu et al. systematically investigated the effects of the probe Rabi frequency, coupling Rabi frequency, and laser beam size. Their results showed that increasing the coupling-field strength accelerated the establishment of atomic coherence, while reducing the beam waist shortened the atomic transit time. Under a varying RF field, atoms already within the laser beams evolve toward the new state, whereas thermal motion continuously introduces new atoms that respond directly to the current field. A sufficiently short transit time therefore allows atomic replacement to dominate the ensemble response, leading to a broader instantaneous bandwidth. After optimization, an instantaneous bandwidth of approximately 6.8 MHz was achieved \cite{Sensing-Bandwidth-Hu2023}. To compensate for the reduction in effective atom number caused by the smaller beam waist, Hu et al. further combined the outputs of two small-beam excitation channels. This increased the bandwidth from $578~\mathrm{kHz}$ to $872~\mathrm{kHz}$ while maintaining a sensitivity of about $0.47~\mu\mathrm{V}\,\mathrm{cm}^{-1}\,\mathrm{Hz}^{-1/2}$. Yuan et al. later introduced an auxiliary microwave field to modify the modulation transfer around the EIT resonance. By jointly optimizing the powers of the two microwave fields, they obtained an instantaneous bandwidth of approximately $220~\mathrm{kHz}$ and a data rate of about $62~\mathrm{kbit/s}$ \cite{Comm-AM/FM-Yuan2024}. This result also showed that auxiliary RF fields can be used to control the dynamic response.

In addition to directly modifying the atomic response, another line of work has focused on high-speed optical readout. Yang et al. used a strong coupling field to enhance one of the sidebands generated through six-wave mixing and obtained an instantaneous response range of approximately $\pm 10.2~\mathrm{MHz}$ at a sensitivity of about $62~\mathrm{nV}\,\mathrm{cm}^{-1}\,\mathrm{Hz}^{-1/2}$ \cite{Comm-BW-Yang2024}. To characterize the bandwidth of this readout mechanism, Chen et al.~\cite{Comm-CB-Chen2026} developed a baseband model and a closed-form two-pole approximation for six-wave-mixing transduction. Their numerical results showed a 3-dB bandwidth of approximately $10~\mathrm{MHz}$ and identified tradeoffs among bandwidth, sensitivity, and linearity. Shylla et al. further separated the positive and negative optical sidebands. After optimization, the $3~\mathrm{dB}$ bandwidth reached $10.5 \pm 0.4~\mathrm{MHz}$ for the negative sideband and $3.4 \pm 0.2~\mathrm{MHz}$ for the positive sideband \cite{Sensing-Sideband-Shylla2025}. To reduce the sensitivity penalty associated with bandwidth enhancement, Manchaiah et al. combined a smaller beam waist and a stronger coupling field to accelerate the atomic response. Optical homodyne detection was used to compensate for the weaker probe signal caused by the reduced beam waist. This substantially increased the photodetector output and reduced the limitations imposed by detector noise and electronic bandwidth. As a result, an $8~\mathrm{MHz}$ response bandwidth was achieved while retaining high sensitivity \cite{Sensing-Bandwidth-Manchaiah2026}.

\subsubsection{Transient Readout}
Transient readout provides an alternative to conventional readout. The key idea is to extract RF information from the transient atomic response before the system reaches a new steady state. By using pulsed probing or repeated sampling of the fast transients, the receiver can detect rapidly varying RF signals without waiting for the full EIT relaxation process, thereby extending the effective instantaneous bandwidth.

Knarr et al. used pulsed probe light to read out the atomic response before steady state was reached, reducing the limitation imposed by the EIT buildup time. They measured transient responses for pulse widths from 10 to 1000 ns and proposed a spatiotemporal multiplexed receiver. Their calibrated model predicted a response above 100 MHz and error-free OOK reception at 100 Mb/s \cite{Comm-CB-Knarr2023}. Bohaichuk et al. studied the transient response of room-temperature cesium atoms to 19.4 GHz RF pulses and analyzed the effects of atomic transit, Rydberg-Rydberg collisions, and ionization. By applying matched filtering directly to the transient signal, they achieved single-pulse detection from 10 $\mu$s down to 50 ns, with an equivalent weak-field sensitivity of about 240 nV$\cdot$cm$^{-1}\cdot$Hz$^{-1/2}$ \cite{Sensing-Radar-Bohaichuk2022}.
\subsubsection{Parallel Multi-band Detection}
When the response bandwidth at a single operating point is insufficient, multiple distinguishable RF channels can be used in parallel. Each channel independently recovers the signal amplitude and phase. Such channels can be implemented through atomic, optical-frequency, or spatial multiplexing.

Early multiband reception was realized by using different atomic species or RF transitions to form parallel channels. Holloway et al. used Cs and Rb in the same system to simultaneously receive signals at $19.626~\mathrm{GHz}$ and $20.644~\mathrm{GHz}$, with AM and FM stereo signals recovered independently \cite{Comm-MB-Holloway2021}. Du et al. coupled a common Rydberg state to two different final states and simultaneously received signals at $12.52~\mathrm{GHz}$ and $39.80~\mathrm{GHz}$ \cite{Comm-MB-Du2022}. Meyer et al. used multiple transitions near the Rb $56D$ state together with off-resonant Stark responses. Five carriers at $1.72$, $12.11$, $27.42$, $65.11$, and $115.75~\mathrm{GHz}$ were simultaneously detected, with their amplitude and phase recovered by electronic filtering \cite{Comm-MB-Meyer2023}. Allinson et al. further used a cascade of high-angular-momentum Rydberg states to cover RF transitions from $128~\mathrm{MHz}$ to $0.61~\mathrm{THz}$. Seven RF fields were applied simultaneously, and five modulated signals were independently recovered \cite{Comm-MB-Allinson2024}. As more Rydberg states are involved, optical multiplexing can be used to address and distinguish these channels in parallel. Dixon et al. generated an electro-optic frequency comb on the $852~\mathrm{nm}$ probe beam, allowing multiple probe detunings to be sampled simultaneously without frequency scanning \cite{Sensing-FrequencyComb-Dixon2023}. Prajapati et al. used a $2235~\mathrm{nm}$ mid-infrared coupling-frequency comb to address neighboring $nP$ Rydberg states. Up to seven comb teeth could address seven target states for RF transitions within $1$--$40~\mathrm{GHz}$ \cite{Comm-MB-Prajapati2025}.

Spatial multiplexing provides another way to separate parallel sensing channels. Zhang et al. used two RF-chip-integrated vapor-cell modules to simultaneously receive signals at $300~\mathrm{MHz}$ and $24~\mathrm{GHz}$ \cite{Comm-MIMO-Zhang2024}. Jiao et al. combined a vapor-cell array with a Stark frequency comb. In sequential measurements emulating 21 array positions, each channel covered about $10~\mathrm{MHz}$, giving a total coverage of $210~\mathrm{MHz}$ with an overall sensitivity of $326.6~\mathrm{nV}\,\mathrm{cm}^{-1}\,\mathrm{Hz}^{-1/2}$ \cite{Eng-Multi-Jiao2025}.

\subsection{Continuous Frequency Tuning}

Rydberg atoms provide access to RF fields over an exceptionally wide carrier-frequency range because many pairs of highly excited states can be coupled by electric-dipole transitions. High-sensitivity electrometry, however, normally relies on resonant or near-resonant coupling between a selected pair of Rydberg states. These resonances occur at discrete frequencies, and the spacing between neighboring transitions can be much larger than the instantaneous bandwidth of an individual resonance. Wide frequency accessibility therefore does not guarantee continuous high-sensitivity coverage. The central problem of frequency tuning is to bridge these gaps while retaining as much of the resonant atomic response as possible.

Frequencies between adjacent resonances can first be accessed through the off-resonant atomic response. Detuned RF fields modify the EIT--AT spectrum and can be inferred from the resulting spectral changes~\cite{Sensing-ContTune-Simons2016,Comm-CB-Song2019}. This principle was subsequently used for continuous-frequency communication and field detection~\cite{Comm-CB-Song2019,Sensing-ContTune-Hu2022}. Hu et al.~\cite{Sensing-ContTune-Hu2022} used a strong off-resonant local-oscillator field to shift the Rydberg level and recover a weak signal through an intermediate-frequency response. A single Rydberg state was used for continuous detection from $2$ to $5$~GHz, with a minimum detectable field of $2.25~\mu\mathrm{V/cm}$. Because the signal is detected away from a bare atomic resonance, the response changes with detuning and the sensitivity is not uniform across the tuning range.

Higher sensitivity in the frequency gaps can be obtained by creating additional resonant pathways. An auxiliary microwave field dresses neighboring Rydberg states and provides a new transition channel for a signal that would otherwise be far detuned from the optically addressed resonance. Jia et al.~\cite{Sensing-ContTune-Jia2021} used microwave dressing to extend the useful frequency range of AT-based electrometry. Simons et al.~\cite{Sensing-ContTune-Simons2021} further detected a detuned signal through a two-photon AT resonance formed with an adjacent Rydberg transition. The resonant two-photon response was substantially stronger than the corresponding off-resonant Stark response and enabled continuous detection between neighboring Rydberg resonances. Cui et al.~\cite{Sensing-ContTune-Cui2023} extended the useful frequency range of this auxiliary-field approach. Berweger et al.~\cite{Sensing-ContTune-Berweger2023} subsequently compared sequential, split, and inverse-sequential tuning in multilevel Rydberg systems. These configurations use different couplings among neighboring states and provide different tuning trajectories, with avoided crossings and Raman features serving as additional sensing resonances.

The atomic resonance itself can also be shifted through the Zeeman effect. A static magnetic field lifts the degeneracy of the magnetic sublevels and changes the frequencies of the corresponding microwave transitions. Li et al.~\cite{Sensing-ContTune-Li2023} characterized the magnetic-field-dependent EIT--AT spectra in a Rydberg vapor cell and resolved the resulting Zeeman structure. Shi et al.~\cite{Sensing-ContTune-Shi2023} applied this mechanism to a microwave mixed receiver and demonstrated more than $100$~MHz of tunable receiver frequency under magnetic-field control. In this case, the target frequency remains close to a shifted atomic resonance rather than being detected through the decreasing response of an off-resonant transition.

As the tuning range increases, the Zeeman structure itself becomes part of the receiver design. Multiple magnetic sublevels can introduce overlapping transitions with different dipole moments and polarization dependences. Schlossberger et al.~\cite{Sensing-ContTune-Schlossberger2024} used a strong magnetic field to define a quantization axis and resolve individual Zeeman-dependent AT spectra, allowing a single transition dipole moment to be isolated. Liu et al.~\cite{Sensing-ContTune-Liu2026} extended this approach to a three-photon Rydberg-EIT system and used a selected Zeeman sublevel for continuously tunable resonant detection near the L band. State selection therefore provides a way to retain a well-defined resonant response as the transition frequency is moved by the external field.

The different tuning strategies address the frequency-gap problem in different ways. Off-resonant detection provides broad coverage without reconstructing the atomic resonance, whereas auxiliary-field and Zeeman schemes retain resonant detection by creating or moving a suitable transition. The practical tuning range is therefore meaningful only together with the sensitivity variation across that range and the ability to maintain a well-defined calibration as the atomic configuration changes.

\subsection{Receiver Engineering}
Receiver engineering is a critical step in translating laboratory demonstrations into deployable instruments. For RAREs, this task is particularly challenging because the atomic response must be reliably generated and read out through the coordinated operation of the vapor cell, optical subsystem, control electronics, and calibration chain. As a result, practical receiver performance depends not only on intrinsic atomic sensitivity but also on system integration, operating-point stability, environmental perturbations, and channel consistency. Existing studies on Rydberg receiver engineering can be broadly classified into three categories: compact system integration, operational robustness, and multichannel architectures.

\subsubsection{Compact system integration}
\begin{figure*}[!t]
    \centering
    \includegraphics[width=\textwidth]{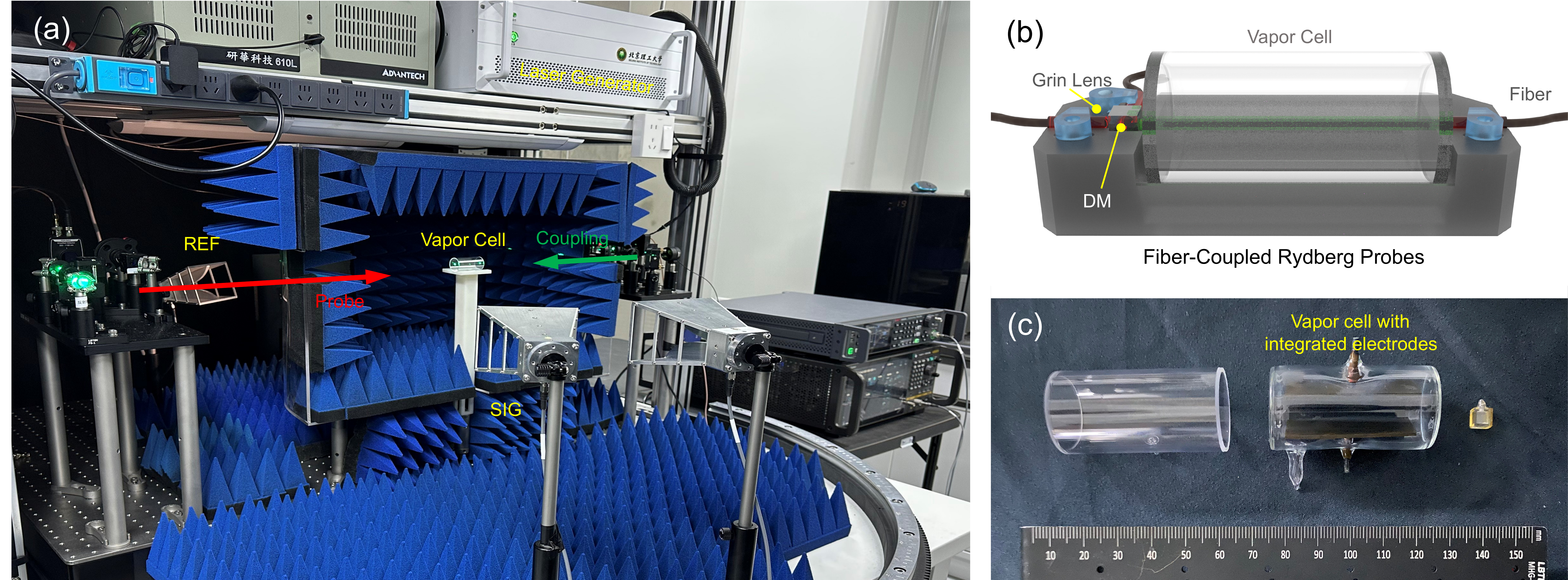}
	\vspace{-2mm}
    \caption{Hardware for Rydberg atomic receivers. (a) Laboratory setup showing the probe and coupling beams, the vapor cell, and the RF signal (SIG) and reference (REF) sources. (b) Schematic of a fiber-coupled probe with a gradient-index (GRIN) lens and a dichroic mirror (DM). (c) Three vapor cells. From left to right, a conventional cylindrical cell, a cell with integrated electrodes, and a miniature cell.}
	\vspace{-3mm}
    \label{fig:receiver_hardware}
\end{figure*}

Compact integration must be evaluated at receiver level rather than from vapor-cell dimensions alone. The laboratory setup in Fig.~\ref{fig:receiver_hardware}(a) shows the optical paths and RF sources around the vapor cell. A deployable RARE combines the cell, probe and coupling lasers, beam delivery, photodetection, frequency stabilization, RF or LO coupling, digitization, thermal management, and supervisory control. The objective is to reduce total size, power, and alignment burden while preserving a known transfer function. A useful architectural starting point is therefore to separate a low-invasiveness sensing head from a remote optical and control chassis.

Fiber-coupled probes implement this partition by placing the vapor cell and beam delivery optics in a compact head, as illustrated in Fig.~\ref{fig:receiver_hardware}(b). Simons et al. delivered counter-propagating beams through fibers and gradient-index lenses to a 10-mm cesium cell, leaving bulky laser and readout hardware remote~\cite{Eng-Compact-Simons2018}. The resulting dielectric head supports antenna scans, near-field measurements, and access to confined test regions~\cite{Eng-Compact-Holloway2018,Eng-Compact-Cardman2020}. Fiberization, however, introduces coupling loss, polarization drift, beam-overlap error, and parasitic etalons. Mao et al. addressed these interface losses with an integrated cell, dichroic optics, collimators, and three fibers, reporting 40.4-percent optical efficiency and approximately $\lambda/6$ field-imaging resolution~\cite{Eng-Compact-Mao2023}. These results show that deterministic optical interfaces, rather than cell size alone, control whether the head remains usable after movement or reassembly.
Once the optical path is integrated, the vapor-cell package becomes the next constraint. Figure~\ref{fig:receiver_hardware}(c) shows a conventional cylindrical cell, a cell with integrated electrodes, and a miniature cell. Miniaturization lowers thermal mass and enables finer sampling and wafer-level manufacture, but shorter paths, wall interactions, and dielectric boundaries can change both the atomic spectrum and the internal RF field. Cell shape, wall thickness, permittivity, loss, incidence angle, and polarization must therefore be included in the receiver model~\cite{Eng-Cell-Song2018,Eng-Cell-Zou2024}. MEMS-cell scattering and internal-field measurements provide experimental checks for RF-transparent designs~\cite{Eng-Cell-Noaman2023}, while all-glass wafer processing offers hermetic sealing and low-loss electrometry~\cite{Eng-Cell-ArtusioGlimpse2025}. Supported structures can nevertheless produce angle-dependent resonances and localized enhancement~\cite{Eng-Cell-Maurya2025}. The cell must consequently be co-designed as an atomic, optical, thermal, and electromagnetic component.
Coherent receivers add a final integration problem: delivering a stable LO without enlarging or loading the sensing head. Demonstrated solutions include a coplanar-waveguide-fed enhancement resonator with an LO port~\cite{Eng-Compact-Mao2024} and a transparent dielectric waveguide that generated the LO inside the receiver and downconverted a 19.62609 GHz signal to a 90 kHz intermediate frequency~\cite{Eng-Compact-Li2025}. These structures consolidate functions and increase usable signal, but their resonant or modal behavior becomes part of calibration. Bandwidth, tuning repeatability, temperature coefficient, polarization dependence, and spatial response must therefore be specified together with sensitivity.
Receiver-level comparisons should finally report sensing-head and remote-unit volume, total optical and electrical power, alignment tolerance, RF invasiveness, uncertainty, bandwidth, dynamic range, and restart repeatability. They should also state whether quoted performance includes resonant enhancement, closed-loop stabilization, and correction factors. Existing work demonstrates the main integration building blocks, but automated startup, manufacturing yield, thermal qualification, and long-duration stability are rarely reported together. This gap leads directly to the second engineering question: whether the compact receiver can preserve its calibrated response during operation.

\subsubsection{Operational robustness}
Operational robustness means preserving a known input-output relationship while internal settings and the environment change. Disturbances from lasers, beam overlap, atomic density, cell boundaries, RF or LO coupling, detector gain, temperature, and stray fields form a control chain rather than independent specifications. Resonance must first be acquired, the operating point held, the output calibrated, and the calibration kept valid under environmental stress.
Laser-frequency stabilization is the first layer because a drifting probe or coupling laser moves the receiver away from the selected atomic resonance. A Rydberg-EIT-derived error signal has locked the coupling laser with about 0.7 MHz fluctuation and a minimum Allan variance of $8.9\times10^{-11}$~\cite{Eng-Lock-Jiao2016}. Cascaded atomic and low-cost cavity references further separate long-term accuracy from short-term noise suppression~\cite{Eng-Lock-Jing2024}. Frequency locking fixes the spectral position, however, not the EIT contrast or readout gain.
The next layer must stabilize the complete operating point. Frequency servos maintain detunings, power servos maintain optical Rabi frequencies, diagnostics track polarization and detector gain, and a supervisory loop restores a calibrated spectral feature or intermediate-frequency gain point. FPGA-based coupling-laser stabilization shows how this control can move from laboratory instruments into compact digital hardware without sacrificing measured stability~\cite{Eng-Lock-Liu2026}. The controller should expose lock error, optical power, detector baseline, and recovery state rather than one binary lock status.
A stable operating point still does not guarantee an accurate field value. Autler-Townes splitting links a measured frequency interval to field amplitude through a calculated dipole moment and provides an SI-traceable atomic reference~\cite{Intro-Holloway2014Broadband}. Complete receiver calibration must additionally correct spatial averaging, optical drift, detector response, and dielectric field distortion~\cite{Eng-Met-Holloway2017,Eng-Met-Simons2018}. Calibration records should therefore be indexed by atomic state, carrier frequency, field regime, geometry, optical powers, and temperature, so that the controller can select a valid transfer model rather than reuse one scalar gain.
That transfer model remains valid only within a specified environment. Dielectric boundaries modify the external-to-internal field relation, while photoionization and deposited charge can create structured static-field gradients that shift or broaden Rydberg resonances~\cite{Eng-Cell-Song2018,Eng-Cell-Patrick2025}. Temperature also changes vapor density, optical depth, collisional broadening, package dimensions, and blackbody-induced shifts. A robust instrument must stabilize the atomic medium and surrounding optoelectronics, monitor auxiliary spectral features, and update or reject a calibration when temperature or field conditions leave its validated range.
Field deployment requires these layers to operate as one autonomous sequence. Startup should acquire atomic references, establish EIT, select the transition, set optical powers, validate RF or LO coupling, and record a calibration check. Continuous monitoring should log lock error, cell temperature, detector baseline, calibration residuals, and fault codes. Qualification should measure warm-up and reacquisition time, unattended restart success, vibration and reconnection sensitivity, temperature cycling, connector aging, and long-term drift. Once one receiver can repeatedly recover a calibrated state, the next problem is reproducing that state across multiple spatial channels.

\subsubsection{Distributed Atomic Sampling Architectures}
Distributed atomic sampling extends a calibrated receiver into a multichannel physical front end. The objective is a set of sampling volumes with known coordinates, matched transfer functions, controlled coupling, and phase-coherent outputs. Architectures include several excitation regions in one cell, independently packaged cells on a frame, or a shared optical chassis feeding fiber-coupled heads. Selection depends on aperture, channel count, invasiveness, and calibration workload rather than on the downstream combination algorithm.
A shared-cell architecture reduces hardware duplication. Otto et al. formed up to four receiver volumes in one 75 mm rubidium cell by crossing one coupling beam with parallel probe beams at about 2 degrees; regions separated by more than 3 mm behaved as distinct channels~\cite{Eng-Multi-Otto2021}. The cell and coupling power are shared, but window size, beam overlap, and Rayleigh range limit aperture. Separate cells provide larger baselines and replaceable elements while adding fibers, detectors, thermal paths, and cell variation.
Geometry then converts the architecture into a surveyed aperture. Linear arrays are easy to package, whereas L-shaped, planar, or nested layouts provide orthogonal and multiple baselines. Yan et al. generated three L-shaped channels and camera-calibrated the baseline to 9.8 mm; beam-displacer tolerance directly produced baseline error~\cite{Eng-Array-Yan2023}. Mao et al. placed two fiber-integrated probes containing 10-mm cubic cesium cells 8.3 cm apart~\cite{Eng-Array-Mao2024}. These examples motivate keyed orientation, stable datums, coordinate surveys, and interchangeable baseline modules before signal combination.
Mechanical baseline is not identical to electromagnetic spacing. Glass walls, holders, electrodes, fibers, and photodetectors can reshape the internal field or scatter the incident wave~\cite{Eng-Cell-Song2018,Eng-Cell-Noaman2023}. Simulation should therefore include the complete fixture and be checked by field mapping after assembly. Low-permittivity supports should remain outside the active aperture, electronics should be displaced from the cell plane, and thermal design should suppress channel differences in vapor density and EIT linewidth.
After geometry is fixed, optical and LO resources must be distributed without destroying channel equivalence. Stabilized lasers should be divided through measured split ratios, polarization-maintaining paths, monitor taps, and repeatable collimators. Wu et al. proposed a six-port fiber-coupled cell array and cascaded gratings, reporting 90.4-percent intensity uniformity and small inter-channel wavefront phase differences, and demonstrated a coherent 2-by-2 probe array~\cite{Eng-Multi-Wu2025}. Optical-power imbalance, LO amplitude and phase, detector delay, and phase zero must be measured for every branch.
Arrays may also assign different operating points to different positions. Jiao et al. proposed a linear cell array in a position-dependent Stark field, with microwave-comb lines addressing successive cells; their proof of principle emulated 21 positions with 0.23 cm minimum spacing and demonstrated two-cell reception~\cite{Eng-Multi-Jiao2025}. This makes electrode-field accuracy, cell position, comb-line power, and laser compensation explicit specifications. Qualification should report aperture, pitch, coordinate error, channel yield, gain and phase residuals, crosstalk, thermal gradient, and replacement repeatability. The deliverable is a calibrated hardware channel vector on which downstream processing can operate.


\section{Rydberg Atomic Communications}

RAREs hold significant promise for next-generation wireless communication systems because of their high sensitivity, wavelength-independent footprint, wide frequency coverage, and intrinsic immunity to mutual coupling. In this section, we survey communication technologies tailored to Rydberg atomic receivers. These technologies are classified into four main categories: waveform design, antenna-array technology, full-spectrum detection, and other emerging {wireless} communication applications.
 Relevant works on modulation waveforms and multiplexing are summarized in Table~\ref{tab:comm_waveform}.

\begin{table*}[!t]
	\centering
	\hypersetup{hidelinks}
	\caption{Summary of modulation waveforms and multiplexing for Rydberg atomic communications.}
	\label{tab:comm_waveform}
	\scriptsize
	\renewcommand{\arraystretch}{1.0}
	\setlength{\tabcolsep}{2pt}
	\setlength{\extrarowheight}{0pt}
	\setlength{\arrayrulewidth}{0.35pt}
	\begin{tabularx}{\textwidth}{|>{\centering\arraybackslash}p{1.2cm}|>{\centering\arraybackslash}p{0.72cm}|>{\raggedright\arraybackslash}p{3.75cm}|>{\raggedright\arraybackslash}X|}
		\hline
		\textbf{Category} & \textbf{Ref.} & \textbf{Quantum approach} & \textbf{Main contribution / result} \\
		\hline
		\multirow{9}{1.2cm}{\centering \textbf{AM/FM}}
		& \cite{Comm-AM/FM-Meyer2018} & EIT--AT & \textbf{(T+E)}~Use optical heterodyne readout of a 17~GHz AM carrier to estimate a photon-shot-noise-limited channel capacity of up to 8.2~Mbit/s. \\
		\cline{2-4}
		& \cite{Comm-AM/FM-Holloway2019} & EIT--AT & \textbf{(E)}~Use Cs/Rb EIT--AT readout to recover two AM-encoded guitar signals simultaneously in real time. \\
		\cline{2-4}
		& \cite{Comm-AM/FM-Anderson2021} & EIT--AT & \textbf{(E)}~Receive AM/FM signals from C-band to Q-band with a 3-dB baseband bandwidth of about 100~kHz. \\
		\cline{2-4}
		& \cite{Comm-AM/FM-Borowka2022} & EIT--AT & \textbf{(T+E)}~Propose a susceptibility-derivative model to explain AM/FM transfer and identify distinct optimal operating points. \\
		\cline{2-4}
		& \cite{Comm-AM/FM-Yang2022} & EIT--AT; AC Stark shift & \textbf{(E)}~Detect 10~kHz amplitude modulation on RF carriers tunable from 2 to 18~GHz. \\
		\cline{2-4}
		& \cite{Comm-AM/FM-Yang2023} & EIT--AT; AC Stark shift; Homodyne & \textbf{(E)}~Use a co-frequency LO to improve AM sensitivity by 14.2~dB on resonance and 20.3~dB off resonance. \\
		\cline{2-4}
		& \cite{Comm-AM/FM-Hao2024} & EIT--AT & \textbf{(E)}~Propose microwave-AM shoulder-spacing readout to detect fields $\sim30$ times weaker than conventional AT splitting. \\
		\cline{2-4}
		& \cite{Comm-AM/FM-Wu2025} & EIT--AT & 
		\textbf{(E)}~Use HMM-based Viterbi decoding to reduce transient-induced OOK detection errors. \\
		\cline{2-4}
		& \cite{Comm-AM/FM-Schlossberger2026} & AC Stark shift; Heterodyne & \textbf{(E)}~Use AC-Stark mixing with an offset LO and lock-in readout to recover handheld-radio FM audio at $462/467$~MHz. \\
		\hline
		\multirow{12}{1.2cm}{\centering \textbf{PM/APSK}}
		& \cite{Comm-PM-Simons2019} & EIT--AT; Heterodyne & 
		\textbf{(E)}~Propose a Rydberg atomic mixer to downconvert RF phase into an optically detected IF beat for phase measurement. \\
		\cline{2-4}
		& \cite{Comm-PM-Holloway2019} & EIT--AT; Heterodyne & \textbf{(E)}~Use an RF-referenced atomic mixer to demodulate BPSK/QPSK and 16/32/64QAM signals. \\
		\cline{2-4}
		& \cite{Comm-PM-Jing2020} & EIT--AT; Heterodyne & \textbf{(E)}~Propose microwave-dressed superheterodyne reception with a sensitivity of $55~\mathrm{nV\,cm^{-1}\,Hz^{-1/2}}$. \\
		\cline{2-4}
		& \cite{Comm-PM-Anderson2020} & Closed-loop interferometry & \textbf{(E)}~Use phase-coherent optical excitation paths as a reference to retrieve RF phase without an RF LO. \\
		\cline{2-4}
		& \cite{Comm-PM-Jia2021} & EIT--AT; Heterodyne; {Zeeman shift} & \textbf{(E)}~Use AC Zeeman modulation in an atomic mixer to map relative RF phase onto beat amplitude. \\
		\cline{2-4}
		& \cite{Comm-PM-Berweger2023} & Closed-loop interferometry & \textbf{(E)}~Use a two-optical/two-RF transition loop to resolve $360^\circ$ RF phase and demodulate four-phase signals. \\
		\cline{2-4}
		& \cite{Comm-PM-Nowosielski2024} & Heterodyne & \textbf{(T+E)}~Use five-level EIA and heterodyne readout near 2.4~GHz to estimate a maximum achievable single-channel capacity of 19.3~Mbit/s with 4-QAM. \\
		\cline{2-4}
		& \cite{Comm-PM-Zhu2025} & EIT--AT; Homodyne & \textbf{(E)}~Use LO-phase-controlled homodyne readout to extract I/Q components and demodulate 4QAM. \\
		\cline{2-4}
		& \cite{Comm-PM-Nowosielski2025} & Closed-loop interferometry & \textbf{(E)}~Use a three-tone microwave transition loop for S-band phase-sensitive reception without a near-frequency RF LO. \\
		\cline{2-4}
		& \cite{Comm-PM-Borowka2025} & Closed-loop interferometry & \textbf{(E)}~Propose phase-compensated optical-bias detection to recover RF phase without a microwave LO at the cell. \\
		\cline{2-4}
		& \cite{Comm-PM-Oliver2026} & EIT--AT; Homodyne & \textbf{(E)}~Use two EIT ladders with quadrature LOs to recover baseband I/Q components simultaneously. \\
		\cline{2-4}
		& \cite{Comm-PM-Wang2026} & EIT--AT & \textbf{(E)}~Recover a QR code transmitted using 16APSK at 13.33~kbit/s with no bit errors observed in the decoded image data. \\
		\hline
		\multirow{6}{1.2cm}[-0.5\baselineskip]{\centering \makecell{\textbf{FDM/}\\\textbf{OFDM}}}
		& \cite{Comm-AI-Liu2022} & EIT--AT; Heterodyne & \textbf{(E)}~Use deep learning to decode atomic-mixer traces, achieving 99.38\% recognition accuracy for four FDM bins. \\
		\cline{2-4}
		& \cite{Comm-FDM-Cai2023} & EIT--AT; Heterodyne & \textbf{(E)}~Use one resonant LO to mix multiple phase-modulated carriers into separate IF channels for FDM demodulation. \\
		\cline{2-4}
		& \cite{Comm-FDM-Wu2025} & EIT--AT & \textbf{(T+E)}~Propose LO-biased OFDM with a clipping/nonlinearity model to predict reception SNR and BER. \\
		\cline{2-4}
		& \cite{Comm-FDM-Tao2026} & EIT--AT & \textbf{(E)}~Use DC-biased OFDM and pilot-based equalization to demonstrate LO-free atomic image transmission. \\
		\cline{2-4}
		& \cite{Comm-OFDM-Chen2026} & Six-wave mixing & \textbf{(T)}~Develop an atomic comb to mitigate ICI by tuning the amplitudes and phases of multiple LO tones. \\
		\cline{2-4}
		& \cite{Comm-FDM-Wang2026} & EIT; Quadrupole coupling & \textbf{(T)}~Propose quadrupole-coupled OAM--FDM to recognize OAM and demodulate FDM in simulations. \\
		\hline
	\end{tabularx}
	\par\vspace{0.5mm}
\par\smallskip\noindent\parbox{\linewidth}{\raggedright\scriptsize\textit{Note.} (E) denotes an experimental result, (T) a result from analysis or simulation, and (T+E) a model or prediction supported in part by measurements.}
\end{table*}

\subsection{Modulation Waveform}
Modulation waveform design is a fundamental physical-layer technique for adapting information-bearing signals to the characteristics of a receiver. 
For RAREs, this task is particularly important because the received RF waveform must be converted into an optical observable through quantum-optical responses such as AT splitting, AC Stark shifts, and 
{the associated modulation of the EIT spectrum.}
As a result, the achievable communication performance depends not only on the modulation format itself, but also on the operating point, reference-field configuration, atomic response bandwidth, and nonlinear transfer characteristics of the RARE. Existing studies on waveform design {and multiplexing} for Rydberg atomic communications can be broadly classified into four categories: amplitude modulation (AM), frequency modulation (FM), phase modulation (PM), and frequency-division multiplexing (FDM).

\subsubsection{Amplitude/frequency modulation}
AM {and} FM are {among} the most direct waveform designs for RARE-based communications~\cite{Comm-AM/FM-Meyer2018,Comm-AM/FM-Yang2022,Comm-AM/FM-Yang2023,Comm-AM/FM-Hao2024,Comm-AM/FM-Wu2025,Comm-PM-Wang2026,Comm-AM/FM-Anderson2021,Comm-AM/FM-Borowka2022,Comm-AM/FM-Holloway2019,Comm-AM/FM-Schlossberger2026}. 
The key idea is to transform {variations in} the amplitude (related to {the RF} Rabi-frequency) or frequency (related to {the RF} detuning) of an RF carrier into 
{variations in the optical EIT spectrum through}
AT splitting and AC Stark shift.
By fixing the probe or coupling laser at an appropriate operating point, the receiver can directly read out the baseband waveform from the probe transmission. 

Meyer et al.~\cite{Comm-AM/FM-Meyer2018} detected amplitude modulation on a 17~GHz microwave carrier
and reported a photon-shot-noise-limited channel capacity of up to 8.2~Mbit/s.
The carrier-frequency coverage of AM reception was later extended in~\cite{Comm-AM/FM-Yang2022}.
They detected a 10 kHz AM signal over RF carriers from 2 GHz to 18 GHz, based on off-resonant AC-Stark shifts and on-resonant AT splitting effects.
Specifically, they estimated data capacities of about 11.2 Mbit/s near the 9.94 GHz resonant carrier and 5.2 Mbit/s at a 15.4 GHz off-resonant carrier.
To further improve the sensitivity of AM reception, Yang et al.~\cite{Comm-AM/FM-Yang2023} introduced an atomic homodyne architecture in which a co-frequency LO field interferes with the AM signal field at the RARE. 
Compared with direct AM detection, the homodyne approach improved the measured sensitivity by 14.2 dB and 20.3 dB in the resonant and off-resonant regions, respectively. 
However, for weak-field detection, AT-splitting is difficult to resolve because the spectral interval becomes smaller than the EIT linewidth.
To address this challenge, Hao et al.~\cite{Comm-AM/FM-Hao2024} modulated the microwave amplitude to obtain a dispersive {signal} and showed that the shoulder spacing of this signal can reduce the minimum measurable microwave field by about 30 times compared with conventional AT-splitting readout. 
For on--off keying (OOK), the finite response time of the atomic receiver affects symbol detection at high rates. When the symbol duration is comparable to the rise or fall time of the optical signal, the response to one symbol extends into the next. The sampled amplitude then depends on both the current symbol and the preceding symbol. Wu et al.~\cite{Comm-AM/FM-Wu2025} measured these transitions and used a hidden Markov model to account for the unequal rise and fall times. In their experiments, Viterbi sequence detection achieved a lower bit error rate than symbol-by-symbol detection, with a larger advantage at higher symbol rates.

FM reception follows a similar principle, whereby carrier-frequency detuning changes the EIT spectrum.
For example, Anderson et al.~\cite{Comm-AM/FM-Anderson2021} 
{demonstrated RARE-based reception of both AM and FM signals}
over carrier frequencies spanning from C-band to Q-band, with a 3-dB baseband bandwidth of about 100 kHz. 
The modulation-transfer mechanism of AM and FM signals was further analyzed in~\cite{Comm-AM/FM-Borowka2022}.
They interpreted the AM and FM responses as derivatives of the susceptibility with respect to the RF Rabi-frequency and detuning, respectively. 
This analysis revealed that AM and FM generally {require} different optimal {operating conditions}, including detunings, Rabi-frequency, and bandwidth.

Beyond laboratory-defined AM/FM waveforms, recent experiments have shown that RAREs can receive real-world communication signals. 
Holloway et al.~\cite{Comm-AM/FM-Holloway2019} used the output of a guitar to AM modulate a microwave carrier and recovered the audio waveform {using a} RARE.
More recently, Schlossberger et al.~\cite{Comm-AM/FM-Schlossberger2026} demonstrated the reception of a real-world FM audio signal using a consumer handheld UHF two-way radio operating on Family Radio Service channels near 462 and 467 MHz. 
Because the AC-Stark response of the chosen Rydberg state varies negligibly over the narrow FM deviation of a channel, an offset LO was introduced to convert the {received} FM signal into a low-frequency amplitude beat note for lock-in demodulation. 


\subsubsection{Phase modulation}
Beyond AM and FM, 
the information can also be encoded in discrete phase states or amplitude-phase constellations.
For {a conventional} RARE, phase information cannot be directly converted into an optical readout; instead, a reference signal is required to obtain the relative phase.
This can be achieved {using an} RF LO, {a closed} atomic transition loop, or an optically generated {reference}, so that the RF phase and amplitude are converted into an optical IF or baseband {response} that can be decoded with standard communication algorithms~\cite{Comm-PM-Simons2019,Comm-PM-Holloway2019,Comm-PM-Jing2020,Comm-PM-Nowosielski2024,Comm-PM-Zhu2025,Comm-PM-Oliver2026,Comm-PM-Jia2021,Comm-PM-Berweger2023,Comm-PM-Nowosielski2025,Comm-PM-Anderson2020,Comm-PM-Borowka2025}. Wang et al.~\cite{Comm-PM-Wang2026} demonstrated the reception of amplitude and phase shift keying (APSK) signals, in which both amplitude and phase distinguish the transmitted symbols.

The first route is LO-assisted heterodyne or superheterodyne reception. 
Simons et al.~\cite{Comm-PM-Simons2019} first developed an atomic mixer by introducing a resonant LO field and a detuned signal field to jointly drive a Rydberg transition. 
The {RARE} naturally {downconverts} the RF signal to an IF beat note{, whose phase carries the relative phase between the signal and LO fields.} 
Any LO phase fluctuations that remain after carrier tracking appear as rotations of the received constellation points.

Holloway et al.~\cite{Comm-PM-Holloway2019} used an atomic mixer to receive BPSK, QPSK, and QAM signals. They measured the error vector magnitude (EVM), which quantifies how far received symbols deviate from their ideal constellation points. At intermediate frequencies of 1 and 2~MHz, the measured BPSK EVM remained below 5\% for symbol rates below 400~ksymbol/s and approached 10\% at about 700~ksymbol/s. The same receiver also resolved 16-QAM, 32-QAM, and 64-QAM. At a given average symbol energy, the closer spacing of higher-order QAM constellation points leaves less tolerance for errors in amplitude and phase. In this experiment, the highest symbol rate at which 64-QAM could be received was lower than that for 16-QAM. The authors also observed that the mixer output decreased at higher intermediate frequencies, making adjacent QAM amplitude levels harder to distinguish.
Jing et al.~\cite{Comm-PM-Jing2020} advanced this idea by developing an atomic superheterodyne receiver based on microwave-dressed Rydberg spectroscopy. 
In this scheme, a strong resonant LO field was introduced to dress the Rydberg states{, enabling the atomic medium to} act as an atomic amplifier and mixer for weak {signals}.
These studies established superheterodyne RAREs as {a widely used architecture} for coherent digital reception.

To enhance the bandwidth and reduce the complexity of IF processing, the homodyne detection was also developed. 
In this scheme, the LO and signal fields share the same carrier frequency, {allowing} the signal {to} be directly downconverted to the baseband.
Specifically, Oliver et al.~\cite{Comm-PM-Oliver2026} implemented a dual-ladder RARE using two spatially overlapped but independent EIT ladders. 
The two arms were isolated using orthogonal optical polarizations together with frequency-shifted laser beams. 
Two LO signals, polarization-matched to the corresponding EIT arms and set $90^\circ$ out of phase, {then enabled the simultaneous homodyne detection} of the I and Q components.

Another branch explores alternative reference signals for phase reception. 
For example, an AC magnetic field was introduced in~\cite{Comm-PM-Jia2021} as an additional reference in a RARE-based mixer. 
By periodically modulating the Rydberg levels through the Zeeman effect, the reference magnetic field is superposed with the RF beat signal inside the atoms, converting the relative phase not only into the beat phase but also into the beat amplitude. 
However, these heterodyne and homodyne receivers {require} a near-frequency or co-frequency RF LO, which may perturb the signal field. 
To relax this requirement, closed atomic transition loops driven by microwave fields at different frequencies were constructed in~\cite{Comm-PM-Berweger2023,Comm-PM-Nowosielski2025}. 
The key idea is that the probe transmission depends on the accumulated phase around the loop, so the RF phase can be retrieved. 
However, these schemes still require auxiliary RF LO fields, which may suffer from practical imperfections including power leakage and multi-path fading and introduce unwanted exposure. 
To further avoid RF reference, Anderson et al.~\cite{Comm-PM-Anderson2020} developed a holographic phase-sensing method in which phase-coherent optical excitation pathways provide the reference for retrieving the RF phase. 
Furthermore, Borowka et al.~\cite{Comm-PM-Borowka2025} 
{developed an optically biased receiver in which multiple optical fields and the RF signal form a closed transition loop. The product of the optical fields acts as an effective LO, enabling recovery of the RF-signal phase without applying a microwave LO to the vapor cell.}

\subsubsection{Frequency-division multiplexing}
FDM improves the spectral efficiency of RARE-based communication by allocating independent data streams to multiple frequency components~\cite{Comm-AI-Liu2022,Comm-FDM-Cai2023,Comm-FDM-Wu2025,Comm-FDM-Tao2026,Comm-FDM-Wang2026}. 
The essential task is therefore to map each frequency component into a distinguishable optical {spectral feature} or {IF} signal, after which the parallel streams can be separated by spectral recognition, IF filtering, or digital demultiplexing.

Early FDM demonstrations mainly relied on multifrequency atomic mixing. 
Liu et al.~\cite{Comm-AI-Liu2022} used a Rydberg atomic mixer to convert FDM-2PSK microwave signals into optical spectral features, and then employed deep learning to {decode} the phase states carried by different frequency bins. 
Cai et al.~\cite{Comm-FDM-Cai2023} realized coherent FDM reception with a heterodyne PM architecture, where a strong LO field mixed with several detuned phase-modulated carriers and separated different channels {according to} their IF frequencies. 
Along another route, an OAM-FDM scheme was proposed in~\cite{Comm-FDM-Wang2026}, in which each FDM subcarrier is associated with an {orbital angular momentum (OAM)} mode and jointly read out from the Rydberg EIT spectrum. 
When FDM carriers arrive at different strengths, recovery of a weak channel depends on how the atomic receiver responds to the stronger carriers. The nonlinear atomic response can generate intermodulation products that interfere with a weak channel if they fall within its detection bandwidth. Gonçalves et al.~\cite{Comm-LinearDynamicRange-gonccalves2026} measured these products in a cesium heterodyne receiver near 9.94~GHz using two equal-amplitude tones separated by 10~kHz or 1.01~MHz. Their measurements characterize the distortion at both spacings. Determining how much stronger a neighboring carrier can be while a weak channel still meets a target BER requires measurements with unequal input levels and a modulated weak signal.
Furthermore, orthogonal frequency-division multiplexing (OFDM) has been explored to adapt RAREs to broadband subcarrier-based waveforms. 
Because the atomic optical readout is nonlinear and amplitude-sensitive, OFDM reception requires additional waveform and operating-point design. 
Inspired by direct current-biased optical orthogonal frequency division multiplexing (DCO-OFDM) technology in optical communication systems, Wu et al.~\cite{Comm-FDM-Wu2025} proposed a LO direct-current-biased OFDM (LODC-OFDM)   waveform.
In this scheme, the frequency-domain OFDM symbols were constructed with Hermitian symmetry to generate the real-valued time-domain signals {that modulate} the amplitude of the RF signal.
{The LO field replaced the DC bias used in conventional DCO-OFDM and shifted the composite RF envelope into a suitable operating region, thereby mitigating nonlinear distortion and clipping associated with the high peak-to-average power ratio (PAPR) of OFDM signals.}
Moreover, Tao et al.~\cite{Comm-FDM-Tao2026} experimentally demonstrated LO-free RARE-based OFDM using {a} similar DCO-OFDM waveform, {together with} pilot-aided channel estimation, equalization, and clipping to support QAM data transmission. These waveform and equalization approaches address the nonlinear response of the atomic receiver. Chen et al.~\cite{Comm-OFDM-Chen2026} proposed an atomic-comb receiver that uses six-wave mixing and a programmable multi-tone LO to shape the RF-to-optical response for OFDM reception. By optimizing the amplitude and phase of each LO tone, the design suppresses inter-carrier interference caused by carrier-frequency offset and phase noise while controlling the effective channel memory to mitigate inter-symbol interference. Simulation results showed reduced inter-carrier leakage and a shorter effective channel memory. 

\subsection{Antenna Array Technology}
\begin{figure*}
	\centering
	\includegraphics[width=6.5in]{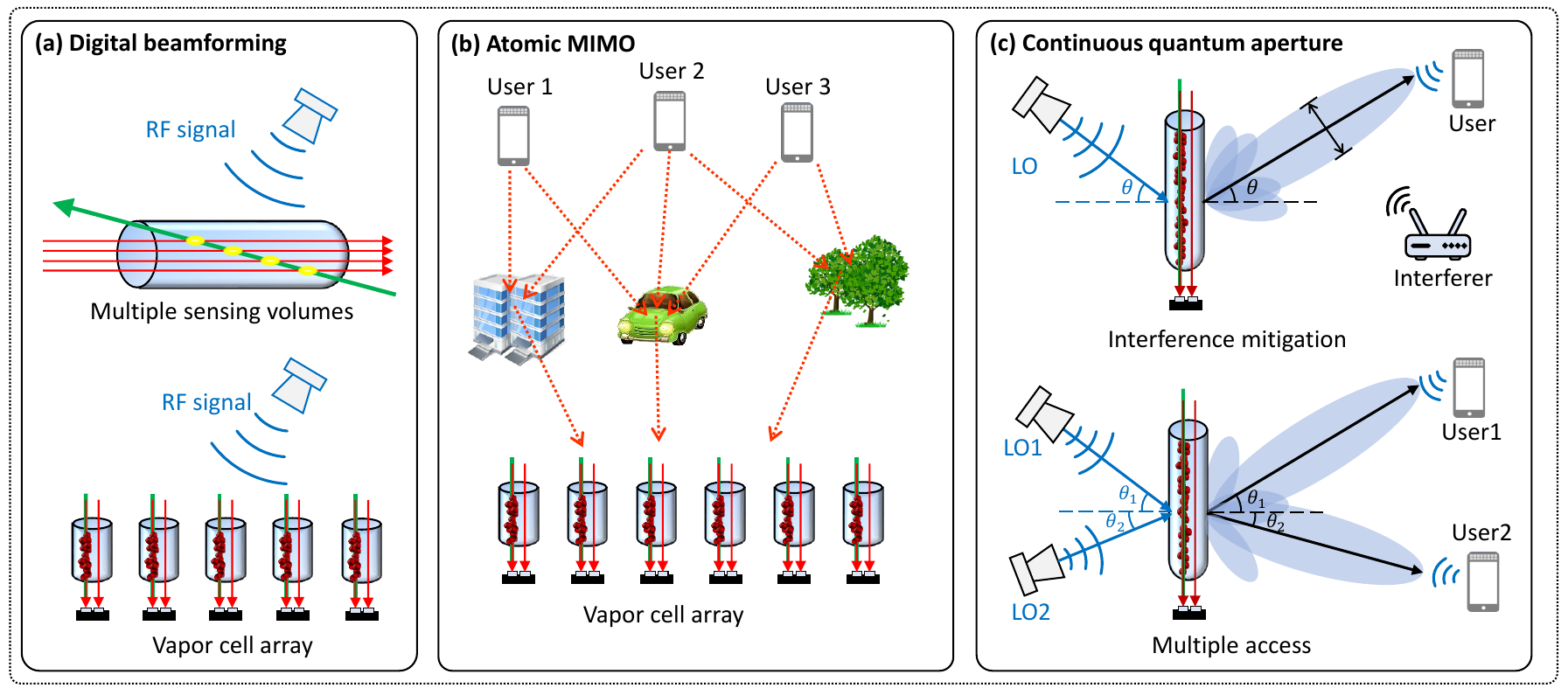}
	\vspace*{-1em}
	\caption{Rydberg atomic antenna array technologies: (a) digital beamforming, (b) atomic MIMO, and (c) continuous quantum aperture.
    } 
	\vspace*{-1em}
	\label{img:array}
\end{figure*}

Large antenna arrays are among the {key enabling} technologies in {5G-and-beyond} wireless communication systems. By shaping electromagnetic wavefronts, antenna arrays provide substantial spatial diversity and multiplexing gains, thereby significantly improving channel capacity. As discussed earlier, Rydberg atomic receivers are naturally compatible with array-based architectures through optical-path topology design. Moreover, Rydberg atomic antenna arrays have the potential to overcome some limitations of conventional antenna arrays, owing to the high spatial resolution of Rydberg probes and their intrinsic immunity to {conventional electromagnetic} mutual coupling. Motivated by these advantages, recent studies have explored this emerging antenna-array paradigm {to improve} communication performance. As shown in Fig.~\ref{img:array}, these efforts can be broadly classified into three categories: digital beamforming (Fig.~\ref{img:array}(a)), atomic multiple-input multiple-output (MIMO) (Fig.~\ref{img:array}(b)), and {the} continuous quantum aperture (Fig.~\ref{img:array}(c)). Table~\ref{tab:comm_array} summarizes the contributions on discrete atomic arrays and continuous quantum apertures.

\begin{table*}[!t]
	\centering
	\hypersetup{hidelinks}
	\caption{Summary on antenna-array technology for Rydberg atomic communications.}
	\label{tab:comm_array}
	\scriptsize
	\renewcommand{\arraystretch}{1.0}
	\setlength{\tabcolsep}{2pt}
	\setlength{\extrarowheight}{0pt}
	\setlength{\arrayrulewidth}{0.35pt}
	\begin{tabularx}{\textwidth}{|>{\centering\arraybackslash}p{2.4cm}|>{\centering\arraybackslash}p{1.35cm}|>{\raggedright\arraybackslash}p{2.9cm}|>{\raggedright\arraybackslash}X|}
		\hline
		\textbf{Category} & \textbf{Ref.} & \textbf{Approach} & \textbf{Main contribution / result} \\
		\hline
		\multirow{3}{2.4cm}{\centering \makecell{\textbf{Digital}\\\textbf{beamforming}}}
		& \cite{Eng-Multi-Otto2021} & Four-channel shared cell & \textbf{(T+E)}~Use four probe-defined channels to verify linear SNR and logarithmic capacity scaling with array size. \\
		\cline{2-4}
		& \cite{Eng-Array-Mao2024} & Two fiber probes & \textbf{(E)}~Use two fiber-coupled probes for digital beamforming at $2.63$~GHz, achieving AoA errors below $1.5^\circ$. \\
		\cline{2-4}
		& \cite{Comm-DBF-Han2026} & Superdirective array & \textbf{(T)}~Use atomic-array modeling to predict $N^2$ end-fire directivity and asymptotic user-channel orthogonality. \\
		\hline
		\multirow{19}{2.4cm}[-7\baselineskip]{\centering \textbf{Atomic MIMO}}
		& \cite{Comm-MIMO-Cui2024b,Comm-MIMO-Cui2025b} & Multiuser phase retrieval & 
		\textbf{(T)}~Establish magnitude-only atomic MIMO model and propose nonlinear signal detection algorithms. \\
		\cline{2-4}
		& \cite{Comm-MIMO-Cui2025a} & IQ-aware precoding & \textbf{(T)}~Propose independent I/Q precoding to achieve atomic-MIMO capacity under a strong-reference model. \\
		\cline{2-4}
		& \cite{Comm-MIMO-Xu2025} & Channel estimation & \textbf{(T)}~Propose projected-gradient estimation to recover channels of 1D/2D atomic arrays in simulations. \\
		\cline{2-4}
		& \cite{Comm-MIMO-Peng2025} & RIS-assisted detection & \textbf{(T)}~Propose RIS-assisted PAM phase alignment to reduce phase ambiguity and BER of atomic MIMO. \\
		\cline{2-4}
		& \cite{Comm-MIMO-Jeon2026b} & RIS-assisted ISAC & \textbf{(T)}~Propose joint beamforming/RIS optimization to improve communication capacity and sensing CRB. \\
		\cline{2-4}
		& \cite{Comm-MIMO-Zhu2025} & Multiuser MPSK & \textbf{(T)}~Propose penalized MM/Wirtinger-flow detection to improve  multiuser MPSK accuracy at lower complexity. \\
		\cline{2-4}
		& \cite{Comm-MIMO-Kang2026} & DeepQ-MIMO & \textbf{(T)}~Propose DeepQ-MIMO to jointly learn transmit/receive processing and reference injection to reduce BER. \\
		\cline{2-4}
		& \cite{Comm-MIMO-Jeon2026a} & RF-lens Quantum-PROBE & \textbf{(T)}~Propose RF-lens power-profile recovery using NN-LASSO or SIC to estimate multiple AoAs. \\
		\cline{2-4}
		& \cite{Comm-MIMO-Atapattu2025} & Unknown-signal detection & \textbf{(T)}~Propose magnitude-domain LRT/CFAR detectors for unknown RF signals. \\
		\cline{2-4}
		& \cite{Comm-MIMO-Atapattu2026} & Multi-shot detection & \textbf{(T)}~Propose phase-averaged multi-shot detection to approach the genie-aided bound with $5$--$10$ shots. \\
		\cline{2-4}
		& \cite{Comm-MIMO-Liu2026} & Phase-rotated spreading & \textbf{(T)}~Propose two-slot symbol spreading to reconstruct a linear model for scalable MIMO detection. \\
		\cline{2-4}
		& \cite{Comm-MIMO-Jeon2026c} & FRIS channel shaping & \textbf{(T)}~Propose joint FRIS port/phase and transmit-beamformer optimization to reduce BER. \\
		\cline{2-4}
		& \cite{Comm-MIMO-Gong2025b} & Superheterodyne uplink & \textbf{(T)}~Propose a quantum-optical baseband model to derive MRC/ZF achievable-rate bounds. \\
		\cline{2-4}
		& \cite{Comm-MIMO-Peng2025b,Comm-MIMO-Peng2026} & Ground--space uplink & \textbf{(T)}~Use onboard RAQ-MIMO with imperfect-CSI MRC/ZF analysis to predict uplink-rate capacity. \\
		\cline{2-4}
		& \cite{Comm-MIMO-Wu2025} & Multiplexed array reuse & \textbf{(T)}~Propose multiplexed array reuse and sparse-channel precoding to reduce duplicated receiver hardware. \\
		\cline{2-4}
		& \cite{Comm-MIMO-Gong2025a} & Multi-target DOA & \textbf{(T)}~Propose RAQ-ESPRIT for multi-target DOA estimation. \\
		\cline{2-4}
		& \cite{Comm-MIMO-Yuan2025} & Electromagnetic modeling & \textbf{(T)}~Use ray/Green-function channel models to reveal single-polarization atomic-MIMO capacity advantages. \\
		\cline{2-4}
		& \cite{Comm-MIMO-Zhang2024} & RF-chip spatial channels & \textbf{(E)}~Use two spatially multiplexed RF-chip modules to receive $300$~MHz/$24$~GHz signals simultaneously. \\
		\cline{2-4}
		& \cite{Comm-MB-Zhu2025} & Multiband RAQ-MIMO & \textbf{(T)}~Propose joint LO/precoder/combiner optimization to improve multiband MIMO capacity. \\
		\hline
		\multirow{4}{2.4cm}[-2\baselineskip]{\centering \makecell{\textbf{Continuous quantum}\\\textbf{aperture}}}
		& \cite{Comm-CA-Cui2026a} & Atomic beamforming theory & \textbf{(T)}~Reveal atomic beamforming gain of an atomic vapor cell and derive the optimal cell length. \\
		\cline{2-4}
		& \cite{Comm-CA-Cui2026b} & Atomic beamforming validation & \textbf{(E)}~Experimentally demonstrate CQA-based beamforming, interference suppression, and multiple access. \\
		\cline{2-4}
		& \cite{Comm-CA-Guo2025} & Single-cell localization & \textbf{(T)}~Propose inner-vapor-interference fitting with particle swarm optimization for single-cell AoA estimation. \\
		\cline{2-4}
		& \cite{Comm-CA-Guo2026} & Satellite localization & \textbf{(T)}~Propose joint AoA/Doppler inference to enable instantaneous single-satellite 2D localization. \\
		\hline
	\end{tabularx}
	\par\vspace{0.5mm}
\par\smallskip\noindent\parbox{\linewidth}{\raggedright\scriptsize\textit{Note.} (E) denotes an experimental result, (T) a result from analysis or simulation, and (T+E) a model or prediction supported in part by measurements.}
\end{table*}

\subsubsection{Digital beamforming}
Digital beamforming (DBF) provides the most direct route for introducing array-level signal processing into Rydberg atomic receivers~\cite{Eng-Multi-Otto2021,Eng-Array-Mao2024,Comm-DBF-Han2026}. 
The key idea is to deploy distributed atomic sensors to form directional beams that spatially enhance desired signals while suppressing interference.
The earliest effort in this direction was reported by Otto et al.~\cite{Eng-Multi-Otto2021}. In that work, the probe laser was split into four beams that interacted with a slightly tilted coupling laser, thereby creating four independent sensing volumes. This configuration effectively formed a single-input multiple-output (SIMO) architecture with four {spatially} distributed Rydberg receive antennas. 
The experimental results showed that the received SNR scales linearly with the number of probe beams $N$, while the data capacity scales logarithmically with $N$, consistent with the scaling law of conventional antenna arrays.  
Subsequently, Mao et al.~\cite{Eng-Array-Mao2024} realized a two-element Rydberg receiving array using fiber-coupled Rydberg field probes and experimentally implemented digital beamforming at 2.63 GHz. 
By combining the two optical readout signals in the digital domain, they {obtained the receive-array beam pattern} and performed angle-of-arrival (AoA) estimation with {a reported angular deviation of less than} $1.5^\circ$.
More recently, {Han and Yin}~\cite{Comm-DBF-Han2026} theoretically investigated Rydberg-atom-based superdirective arrays. By exploiting negligible {conventional electromagnetic} mutual coupling, they showed that such arrays can, in principle, achieve {directivity scaling as} $N^2$ in the end-fire direction and asymptotically orthogonal user channels.

\subsubsection{Atomic MIMO}
Atomic MIMO, also referred to as Rydberg atomic quantum MIMO (RAQ-MIMO), employs multiple Rydberg atomic sensors to receive RF signals transmitted by multiple classical antennas. 
By exploiting the spatially varying optical responses across the atomic array, this architecture supports parallel data-stream multiplexing and can substantially enhance spectral efficiency~\cite{Comm-MIMO-Cui2025b,Comm-MIMO-Cui2024b,Comm-MIMO-Atapattu2026,Comm-MIMO-Cui2025a,Comm-MIMO-Liu2026,Comm-MIMO-Xu2025,Comm-MIMO-Peng2025,Comm-MIMO-Jeon2026b,Comm-MIMO-Zhu2025,Comm-MIMO-Kang2026,Comm-MIMO-Jeon2026a,Comm-MIMO-Atapattu2025,Comm-MIMO-Jeon2026c,Comm-MIMO-Gong2025b,Comm-MIMO-Peng2025b,Comm-MIMO-Peng2026,Comm-MIMO-Wu2025,Comm-MIMO-Gong2025a,Comm-MIMO-Yuan2025,Comm-MIMO-Zhang2024,Comm-MB-Zhu2025}.
This line of research can be broadly divided into two main directions: EIT-AT-enabled atomic MIMO and superheterodyne-enabled atomic MIMO.

The first direction uses the EIT-AT scheme to measure {the spatial variation in the Rabi frequency of the incident RF field across the atomic array. This measurement} is SI-traceable and well suited to medium- and strong-field regimes. Research along this line has mainly focused on signal-processing algorithms tailored to atomic MIMO systems.
Specifically, Cui et al.~\cite{Comm-MIMO-Cui2025b} introduced atomic MIMO receivers and showed that their input-output relationship is inherently nonlinear, because the measured Rabi frequency captures only the amplitude of the incident signals. As a result, multiuser signal detection becomes a nonlinear phase-retrieval problem rather than the linear signal recovery problem typically encountered in conventional MIMO systems. On this basis, the authors proposed biased Gerchberg-Saxton and expectation-maximization Gerchberg-Saxton algorithms for multiuser signal recovery. 
This framework was subsequently extended in several directions. 
Cui et al.~\cite{Comm-MIMO-Cui2025a} investigated MIMO precoding for Rydberg receivers and proposed an IQ-aware precoding method to improve channel capacity. 
Xu et al.~\cite{Comm-MIMO-Xu2025} addressed the channel estimation problem and developed a projection gradient descent {(PGD)} method applicable to both one-dimensional and two-dimensional atomic arrays. 
RIS-assisted atomic MIMO was also investigated to reduce the bit error rate~\cite{Comm-MIMO-Peng2025} and support ISAC~\cite{Comm-MIMO-Jeon2026b}. 
Beyond these fundamental transceiver functions, researchers have explored advanced tasks such as multiuser MPSK detection~\cite{Comm-MIMO-Zhu2025}, deep-learning-aided MIMO detection~\cite{Comm-MIMO-Kang2026}, multi-AoA estimation with RF lenses~\cite{Comm-MIMO-Jeon2026a}, detection of unknown RF signals~\cite{Comm-MIMO-Atapattu2025}, and {fluid-reconfigurable-intelligent-surface (FRIS)}-enhanced atomic MIMO~\cite{Comm-MIMO-Jeon2026c}.

The second direction considers superheterodyne-based atomic MIMO, in which LO dressing is introduced to facilitate signal downconversion and recovery.
Although this approach {does not retain the direct SI-traceable calibration provided by EIT-AT splitting because it depends on an engineered LO field}, it is phase-sensitive and well suited to weak RF signal reception. 
Existing studies along this line have mainly focused on the theoretical characterization of communication performance.
Specifically, in~\cite{Comm-MIMO-Gong2025b}, Gong et al. developed a baseband signal model for atomic MIMO. They derived the asymptotic channel capacities achieved by maximum-ratio combining (MRC) and zero-forcing (ZF) receivers under both correlated and uncorrelated fading. Their analysis showed that the performance of atomic MIMO jointly depends on the macroscopic array dimensions and the microscopic quantum-optical parameters {and predicted that it} can substantially outperform conventional massive MIMO under both the standard quantum limit (SQL) and the {photon shot limit (PSL)}. Building on this framework, Peng et al.~\cite{Comm-MIMO-Peng2025b,Comm-MIMO-Peng2026} further analyzed the achievable rate of atomic MIMO for direct ground-to-space uplink access under imperfect channel state information (CSI). Further advances include a low-complexity array-reuse design~\cite{Comm-MIMO-Wu2025} and multi-target AoA estimation~\cite{Comm-MIMO-Gong2025a}.

\subsubsection{Continuous quantum aperture}
The preceding two research branches increase the number of antennas primarily by {adding spatially separated antennas and thereby} enlarging the array aperture. 
Beyond this design philosophy, another route toward massive antenna configurations is to reduce the antenna spacing. In the limiting case, this leads to a continuous aperture with {effectively} zero element spacing, which offers the highest degrees of freedom (DoFs) for electromagnetic field manipulation. 
For conventional conductor-based antenna arrays, however, realizing such continuous apertures is highly challenging because mutual coupling between adjacent antenna elements can severely distort the beam pattern and degrade beamforming gain. 
By contrast, {because the atomic medium does not rely on densely packed conducting elements or discrete RF ports}, a Rydberg-atom vapor cell can naturally {realize} a continuous aperture {without conventional interelement mutual coupling}. 
This observation has recently motivated the concept of the continuous quantum aperture (CQA)~\cite{Comm-CA-Cui2026a,Comm-CA-Cui2026b}, in which beamforming is realized by a continuous atomic medium (a single vapor cell). 

In~\cite{Comm-CA-Cui2026b}, Cui et al. established a general theoretical framework for CQA-based beamforming and provided an experimental validation. 
They showed that, for a superheterodyne receiver, the interference between the LO and signal fields induces spatially varying quantum coherence along the Rydberg-atom vapor cell. As the probe laser propagates through the cell, it thereby experiences a position-dependent attenuation, the accumulation of which gives rise to the angular response pattern {as a function of the incident direction} $\theta$:
\begin{align}
    F(\theta) \propto \left| \int_0^L \frac{\partial {\rm Im}(\rho_{12})}{\partial |E_l(y)|} \exp[j(ky\cos\theta- \angle E_{l}(y))]{dy}
    \right|. 
\end{align}
Here, $E_l(y)$ denotes the LO field at position $y\in[0,L]$. The phase profile $\angle E_{l}(y)$ acts as a virtual continuous {phased} array for processing the incident EM field. Consequently, by programming the LO field distribution, diverse beamforming patterns can be synthesized. For instance, multi-peak beamforming with adjustable peak amplitudes can be achieved by controlling the number of LO sources and their transmit powers, while multiband beamforming can be implemented via multiband LO dressing.
More than 100 beam patterns were experimentally measured in~\cite{Comm-CA-Cui2026b}, showing strong agreement between theory and experiment. Moreover, the capabilities of CQA {in} interference mitigation, multiple access, and multiband multiple access were also validated in communication systems. In~\cite{Comm-CA-Cui2026a}, it was further shown that the beamforming gain of a continuous quantum aperture does not increase monotonically with vapor-cell length, because{, when the cell becomes excessively long, optical attenuation causes the signal energy to decay faster than the photon-shot-noise power}. 
Accordingly, the optimal cell length for maximizing beamforming gain was derived in~\cite{Comm-CA-Cui2026a}. 
{For the representative parameters considered in that study, increasing the cell length from 1~cm to its optimal value of 21~cm yielded an SNR improvement of approximately 10~dB.}
In addition, the continuous quantum aperture has also been applied to angle-of-arrival estimation and LEO localization in~\cite{Comm-CA-Guo2025,Comm-CA-Guo2026}.

Beyond purely continuous or discrete configurations, future Rydberg atomic antenna arrays are expected to integrate these two architectures. For example, multiple probe lasers may propagate through a long vapor cell to generate multiple continuous quantum apertures. Such a hybrid discrete-continuous architecture is promising for further enhancing the communication performance of {RAREs}.

\subsection{Full-Spectrum Detection}
By exploiting a range of quantum phenomena, including the AC Stark shift, {the} Zeeman effect, and multilevel electronic transitions, Rydberg atoms can sense electric fields over an extremely broad spectral range. Researchers have leveraged this capability to
{detect communication signals across a broad spectrum}
using compact receiver architectures. Existing efforts in this direction can be broadly categorized into four branches: low-frequency communication, broadband communication, multiband communication, and frequency-hopping communication. Table~\ref{tab:comm_spectrum} summarizes the contributions on full-spectrum reception. 

\begin{table*}[!t]
	\centering
	\hypersetup{hidelinks}
	\caption{Summary on full-spectrum detection for Rydberg atomic communications.}
	\label{tab:comm_spectrum}
	\scriptsize
	\renewcommand{\arraystretch}{1.0}
	\setlength{\tabcolsep}{2pt}
	\setlength{\extrarowheight}{0pt}
	\setlength{\arrayrulewidth}{0.35pt}
	\begin{tabularx}{\textwidth}{|>{\centering\arraybackslash}p{2.45cm}|>{\centering\arraybackslash}p{0.72cm}|>{\raggedright\arraybackslash}p{3.15cm}|>{\raggedright\arraybackslash}X|}
		\hline
		\textbf{Category} & \textbf{Ref.} & \textbf{Approach} & \textbf{Main contribution / result} \\
		\hline
		\multirow{2}{2.45cm}{\centering \makecell{\textbf{Low-frequency}\\\textbf{communications}}}
		& \cite{Comm-LF-Xie2025} & Electrode-assisted Stark & \textbf{(E)}~Use antenna-fed in-cell electrodes and AC-Stark readout to receive $100$~kHz 2FSK/OOK/BPSK at $2$--$8$~kbit/s. \\
		\cline{2-4}
		& \cite{Comm-LF-Shi2025} & Phase-sensitive LF link & \textbf{(E)}~Use phase-sensitive AC-Stark readout to transmit grayscale images at $849.1$~kbit/s on a $100$~kHz carrier. \\
		\hline
		\multirow{4}{2.45cm}[-0.5\baselineskip]{\centering \makecell{\textbf{Broadband}\\\textbf{communications}}}
		& \cite{Comm-CB-Song2019} & Continuous carrier tuning & \textbf{(E)}~Use off-resonant EIT--AT readout for $500$~kbit/s digital reception across a $200$~MHz carrier-tuning range. \\
		\cline{2-4}
		& \cite{Comm-CB-Li2022} & Stark-based AM & 
		\textbf{(E)}~Use LO-biased AC-Stark readout to tune carriers over $\sim5$~GHz and transmit color images at $238$~kbit/s. \\
		\cline{2-4}
		& \cite{Comm-CB-Liu2022} & Multilevel heterodyne & \textbf{(E)}~Use two off-resonant microwave references for multilevel heterodyne reception over a $>1$~GHz tuning range. \\
		\cline{2-4}
		& \cite{Comm-CB-Knarr2023} & Spatiotemporal multiplexing & \textbf{(T)}~Propose spatiotemporal multiplexing to predict a $>100$~MHz response bandwidth. \\
		\hline
		\multirow{13}{2.45cm}[-3\baselineskip]{\centering \makecell{\textbf{Multiband}\\\textbf{communications}}}
		& \cite{Comm-MB-Holloway2021} & Cs/Rb species diversity & \textbf{(E)}~Use Cs/Rb species diversity in one cell to receive AM/FM stereo simultaneously at $19.626/20.644$~GHz. \\
		\cline{2-4}
		& \cite{Comm-MB-Du2022} & Multiple final states & \textbf{(E)}~Use a common initial state and two final states for concurrent analog/digital reception at $12.52/39.80$~GHz. \\
		\cline{2-4}
		& \cite{Comm-MB-Meyer2023} & Parallel demodulation & \textbf{(E)}~Use parallel RF transitions to demodulate OOK simultaneously at $12.0$, $27.5$, $65.6$, and $116.1$~GHz. \\
		\cline{2-4}
		& \cite{Comm-MB-You2023} & Coupled two-band gain & \textbf{(E)}~Use two-band EIA to reveal mutually exclusive peak sensitivities of coupled bands. \\
		\cline{2-4}
		& \cite{Comm-MB-Elgee2023} & High-$\ell$ dressing & \textbf{(E)}~Use dual-optical/dual-microwave dressing of high-$\ell$ states to detect satellite-radio signals near $2.3$~GHz. \\
		\cline{2-4}
		& \cite{Comm-MIMO-Zhang2024} & RF-chip spatial channels & \textbf{(E)}~Use two spatially multiplexed RF-chip modules to receive $300$~MHz/$24$~GHz signals simultaneously. \\
		\cline{2-4}
		& \cite{Comm-MB-Zhang2024} & Three-band imaging link & \textbf{(E)}~Use three RF transitions from one state to transmit RGB channels at $3.78$, $5.106$, and $7.91$~GHz. \\
		\cline{2-4}
		& \cite{Comm-MB-Allinson2024} & Cascaded high-$\ell$ states & \textbf{(E)}~Use cascaded high-$\ell$ states for simultaneous seven-band ($128$~MHz--$0.61$~THz) detection and five-band AM reception. \\
		\cline{2-4}
		& \cite{Comm-MB-Prajapati2025} & Optical-comb addressing & \textbf{(E)}~Use rapid optical-comb switching among seven Rydberg states to detect RF channels over $1$--$40$~GHz. \\
		\cline{2-4}
		& \cite{Comm-MB-Cui2026} & Quantum attention model  & \textbf{(T+E)}~Propose a multiband response model revealing Rabi-controlled sensitivity allocation, validated at $3.2/30.6$~GHz. \\
		\cline{2-4}
		& \cite{Comm-MB-Zhu2025} & Multiband MIMO & \textbf{(T)}~Propose  joint LO/precoder/combiner optimization to improve multiband MIMO capacity. \\
		\cline{2-4}
		& \cite{Comm-ISAC-Cui2026} & Multiband ISAC & \textbf{(T)}~Propose the integrated lower-band communication/higher-band sensing principle. \\
		\cline{2-4}
		& \cite{Comm-MB-Shyamal2026} & Hybrid six-level receiver & \textbf{(T)}~Propose hybrid cascade/parallel six-level coupling for four-band reception. \\
		\hline
		\multirow{4}{2.45cm}[-0.5\baselineskip]{\centering \makecell{\textbf{Frequency hopping}\\\textbf{communications}}}
		& \cite{Comm-FH-Wen2024} & Five-level dual-band hopping & \textbf{(E)}~Propose a five-level CPT receiver to demonstrate $3.05/15.29$~GHz hopping at $20{,}000$~hops/s. \\
		\cline{2-4}
		& \cite{Comm-FH-Tong2026} & Cascaded high-$\ell$ states & \textbf{(E)}~Use high-$\ell$ heterodyne reception for $315$~MHz/$1.7$~GHz/$29.74$~GHz hopping at $60{,}000$~hops/s. \\
		\cline{2-4}
		& \cite{Comm-FH-Qiu2026} & Electrically small antenna & \textbf{(E)}~Use off-resonant Stark heterodyne readout to demonstrate single-cell frequency hopping over $4$--$30$~MHz. \\
		\cline{2-4}
		& \cite{Comm-FH-Ding2025} & Waveguide Stark mixer & \textbf{(E)}~Use waveguide-enhanced AC-Stark heterodyne detection for $100$~kHz--$20$~GHz FHSS, with up to $100{,}000$~hops/s and \textbf{(T)}~$>150{,}000$ candidate channels. \\
		\hline
	\end{tabularx}
	\par\vspace{0.5mm}
\par\smallskip\noindent\parbox{\linewidth}{\raggedright\scriptsize\textit{Note.} (E) denotes an experimental result, (T) a result from analysis or simulation, and (T+E) a model or prediction supported in part by measurements.}
\end{table*}

\subsubsection{Low-frequency communications} 
Low-frequency (LF, $30\sim 300$ kHz) fields play a key role in underwater and through-the-Earth communications because of their long {wavelengths} and strong penetration capability. 
Recent studies have begun to extend RAREs from low-frequency field sensing to low-frequency communications~\cite{Comm-LF-Xie2025,Comm-LF-Shi2025}. A low-frequency communication system was developed in \cite{Comm-LF-Xie2025}. In this implementation, two parallel electrode plates, one of which was connected to an external whip antenna, were embedded into a Cs vapor cell to mitigate the shielding effect of the alkali-metal vapor container on low-frequency fields. 
Based on the AC Stark effect, the receiver successfully detected 2FSK-, OOK-, and BPSK-modulated signals at 100 kHz. Shi et al.~\cite{Comm-LF-Shi2025} adopted a similar setup to achieve image transmission at 100 kHz with a data rate of 849.1 kbps.
These results {demonstrate} that Rydberg-atom receivers can support high-sensitivity low-frequency communication {using compact receiver architectures}.  
 

\subsubsection{Broadband communications}
Leveraging the bandwidth-extension techniques introduced in Section III, RAREs can be adapted for broadband communication applications. Consistent with the two definitions of {receiver} bandwidth, existing studies in this area can be divided into two categories. The first focuses on enabling communication with a wide and continuously tunable carrier frequency, whereas the second aims to enlarge the instantaneous bandwidth of RAREs to improve the achievable channel capacity.

For the first category, bandwidth extension {based on the AC Stark effect} has been widely adopted~\cite{Comm-CB-Song2019,Comm-CB-Li2022}. Specifically, Song et al.~\cite{Comm-CB-Song2019} demonstrated reliable digital communication at a rate of 500 kbps within a continuously tunable frequency range of 200 MHz. Li et al.~\cite{Comm-CB-Li2022} further expanded the tunable range to approximately 5 GHz and demonstrated color image transmission at a data rate of 238 kbps. Beyond the {AC Stark} effect, Liu et al.~\cite{Comm-CB-Liu2022} employed a resonant two-photon microwave transition scheme to achieve a tunable frequency range of 1 GHz while demonstrating QPSK demodulation at a data rate of 75 kbps.
For the second category, Knarr et al.~\cite{Comm-CB-Knarr2023} {proposed} a spatiotemporal multiplexing technique 
{that was predicted to provide a response bandwidth exceeding 100~MHz and enable error-free OOK transmission at data rates of up to 100~Mbps.}
At the time of writing, other bandwidth-extension techniques, such as the Zeeman effect for extending the tunable carrier range and the Stark comb for enlarging the instantaneous bandwidth{, have not been demonstrated for broadband data transmission using RAREs.} Nevertheless, these techniques are expected to further improve the broadband communication {capabilities} of RAREs.




\subsubsection{Multiband communications}
\begin{figure*}
	\centering
	\includegraphics[width=6.5in]{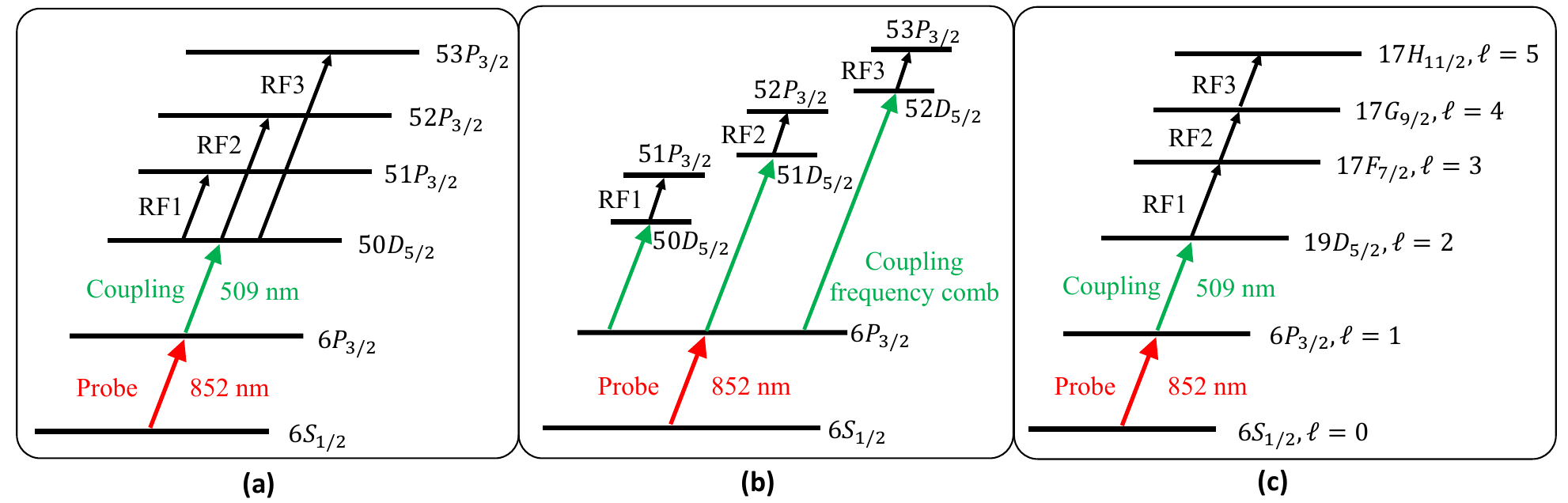}
	\vspace*{-1em}
	\caption{Electron transition structures for realizing multiband communications: (a) the final-state transition scheme, (b) the initial-state transition scheme, and (c) the high-orbital-angular-momentum state transition scheme. 
    } 
	\vspace*{-1em}
	\label{img:multiband}
\end{figure*}

Multiband communication exploits the rich Rydberg-state manifold to enable resonant detection of signals across discrete frequency bands {with frequency separations extending to} the sub-THz scale
\cite{Comm-MB-Du2022,Comm-MB-Meyer2023,Comm-MB-You2023,Comm-MB-Holloway2021,Comm-MIMO-Zhang2024,Comm-MB-Zhang2024,Comm-MB-Cui2026,Comm-MB-Zhu2025,Comm-ISAC-Cui2026,Comm-MB-Prajapati2025,Comm-MB-Allinson2024,Comm-MB-Elgee2023,Comm-MB-Shyamal2026}.
This operation differs fundamentally from conventional broadband communications, where the receiver processes signals over a continuous spectrum, {typically spanning the MHz or GHz scale}. As a distinctive capability of RAREs, multiband operation enables the simultaneous {reception of signals from} multiple heterogeneous devices, such as an S-band mobile user and a W-band satellite link, within a unified receiver architecture.
Existing studies typically adopt three types of Rydberg transition structures for multiband communication: the final-state transition scheme in Fig.~\ref{img:multiband}(a), the initial-state transition scheme in Fig.~\ref{img:multiband}(b), and the cascaded high-orbital-angular-momentum (high-OAM) transition scheme in Fig.~\ref{img:multiband}(c).

The final-state transition {scheme} realizes multiband reception through transitions from a common initial Rydberg state to multiple final Rydberg states ($\ket{3}\rightarrow\ket{n}, \forall n$), as shown in Fig.~\ref{img:multiband}(a). This concept has been experimentally demonstrated in~\cite{Comm-MB-Du2022,Comm-MB-Meyer2023,Comm-MIMO-Zhang2024,Comm-MB-Zhang2024,Comm-MB-Cui2026}. For example, OOK symbols at 12.0, 27.5, 65.6, and 116.1 GHz were concurrently detected via the transitions from $56D_{5/2}$ to $57P_{3/2}$, $54F_{7/2}$, $52F_{7/2}$, and $59P_{3/2}$, respectively~\cite{Comm-MB-Meyer2023}. In addition, color-image transmission was demonstrated in~\cite{Comm-MB-Zhang2024}, where the R, G, and B channels were modulated onto frequency bands at 3.78, 5.106, and 7.91 GHz, respectively.
From the perspective of communication signal processing, a comprehensive transmission model for multiband RAREs was established by Cui et al.~\cite{Comm-MB-Cui2026}. Based on this model, {a Rabi-attention} mechanism was revealed and experimentally validated at 3.2 and 30.6 GHz.
Specifically, because the EIT-AT spectrum is jointly determined by multiple RF signals, the sensitivities of a RARE to different frequency bands are inherently coupled and can therefore influence one another. More importantly, these band-specific sensitivities can be flexibly allocated through the attention coefficient $\alpha_n=\frac{\Omega_n^2}{\sum_{n'=1}^{N} \Omega_{n'}^2}$, which characterizes the distribution of {the reference-field} Rabi frequencies across all frequency bands {$\{\Omega_n\}_{n=1}^{N}$}. By leveraging this attention mechanism, 
{the sensitivity allocation was optimized to maximize the spectral efficiency}
of multiband communication~\cite{Comm-MB-Cui2026}. Zhu et al.~\cite{Comm-MB-Zhu2025} further integrated multiband reception with MIMO technology, and jointly optimized the transmitter precoding and receiver sensitivity allocation for capacity maximization. Further advances include multiband ISAC, where lower bands are used for long-range communication and higher bands for fine-grained sensing~\cite{Comm-ISAC-Cui2026}.

A limitation of the final-state transition strategy is that the number of strongly sensitive Rydberg transitions is typically fewer than 10~\cite{Comm-MB-Cui2026}.
This limitation arises because the transition dipole moment decreases rapidly as the energy separation between Rydberg states increases. 
For example, the transition $60D_{5/2}\rightarrow 61P_{3/2}$ at 3.2 GHz has a dipole moment of $2409qa_0$, whereas {the dipole moment} of $60D_{5/2}\rightarrow 63P_{3/2}$ at 62.8 GHz decreases to $229qa_0$. 
To overcome this limitation, the final-state transition scheme is often complemented by the initial-state transition scheme, in which atoms are {optically} excited to different initial Rydberg states to sense multiband signals, as presented in Fig.~\ref{img:multiband}(b). Specifically, in~\cite{Comm-MB-Prajapati2025}, an optical frequency comb {was} used as the coupling laser to simultaneously excite atoms to seven initial Rydberg states ($72P\sim78P$). This configuration consequently enables sensitive transitions to a large number of final Rydberg states, such as $70S\sim82S$ and $68D\sim80D$.

A third approach to multiband reception is to employ cascaded transitions through  high-OAM states~\cite{Comm-MB-Allinson2024,Comm-MB-Elgee2023,Comm-MB-Shyamal2026}, as illustrated in Fig.~\ref{img:multiband}(c). In this case, the multiband signals are coupled through successively higher-OAM states, such that the energy separations between adjacent states become progressively smaller. For instance, in~\cite{Comm-MB-Allinson2024}, the consecutive high-OAM transitions, $19D_{5/2}\rightarrow 17F_{7/2}\rightarrow 17G_{9/2}\rightarrow 17H_{11/2}\rightarrow 17I_{13/2}\rightarrow\cdots$, enabled the simultaneous detection of RF signals from 128 MHz to 0.61 THz. 
Overall, multiband communication highlights a distinctive advantage of RAREs, which provides a promising foundation for future heterogeneous communication systems. 








\subsubsection{Frequency-hopping communications}
Another promising application of RAREs is frequency-hopping communication~\cite{Comm-FH-Wen2024,Comm-FH-Tong2026,Comm-FH-Qiu2026,Comm-FH-Ding2025}. In a frequency-hopping spread-spectrum (FHSS) system, the carrier frequency rapidly switches among a predefined set of channels, thereby enhancing both anti-jamming performance and communication security. Conventional FHSS implementations 
{covering extremely wide frequency ranges may require multiple RF front ends and frequency-agile mixers, increasing system complexity and potentially limiting the hopping bandwidth and rate.}
For example, GSM operates between 900 MHz and 1800 MHz with a hopping rate of 217 hops/s, while Bluetooth operates over 2.400–2.4835 GHz with 1600 hops/s. 
By contrast, RAREs are naturally well suited to FHSS systems because of their inherent continuous-band and multi-band detection capabilities.

A Rydberg-atom-based frequency-hopping receiver was first proposed and experimentally demonstrated in~\cite{Comm-FH-Wen2024}. In this work, a five-level quantum system was constructed to support resonant reception of carrier frequencies at 15.29 GHz and 3.05 GHz through different transition pathways. Based on this design, dual-channel FHSS was achieved with a hopping rate of 20,000 hops/s. This initial demonstration, although limited to two channels, established the broadband reception capability and high-speed hopping potential of RAREs for FHSS applications. 
Subsequently, Tong et al.~\cite{Comm-FH-Tong2026} expanded the hopping bandwidth to span six octaves. They employed three LO fields to excite 
high-OAM Rydberg transitions, i.e., $48D_{5/2}\rightarrow47F_{7/2}\rightarrow 46G_{9/2}\rightarrow 46H_{11/2}$. This scheme enabled resonant detection of {frequency-hopping} RF hopping signals at 315 MHz, 1.7 GHz, and 29.74 GHz with a hopping rate of 60,000 hops/s. 
Moreover, frequency-hopping communication based on Rydberg atoms has also been demonstrated {across} the shortwave band (4–30 MHz)~\cite{Comm-FH-Qiu2026}. A significant advance was reported by Nan et al.~\cite{Comm-FH-Ding2025}. Their work {combined} the AC-Stark shift with heterodyne detection {in a waveguide-coupled receiver} to enable BPSK frequency hopping over an ultra-broad continuous spectrum, spanning from 100 kHz to 20 GHz (17 octaves). The resulting FHSS system supports more than 150,000 {candidate hopping} channels, which is approximately three orders of magnitude higher than that of conventional GSM and Bluetooth systems, together with a hopping rate of 100,000 hops/s. Collectively, these results highlight the strong potential of RAREs for future secure communication systems and jamming-resistant networks.

\subsection{Other Communication Technologies}

In addition to the research directions discussed above, several advanced communication applications of RAREs have also been explored, which are summarized below. 

\subsubsection{Radio-over-fiber} Radio-over-fiber (RoF) refers to a technique in which RF signals are transported over optical fibers by first converting the RF waveform into the optical domain and then recovering it after fiber transmission. Rydberg-atom-based RoF is particularly attractive because Rydberg atoms can directly convert RF signals into optical signals without requiring additional {electronic} RF-to-optical encoding hardware. This concept was experimentally demonstrated in~\cite{Comm-RoF-Deb2018}, where a Rydberg-atom-based RoF link with a {signal} bandwidth exceeding 1 MHz was achieved. Holloway et al.~\cite{Eng-Compact-Holloway2018} further developed a movable fiber-coupled RARE and demonstrated its capability for near-field imaging and antenna {radiation-pattern} measurements.

\subsubsection{AI-enhanced Rydberg communications} 
Artificial intelligence (AI) has emerged as a promising tool for improving the communication performance of RAREs, especially in noisy and complex electromagnetic environments. Specifically, Liu et al.~\cite{Comm-AI-Liu2022} employed deep learning to directly decode {frequency-division-multiplexed microwave signals} without explicitly solving the Lindblad master equation. Their method achieved higher demodulation accuracy than master-equation fitting that neglected higher-order terms and noise, particularly for noisy signals. Gao et al.~\cite{Comm-ISAC-Gao2025} further introduced an AI-assisted solver to mitigate dynamic interference caused by radar signals in Rydberg atomic communication systems. More recently, Xie et al.~\cite{Comm-AI-Xie2025} proposed a {CNN-Transformer} framework for multifrequency recognition, broadband spectrum monitoring, and automatic modulation {recognition}.


\section{Rydberg Atomic Sensing}

Beyond direct electric-field measurement and wireless communications, RAREs also provide a promising platform for extracting higher-level information from received signals. Different sensing quantities can be estimated from the spatial, spectral, temporal, or phase-dependent response of the atoms. In this section, we survey existing Rydberg atomic sensing technologies in four categories: field imaging, source localization, radar sensing, and other emerging sensing applications.

\subsection{Field Imaging}
\begin{table*}[!t]
    \centering
    \caption{Summary of field-imaging techniques using Rydberg atomic sensors.}
    \label{tab:sensing_field_imaging}

    \scriptsize
    \setlength{\tabcolsep}{2pt} 
    \setlength{\extrarowheight}{0pt} 
    \renewcommand{\arraystretch}{1.0} 
    \renewcommand{\tabularxcolumn}[1]{m{#1}} 

    \begin{tabularx}{\textwidth}{|>{\centering\arraybackslash}m{2.20cm}|>{\centering\arraybackslash}m{0.75cm}|>{\centering\arraybackslash}m{3.35cm}|>{\raggedright\arraybackslash}X|}
        \hline
        \textbf{Category} & \textbf{Ref.} & \textbf{Approach} & \multicolumn{1}{c|}{\textbf{Main contribution}} \\
        \hline
        \multirow{7}{2.20cm}{\centering\textbf{Scanning imaging}}& \cite{Sensing-FI-Anderson2018} & Scanned vapor-cell probe & \textbf{(E)}~Map the near field of a $13.49$~GHz horn antenna with a spatial resolution of approximately $\lambda/10$. \\
        \cline{2-4}
        & \cite{Eng-Compact-Cardman2020} & Portable AC Stark probe & \textbf{(E)}~Map the two-dimensional electric-field amplitude around a $2.5$~GHz Yagi--Uda antenna. \\
        \cline{2-4}
        & \cite{Eng-Compact-Mao2023} & Fiber-coupled EIT probe & \textbf{(E)}~Image the $4.48$~GHz field with approximately $\lambda/6$ resolution using a probe with $40.4\%$ optical efficiency. \\
        \cline{2-4}
        & \cite{Sensing-FI-Hu2025} & Single-ended fiber probe & \textbf{(E)}~Resolve reactive near-field features at $8.556$~GHz with a spatial resolution of approximately $\lambda/56$. \\
        \cline{2-4}
        & \cite{Sensing-FI-Shi2023} & Atomic superheterodyne readout & \textbf{(E)}~Reconstruct a far-field antenna pattern from near-field amplitude and phase measurements at $2.389$~GHz, with longitudinal phase errors below $1.7\%$. \\
        \cline{2-4}
        & \cite{Sensing-SA-Anderson2023} & Coherent probe scanning & \textbf{(E)}~Map field amplitude and phase over extended apertures at $2.5$ and $100$~GHz. \\
        \cline{2-4}
        & \cite{Sensing-SA-Prajapati2023} & Fiber-probe synthetic aperture & \textbf{(E)}~Reconstruct transmitter images at $28$~GHz from field magnitude and phase sampled across a plane. \\
        \hline
        \multirow{6}{2.20cm}{\centering\textbf{Full-field imaging}}& \cite{NF2} & Probe-transmission imaging & \textbf{(E)}~Map the $6.9$~GHz electric field with $66~\mu\mathrm{m}$ spatial resolution. \\
        \cline{2-4}
        & \cite{Sensing-FF-Wade2016} & Raman fluorescence imaging & \textbf{(E)}~Image $0.634$~THz standing waves at approximately $25$~frames/s. \\
        \cline{2-4}
        & \cite{Sensing-FF-Downes2020} & THz-induced fluorescence & \textbf{(E)}~Record full-field $0.55$~THz images at $3000$~frames/s. \\
        \cline{2-4}
        & \cite{Sensing-FF-Schlossberger2024} & Light-sheet fluorescence & \textbf{(E)}~Map electric and static magnetic fields in a plane with approximately $160~\mu\mathrm{m}$ spatial resolution. \\
        \cline{2-4}
        & \cite{Sensing-FF-Pati2026} & AT-induced fluorescence & \textbf{(E)}~Image $91.4$~GHz fields over $6.5~\mathrm{cm}^{2}$ with a $16.4~\mu\mathrm{s}$ response time. \\
        \cline{2-4}
        & \cite{Sensing-FF-Ko2026} & AT-calibrated fluorescence & \textbf{(E)}~Map the absolute millimeter-wave field amplitude from spatially resolved fluorescence spectra. \\
        \hline
    \end{tabularx}
    \par\smallskip\noindent\parbox{\linewidth}{\raggedright\scriptsize\textit{Note.} (E) denotes an experimental result, (T) a result from analysis or simulation, and (T+E) a model or prediction supported in part by measurements.}
\end{table*}

\subsubsection{Scanning Imaging}

Among the techniques summarized in Table~\ref{tab:sensing_field_imaging}, scanning imaging reconstructs the spatial field distribution from local measurements acquired at different positions. A small vapor cell or fiber-coupled atomic probe is moved across the region of interest, and the atomic response at each position is used to determine the local field. The spatial resolution is mainly determined by the effective sensing volume and the sampling interval. This approach is particularly useful for measuring fields with strong spatial variation around antennas and microwave devices.

Early studies established the basic capability of Rydberg atoms for scanned field mapping. Anderson et al.~\cite{Sensing-FI-Anderson2018} used a small Rb vapor cell to image the near field of a $13.49$~GHz horn antenna and achieved a spatial resolution of approximately $\lambda/10$. Cardman et al.~\cite{Eng-Compact-Cardman2020} later developed a portable Rydberg instrument and measured the two-dimensional electric-field distribution around a Yagi--Uda antenna. Subsequent work focused on making the atomic probe more compact and less invasive. Mao et al.~\cite{Eng-Compact-Mao2023} developed a fiber-coupled sensing head with an integrated-probe efficiency of $40.4\%$ and demonstrated field imaging at $4.48$~GHz with a spatial resolution of approximately $\lambda/6$. Hu et al.~\cite{Sensing-FI-Hu2025} further developed a single-ended fiber-integrated probe for reactive near-field imaging at $8.556$~GHz. A spatial resolution of approximately $\lambda/56$ was reported, together with reduced disturbance to the measured field.

Field amplitude alone is insufficient for applications that require the complex spatial field, such as near-to-far-field transformation and coherent aperture reconstruction. Phase-resolved scanning therefore became an important extension of Rydberg field imaging. Shi et al.~\cite{Sensing-FI-Shi2023} introduced an auxiliary RF field as a local oscillator and used atomic superheterodyne detection to measure both the amplitude and phase near a $2.389$~GHz antenna. The measured complex field was then used for near-to-far-field transformation, with a reported longitudinal phase error below $1.7\%$.

Phase-coherent measurements at different spatial positions can also be combined to form a synthetic receiving aperture. Anderson et al.~\cite{Sensing-SA-Anderson2023} demonstrated wide-area amplitude and phase imaging using movable Rydberg probes, including measurements at $2.5$ and $100$~GHz. Prajapati et al.~\cite{Sensing-SA-Prajapati2023} used a fiber-coupled Rydberg probe on a two-dimensional translation stage to construct a synthetic aperture at $28$~GHz. A radiated local-oscillator field enabled magnitude and phase measurements at each sampling position for transmitter imaging and channel sounding. However, the direction-dependent response of the LO-dressed vapor cell~\cite{Comm-CA-Cui2026b} also becomes relevant in coherent scanning. Changes in the incident direction can modify the measured amplitude and phase, which may therefore need to be mitigated through receiver calibration or appropriate beamforming design when accurate field reconstruction is required.

\subsubsection{Full-Field Imaging}

The scanning approaches discussed above effectively use the vapor cell as a single spatial measurement point. In fact, atoms at different positions in the cell interact with the local electromagnetic field at those positions. However, conventional optical readout combines these responses into a single output and therefore loses the spatial information within the cell. To enable full-field imaging, recent studies have introduced spatially resolved optical readout to distinguish the atomic responses at different locations.

Early work demonstrated this concept by replacing single-channel detection with spatially resolved imaging. B"ohi and Treutlein~\cite{Sensing-FI-Bohi2012} used camera-based optical readout to obtain single-shot two-dimensional images of microwave magnetic fields with a spatial resolution of approximately $350~\mu$m. Fan et al.~\cite{NF2} extended spatially resolved imaging to Rydberg microwave electric-field measurements. The transmitted probe beam was imaged onto a camera, and the local atomic spectra at different transverse positions were used to reconstruct the field distribution. A spatial resolution of approximately $66~\mu$m was achieved.

Fluorescence provides another route to parallel spatial readout and is particularly suitable for dynamic imaging. Wade et al.~\cite{Sensing-FF-Wade2016} imaged the spatial fluorescence from Rydberg atoms exposed to a $0.634$~THz field and achieved an imaging rate of approximately $25$ frames/s. Downes et al.~\cite{Sensing-FF-Downes2020} further increased the imaging rate to the kilohertz scale using fast fluorescence detection. These experiments showed that the spatial atomic response can be recorded directly without mechanically scanning a local probe.

Recent studies have further improved the spatial control and sensing area of fluorescence imaging. Schlossberger et al.~\cite{Sensing-FF-Schlossberger2024} used light-sheet excitation to define a two-dimensional sensing plane and imaged electric and magnetic fields with a spatial resolution of approximately $160~\mu$m. Pati et al.~\cite{Sensing-FF-Pati2026} demonstrated millimeter-wave imaging at $91.4$~GHz over a $6.5$~cm$^2$ sensing area and reported a response time of $16.4~\mu$s.
Fluorescence imaging can also provide an absolute field measurement at each image location. Ko et al.~\cite{Sensing-FF-Ko2026} used a fluorescence channel that remains dark in the absence of the millimeter-wave field and obtained the local field amplitude from the Autler--Townes splitting.

\subsection{Source Localization}
\begin{table*}[!t]
    \centering
    \caption{Summary of source-localization techniques using Rydberg atomic receivers.}
    \label{tab:sensing_source_localization}

    \scriptsize
    \setlength{\tabcolsep}{2pt}
    \setlength{\extrarowheight}{0pt}
    \renewcommand{\arraystretch}{1.0}
    \renewcommand{\tabularxcolumn}[1]{m{#1}}

    \begin{tabularx}{\textwidth}{|>{\centering\arraybackslash}m{2.20cm}|>{\centering\arraybackslash}m{0.75cm}|>{\centering\arraybackslash}m{3.35cm}|>{\raggedright\arraybackslash}X|}
        \hline
        \textbf{Category} & \textbf{Ref.} & \textbf{Approach} & \multicolumn{1}{c|}{\textbf{Main contribution}} \\
        \hline
        \multirow{8}{2.20cm}{\centering\textbf{Phase-based methods}}& \cite{Sensing-LOC-Robinson2021} & Two-point heterodyne readout & \textbf{(E)}~Estimate the AoA of a $19.18$~GHz source from the phase difference between two positions in one vapor cell. \\
        \cline{2-4}
        & \cite{Eng-Array-Yan2023} & L-shaped atomic array & \textbf{(E)}~Use two orthogonal baselines to localize a source in three dimensions and estimate its velocity. \\
        \cline{2-4}
        & \cite{Eng-Array-Mao2024} & Two fiber-coupled probes & \textbf{(E)}~Estimate AoA at $2.63$~GHz by digitally combining the phase-coherent probe outputs. \\
        \cline{2-4}
        & \cite{Sensing-LOC-Gill2026} & Scanned heterodyne probe & \textbf{(E)}~Estimate the arrival angles of three microwave sources from their measured phase fronts. \\
        \cline{2-4}
        & \cite{Sensing-LOC-KimMUSIC2025} & Quantum-MUSIC & \textbf{(T)}~Recover the channel matrix from magnitude-only observations and estimate multiple source directions in simulations. \\
        \cline{2-4}
        & \cite{Comm-MIMO-Gong2025a} & Gain-aware ESPRIT & \textbf{(T)}~Estimate multiple source directions while accounting for gain mismatch induced by the local oscillator. \\
        \cline{2-4}
        & \cite{Sensing-LOC-KimMultiBand2025} & Multiband steering-vector matching & \textbf{(T)}~Retrieve channel phase from magnitude measurements and estimate user AoAs across frequency bands. \\
        \cline{2-4}
        & \cite{Sensing-LOC-YanDFS2025} & Dual-channel heterodyne readout & \textbf{(E)}~Measure AoA and Doppler shift jointly, with AoA error below $0.5^\circ$ for angles from $-30^\circ$ to $35^\circ$. \\
        \hline
        \multirow{2}{2.20cm}{\centering\textbf{Amplitude-based methods}}& \cite{Sensing-LOC-ChenBeam2025} & Four-probe beam comparison & \textbf{(E)}~Estimate AoA at $8.567$~GHz from the relative amplitudes of the receive beams. \\
        \cline{2-4}
        & \cite{Comm-MIMO-Jeon2026a} & RF-lens power profiles & \textbf{(T)}~Estimate the AoAs of multiple sources from their spatial power profiles using magnitude-only sparse recovery. \\
        \hline
        \multirow{6}{2.20cm}{\centering\textbf{Atomic-response methods}}& \cite{Comm-CA-Guo2025} & In-cell RF interference & \textbf{(T)}~Infer AoA from the modeled variation of the integrated atomic response with incidence angle. \\
        \cline{2-4}
        & \cite{Sensing-LOC-Talashila2025} & Boundary-induced standing waves & \textbf{(E)}~Measure AoA from standing-wave amplitudes with $1.7^\circ$ angular resolution over $0^\circ$--$60^\circ$. \\
        \cline{2-4}
        & \cite{Sensing-LOC-Schlossberger2025} & Light-sheet fluorescence & \textbf{(E)}~Estimate AoA from standing-wave images over nearly $4\pi$ steradians, with uncertainty on the order of $1^\circ$ for in-plane incidence. \\
        \cline{2-4}
        & \cite{Comm-CA-Han2026} & Fluorescence model and Prony analysis & \textbf{(T)}~Separate multiple incident directions in simulations using spatial frequencies in the modeled fluorescence pattern. \\
        \cline{2-4}
        & \cite{Sensing-LOC-Elgee2024} & Three-axis RF polarimetry & \textbf{(E)}~Reconstruct the three-dimensional RF polarization ellipse from three orthogonal heterodyne signals. \\
        \cline{2-4}
        & \cite{Sensing-LOC-ChenPolarization2025} & Electric and magnetic dipole readout & \textbf{(T)}~Estimate source direction from modeled electric- and magnetic-field orientations in a single vapor cell. \\
        \hline
    \end{tabularx}
    \par\smallskip\noindent\parbox{\linewidth}{\raggedright\scriptsize\textit{Note.} (E) denotes an experimental result, (T) a result from analysis or simulation, and (T+E) a model or prediction supported in part by measurements.}
\end{table*}

Source localization estimates the direction or position of a radiating source from the received electromagnetic field. Conventional direction finding commonly relies on phase differences between spatially separated channels or on angle-dependent amplitude responses. Rydberg atomic receivers provide a new physical front end for these established approaches. Their subwavelength sensing volumes and flexible spatial channels support compact spatial sampling without conventional metallic antenna elements and their associated mutual coupling. Beyond these classical approaches, the atomic medium can itself produce direction-dependent spatial or vector responses that provide additional information for localization. Table~\ref{tab:sensing_source_localization} groups these studies into phase-based, amplitude-based, and atomic-response methods.

\begin{figure*}
    \centering
    \includegraphics[width=6.5in]{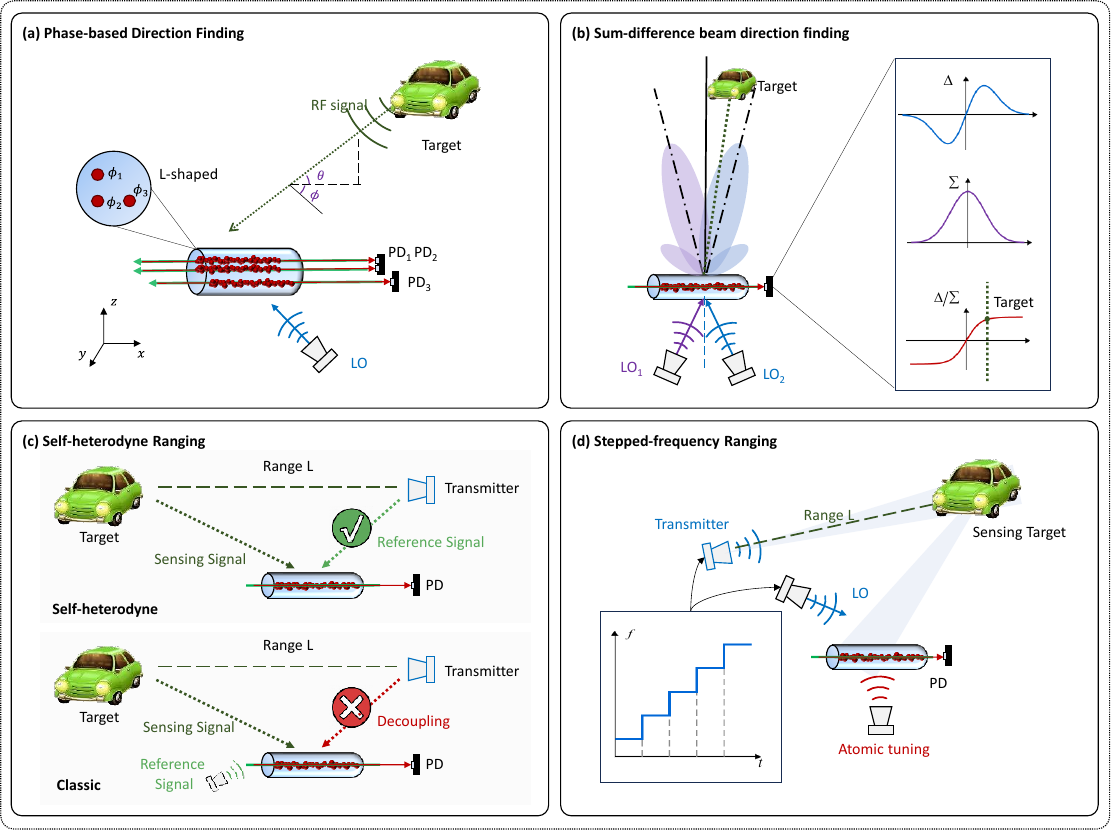}
    \caption{Rydberg atomic radar sensing technologies: (a) phase-based direction finding, (b) sum-difference beam direction finding, (c) self-heterodyne ranging, and (d) stepped-frequency ranging.}
    \label{img:radar_sensing}
\end{figure*}

\subsubsection{Phase-Based Methods}

Phase-based methods estimate the incident direction from phase differences measured at spatially separated receiving channels. In Rydberg atomic receivers, coherent heterodyne or superheterodyne detection transfers the relative RF phase to an optically detected intermediate-frequency signal. Phase measurements from different atomic sensing volumes can then be compared over a known baseline, as illustrated in Fig.~\ref{img:radar_sensing}(a). This is the same basic principle as in conventional interferometric direction finding.

Robinson et al.~\cite{Sensing-LOC-Robinson2021} provided an early experimental demonstration of this approach. They measured the heterodyne phase at two positions inside a Cs vapor cell and estimated the AoA of a $19.18$~GHz source from the resulting phase difference. This experiment established the basic link between spatially resolved atomic phase measurements and direction estimation. The approach was subsequently extended from one-dimensional AoA sensing to array-based localization. Yan et al.~\cite{Eng-Array-Yan2023} constructed a Ku-band L-shaped array with two orthogonal baselines and combined phase-difference and AoA measurements for three-dimensional localization. The reported theoretical maximum location tolerance was $5.7$~mm, and the same system estimated the velocity of a moving source with errors below $0.5$~mm/s. Mao et al.~\cite{Eng-Array-Mao2024} further used two fiber-coupled Rydberg probes to implement digital beamforming at $2.63$~GHz. Their phase-coherent outputs were combined to form the receive beam pattern, with a reported AoA variance below $1.5^\circ$.
Gill et al.~\cite{Sensing-LOC-Gill2026} scanned a single all-dielectric probe through the field of three simultaneously transmitting microwave sources. For phase-stable signals in the far field, the measured phase fronts gave the angle of arrival of each signal.

With phase-coherent atomic arrays established, later studies extended the same measurement principle to more complex localization tasks. Quantum-MUSIC and ESPRIT-based methods have been investigated for multi-source and multi-target DoA estimation~\cite{Sensing-LOC-KimMUSIC2025,Comm-MIMO-Gong2025a}. The latter also incorporated LO-induced gain mismatch into the array model, accounting for channel inconsistency in phase-sensitive processing. Frequency diversity has also been introduced through multi-band phase measurements and steering-vector matching~\cite{Sensing-LOC-KimMultiBand2025}. Beyond angle estimation alone, Yan et al.~\cite{Sensing-LOC-YanDFS2025} combined heterodyne phase measurement with Doppler-frequency sensing in a single atomic receiver. The experiment achieved an AoA error below $0.5^\circ$ over $-30^\circ$ to $35^\circ$, together with a Doppler-frequency error within $\pm0.5$~Hz over an $11$--$18$~GHz tuning range.

\subsubsection{Amplitude-Based Methods}

Amplitude-based methods estimate the incident direction from angle-dependent differences in the received signal strength. Instead of comparing the RF phase across spatial channels, these methods construct a set of amplitude responses that varies with the source direction. The measured response can then be matched to a calibrated angular pattern to determine the AoA, as illustrated by the sum and difference beams in Fig.~\ref{img:radar_sensing}(b). In Rydberg atomic receivers, such responses can be formed by multiple receiving channels or by spatially shaping the incident field before atomic detection.

Chen et al.~\cite{Sensing-LOC-ChenBeam2025} demonstrated amplitude-based direction finding with a four-element Rydberg probe array at $8.567$~GHz. The array formed multiple receiving beams with different angular responses. Comparing their measured amplitudes therefore provided a direct estimate of the incident direction without recovering the RF phase at each channel. This experiment established a Rydberg implementation of the conventional amplitude-comparison principle and showed that directional information can be obtained from independently detected atomic channels.

A further development is to encode the incident angle into a spatial power pattern before atomic readout. The Quantum-PROBE architecture~\cite{Comm-MIMO-Jeon2026a} places an RF lens in front of multiple Rydberg receivers. Different incident directions are focused into distinct power profiles across the receiver array, which enables magnitude-only AoA estimation without explicit phase reconstruction. More importantly, the observed profile can be represented as a superposition of angle-dependent power patterns when multiple sources are present. The resulting sparse-recovery framework therefore extends amplitude-based Rydberg localization from single-source direction finding to multi-source AoA estimation.

\subsubsection{Atomic-Response Methods}

Atomic-response methods obtain directional information from spatial or vector responses formed within the atomic medium itself. Unlike the phase- and amplitude-comparison approaches discussed above, these methods do not primarily rely on differences between several discrete receiving channels. Instead, the incident direction changes the field distribution or polarization-dependent response inside or near the vapor cell, providing additional information for source localization.

One route exploits the spatial field pattern produced by interference or reflection within the sensing region. Guo et al.~\cite{Comm-CA-Guo2025} proposed a single-receiver AoA method based on the direction-dependent field distribution formed inside a vapor cell. The incident angle changes the internal interference pattern and therefore modifies the integrated atomic response, enabling direction estimation without a conventional receiving array. Talashila et al.~\cite{Sensing-LOC-Talashila2025} used a related principle and demonstrated amplitude-only AoA sensing from a subwavelength standing wave near a conducting boundary. The method does not require a strong RF phase reference and achieved an angular resolution of approximately $1.7^\circ$ over $0^\circ$--$60^\circ$.

Spatially resolved optical readout provides more direct access to these interference patterns. Schlossberger et al.~\cite{Sensing-LOC-Schlossberger2025} used light-sheet fluorescence imaging to map RF standing waves inside a vapor cell and estimated the incident direction from their spatial structure. The experiment covered nearly $4\pi$ steradians, with an AoA uncertainty on the order of $1^\circ$ for in-plane incidence. Han et al.~\cite{Comm-CA-Han2026} further extended this spatial-response approach to multiple sources. Prony spectral analysis was applied to the fluorescence pattern to separate the spatial frequencies associated with different incident directions.

Another route uses the vector sensitivity of Rydberg transitions. Elgee et al.~\cite{Sensing-LOC-Elgee2024} used three orthogonally polarized LO fields to reconstruct the three-dimensional polarization ellipse of an incident RF field, providing the vector-field information needed for directional inference. Chen et al.~\cite{Sensing-LOC-ChenPolarization2025} proposed polarization-aware DoA estimation using a single Rydberg atomic receiver. Their scheme uses sequential EIT measurements to infer the orientations of the electric and magnetic fields and reconstruct the incident direction. Its performance was evaluated through simulations.

\subsection{Radar Sensing}
Radar sensing extracts target range and motion from the echo of a transmitted RF waveform. In Rydberg atomic receivers, transmitter leakage or an auxiliary RF field can provide a coherent reference for the weak echo. Their interaction in the atomic medium converts delay- or Doppler-dependent differences into optically detected signals that can be processed with conventional radar algorithms. The studies summarized in Table~\ref{tab:sensing_radar} cover ranging, radar imaging, and Doppler-based velocity sensing.
\begin{table*}[!t]
    \centering
    \caption{Summary of radar-sensing techniques using Rydberg atomic receivers.}
    \label{tab:sensing_radar}

    \scriptsize
    \setlength{\tabcolsep}{2pt}
    \setlength{\extrarowheight}{0pt}
    \renewcommand{\arraystretch}{1.0}
    \renewcommand{\tabularxcolumn}[1]{m{#1}}

    \begin{tabularx}{\textwidth}{|>{\centering\arraybackslash}m{2.20cm}|>{\centering\arraybackslash}m{0.75cm}|>{\centering\arraybackslash}m{3.35cm}|>{\raggedright\arraybackslash}X|}
        \hline
        \textbf{Category} & \textbf{Ref.} & \textbf{Approach} & \multicolumn{1}{c|}{\textbf{Main contribution}} \\
        \hline
        \multirow{7}{2.20cm}{\centering\textbf{Ranging and imaging}}& \cite{Sensing-Radar-Bohaichuk2022} & Time-resolved optical readout & \textbf{(E)}~Detect individual RF pulses as short as $50$~ns and measure the broadening of their optical response. \\
        \cline{2-4}
        & \cite{Sensing-Radar-CuiSelfHet2025} & Self-heterodyne FMCW & \textbf{(T)}~Estimate range in two stages using transmitter leakage as the reference for the delayed echo. \\
        \cline{2-4}
        & \cite{Sensing-Radar-CuiSelfHet2026} & Self-heterodyne power control & \textbf{(T)}~Optimize the transmit-power trajectory to improve the predicted ranging sensitivity. \\
        \cline{2-4}
        & \cite{Sensing-Radar-Watterson2025} & Off-resonant Stark FMCW & \textbf{(E)}~Use a $3.2$~GHz sweep to image targets up to $5$~m away with $4.7$~cm range resolution. \\
        \cline{2-4}
        & \cite{Sensing-Radar-GPR2026} & Ground-penetrating FMCW & \textbf{(E)}~Demonstrate ground-penetrating radar imaging with a Rydberg atomic receiver. \\
        \cline{2-4}
        & \cite{Sensing-Radar-YanSAR2026} & Atomic synthetic-aperture radar & \textbf{(E)}~Reconstruct broadband microwave images with millimeter-scale spatial resolution approaching the Rayleigh criterion. \\
        \cline{2-4}
        & \cite{Sensing-Radar-Chen2025} & Stepped-frequency homodyne readout & \textbf{(E)}~Achieve centimeter-level ranging accuracy and distinguish two targets separated by $15$~cm. \\
        \hline
        \multirow{4}{2.20cm}{\centering\textbf{Velocity sensing}}& \cite{Sensing-Radar-Doppler2025} & Atomic superheterodyne readout & \textbf{(E)}~Distinguish approaching and receding targets by the sign of the measured Doppler shift. \\
        \cline{2-4}
        & \cite{Sensing-LOC-YanDFS2025} & Dual-channel heterodyne readout & \textbf{(E)}~Measure signed Doppler shifts together with AoA, with Doppler errors within $\pm0.5$~Hz. \\
        \cline{2-4}
        & \cite{Sensing-Radar-Xu2026} & Atomic microwave mixer & \textbf{(E)}~Measure target velocities from $0$ to $1.2$~m/s and resolve $62.4~\mu\mathrm{m/s}$ with extended coherent acquisition. \\
        \cline{2-4}
        & \cite{Sensing-Radar-Banerjee2026} & Invariant-function estimator & \textbf{(T)}~Estimate radial velocity from Doppler shifts and evaluate the root-mean-square error in simulations. \\
        \hline
        \textbf{Joint range--velocity}& \cite{Comm-ISAC-Kim2026} & Dual-chirp AFDM & \textbf{(T)}~Separate target delay and Doppler shift in simulations using two distinct chirp rates. \\
        \hline
    \end{tabularx}
    \par\smallskip\noindent\parbox{\linewidth}{\raggedright\scriptsize\textit{Note.} (E) denotes an experimental result, (T) a result from analysis or simulation, and (T+E) a model or prediction supported in part by measurements.}
\end{table*}

\subsubsection{Range Sensing}

Range resolution is fundamentally determined by the effective bandwidth of the transmitted and received waveform. For a bandwidth $B$, the range resolution is approximately $\Delta R=c/(2B)$. Conventional radar therefore relies on short wideband pulses or broadband frequency sweeps to obtain fine range resolution. This requirement is more difficult to satisfy with a Rydberg receiver. A selected resonant transition provides high field sensitivity but only over a relatively narrow instantaneous bandwidth. At the same time, the large number of accessible Rydberg transitions gives the receiver a much wider tunable frequency range. Existing ranging studies have mainly explored two ways to exploit these different bandwidth characteristics: obtaining a broad continuous atomic response for direct wideband ranging and synthesizing a large effective bandwidth from measurements at multiple resonant frequencies.

The finite temporal response of the atomic receiver first places a constraint on direct pulsed ranging. Bohaichuk et al.~\cite{Sensing-Radar-Bohaichuk2022} investigated the transient response of a Rydberg electrometer to short RF pulses and showed that the detected waveform is affected by optical pumping and atomic dynamics. Nanosecond-scale RF pulses could be detected, but their optical responses were broadened by the finite atomic response time. Since finer pulse-based range resolution requires shorter pulses and a correspondingly larger instantaneous bandwidth, this temporal response becomes an important consideration for direct pulsed radar.

FMCW ranging provides another route by converting propagation delay into a beat frequency rather than resolving the delay directly in time. Cui et al.~\cite{Sensing-Radar-CuiSelfHet2025,Sensing-Radar-CuiSelfHet2026} theoretically investigated this approach using a self-heterodyne Rydberg receiver. Transmitter leakage served as the reference field and mixed with the delayed echo in the atomic medium, as shown in Fig.~\ref{img:radar_sensing}(c). The resulting optical beat signal therefore contained the target-delay information. A two-stage estimator first obtained a coarse range estimate and then refined it using the nonlinear self-heterodyne response, establishing a theoretical framework for atomic FMCW ranging with transmitter leakage as the reference.

Watterson et al.~\cite{Sensing-Radar-Watterson2025} subsequently demonstrated broadband FMCW ranging experimentally. A conventional resonant Rydberg receiver cannot remain near resonance during a gigahertz-scale continuous sweep. Their system instead used the off-resonant Stark response of a single Rydberg state, enabling continuous reception from $800$~MHz to $4$~GHz without switching among resonant transitions. Transmitter leakage mixed with the delayed target echo inside the vapor cell, and the resulting optical beat frequency was used to estimate range. The $3.2$~GHz sweep bandwidth provided a range resolution of approximately $4.7$~cm, with targets measured at distances up to $5$~m. Transverse scanning further enabled two-dimensional radar imaging. The same broadband atomic FMCW architecture was later extended to ground-penetrating radar imaging~\cite{Sensing-Radar-GPR2026}. In a separate broadband imaging experiment, Yan et al.~\cite{Sensing-Radar-YanSAR2026} used polarization-independent atomic readout for synthetic-aperture radar imaging with millimeter-scale spatial resolution approaching the Rayleigh limit.

The broadband FMCW approach gains a large continuous bandwidth from the off-resonant atomic response, but this response is weaker than resonant EIT-AT detection. In stepped-frequency ranging (Fig.~\ref{img:radar_sensing}(d)), the receiver is tuned across a sequence of carrier frequencies while retaining resonant detection. Chen et al.~\cite{Sensing-Radar-Chen2025} measured the target response at a sequence of carrier frequencies and combined these measurements to synthesize a broad range bandwidth. Different Rydberg transitions provided coarse frequency coverage, and field-induced tuning bridged the remaining frequency intervals for non-uniform stepped-frequency synthesis. Measurements over $2.6$--$3.6$~GHz were then combined through compressive reconstruction. The experiment achieved centimeter-level ranging accuracy and resolved two targets separated by $15$~cm. In this way, the wide tunable frequency range of the Rydberg receiver was converted into range resolution without requiring a gigahertz-scale instantaneous atomic bandwidth.

This frequency-stepped architecture introduces a different constraint on measurement speed. Accessing different resonant frequency points requires corresponding changes in the optical excitation and, in particular, the coupling-laser frequency. The achievable update rate is therefore linked to the frequency agility of the optical system. Faster tunable lasers and multi-frequency optical addressing, including optical-frequency-comb techniques, may help reduce this switching overhead.

\subsubsection{Velocity Sensing}

Velocity sensing relies on the Doppler frequency shift introduced by target motion. For a monostatic radar, the radial velocity is related to the Doppler frequency by $f_{\mathrm{D}}=2v_{\mathrm{r}}/\lambda$. Velocity resolution therefore depends on how accurately this small frequency offset can be measured. This requirement differs from range sensing, where fine resolution generally requires a large signal bandwidth. Doppler sensing instead benefits from coherent narrowband measurement over a sufficiently long observation time. In a Rydberg superheterodyne receiver, the frequency difference between the reference field and the Doppler-shifted echo is transferred to a low-frequency optical beat signal. Its frequency resolution determines the minimum resolvable velocity difference, while the atomic detection bandwidth sets the accessible Doppler range.

Early studies focused on recovering the Doppler frequency shift from the atomic readout. Tian et al.~\cite{Sensing-Radar-Doppler2025} used Rydberg atomic superheterodyne detection to measure the frequency offset carried by a Doppler-shifted microwave signal. Yan et al.~\cite{Sensing-LOC-YanDFS2025} extended this capability to joint Doppler and directional sensing in one atomic receiver. Their heterodyne system operated over $11$--$18$~GHz and measured the Doppler shift with an error within $\pm0.5$~Hz. A relative frequency reference was introduced to distinguish positive and negative frequency shifts, so the direction of motion could also be identified.

Direct velocity estimation requires the measured frequency shift to be linked to the echo from a moving target. Xu et al.~\cite{Sensing-Radar-Xu2026} used a Rydberg microwave mixer at approximately $24$~GHz to measure reflections from a moving metallic target. Increasing the coherent acquisition time improved the frequency resolution to the millihertz level. The experiment covered target velocities from rest to $1.2$~m/s and achieved a velocity resolution of $62.4~\mu\mathrm{m/s}$. The available atomic detection bandwidth further determines the upper measurable Doppler shift and therefore the velocity span of the receiver.

At the radar processing stage, Banerjee and Kundu~\cite{Sensing-Radar-Banerjee2026} derived an invariant-function estimator for Doppler frequency and radial velocity from a Rydberg radar signal model and compared its velocity error with that of a conventional receiver in simulations. When range and velocity are both unknown, a single chirp leaves the delay and Doppler contributions mixed in the same optical beat frequency. Kim et al.~\cite{Comm-ISAC-Kim2026} used an affine-frequency-division-multiplexing waveform with two post-chirp rates to obtain independent measurements for joint range and velocity estimation.

\subsection{Other Sensing Applications}
Beyond field imaging, source localization, and radar sensing, RAREs have also been extended to sensing tasks that extract information beyond the spatial and kinematic properties of RF sources and targets. One direction focuses on the received signal itself, using the atomic response to recover its spectral or waveform characteristics. Other applications treat changes in the RF field as signatures of external physical quantities, such as motion, material properties, or environmental conditions, and this principle can be further extended to the diagnosis of device or system states. We review these developments in three categories: spectrum sensing, physical sensing, and diagnostic sensing.

\subsubsection{Spectrum Sensing}

Spectrum sensing identifies the frequency components present in an unknown RF environment. In Rydberg receivers, selecting or tuning atomic responses allows access to widely separated carrier frequencies, while the instantaneous bandwidth determines how much spectrum can be observed at a fixed setting. Meyer et al.~\cite{PhysRevApplied.15.014053} illustrated this distinction with a waveguide-coupled spectrum analyzer that covered DC to $20$~GHz continuously, with a $4$~MHz instantaneous bandwidth and more than $80$~dB of linear dynamic range. The system detected ambient AM, FM, Wi-Fi, and Bluetooth signals.

Access to several spectral components can also be obtained through the atomic-state structure itself. Zhang et al.~\cite{Sensing-Spec-Zhang2022} introduced a microwave-frequency-comb spectrometer based on multiple-microwave dressing. The resulting comb maps an unknown microwave frequency onto a set of atomic spectral features and enabled real-time absolute-frequency measurement over a $125$~MHz range. When several comb teeth are consistent with the same readout, the inferred RF frequency can be ambiguous. Wang et al.~\cite{Sensing-Spec-WangTriComb2026} resolved this ambiguity by combining three microwave frequency combs in a hierarchical search and reported a mean absolute frequency error of $0.791$~Hz in experiments. Allinson et al.~\cite{Comm-MB-Allinson2024} used cascaded transitions between high-orbital-angular-momentum Rydberg states to detect multiple RF fields simultaneously, with carrier frequencies extending from $128$~MHz to $0.61$~THz. This span represents simultaneous access to widely separated frequency bands rather than a continuous instantaneous bandwidth. With multiple RF components encoded into the measured atomic spectrum, signal identification can also be performed from the resulting spectral pattern. Liu et al.~\cite{Comm-AI-Liu2022} used deep learning for this purpose and demonstrated recognition of multifrequency microwave signals from the atomic spectrum.

A more recent direction is to recover broadband spectral information without requiring a correspondingly broad atomic output. In~\cite{Sensing-Spec-Compress2026}, a frequency-modulated local oscillator generated multiple parallel sensing channels that mapped different input-frequency components onto narrowband intermediate-frequency responses. More than $640$~MHz of input spectral information was compressed into an atomic output bandwidth of approximately $126$~kHz. The resulting physical multiplexing provides a route to broadband spectrum reconstruction while retaining the intrinsically narrow response of the atomic receiver.

\subsubsection{Physical Sensing}

Physical sensing uses the received RF field as a probe of an external physical quantity rather than as the final object of measurement. Motion changes the propagation path and therefore the received field, while material and environmental properties modify the field through their dielectric or scattering response. These variations can be measured by a Rydberg receiver and related to the underlying physical quantity.

Motion sensing provides a direct example of this principle. Zhang et al.~\cite{Sensing-Motion-Zhang2023} developed a quantum wireless sensing system operating at $2.4$, $5$, and $28$~GHz. Target motion was inferred from the resulting variations in the received RF field. Compared with conventional wireless sensing, the Rydberg receiver improved the displacement granularity of Wi-Fi signals from the millimeter to the submillimeter scale. Measurements at $28$~GHz further resolved micrometer-scale vibrations.

Material properties can also be encoded in the frequency-dependent response of the RF field. Hu et al.~\cite{Sensing-Material-Jiao2024} used a Rydberg receiver to measure the attenuation signatures produced by different liquids. Because the measured amplitude also varies with propagation distance, liquid volume, and container geometry, measurements at several frequencies were combined through a double-ratio representation to suppress these effects while retaining the material-dependent response. The resulting system distinguished 17 liquids with an accuracy above $99.9\%$ and measured milk pH with a granularity of $0.1$. The use of multiple carrier frequencies was important because the contrast between different liquids varied with frequency.
Reflected terahertz fields also carry information about the sample. Li et al.~\cite{Sensing-Material-LiTHz2026} converted these fields into fluorescence images in a cesium vapor cell and used the resulting images to distinguish polar liquids.

Remote sensing can use signals from distant transmitters or place the sensing head away from the optical instrumentation. Arumugam et al.~\cite{Sensing-Remote-Arumugam2024} used XM satellite transmissions to estimate soil moisture. By correlating the direct and ground-reflected signals, the receiver recovered the satellite waveform below the atomic readout noise and related the reflected response to the dielectric properties of the soil. The method was tested with emulated and actual satellite signals, including outdoor measurements over natural terrain. For remote optical readout, Otto et al.~\cite{Sensing-Remote-Otto2023} paired a vapor cell with a corner-cube reflector to form a passive sensing head that could be interrogated from approximately $30$~m away.

\subsubsection{Diagnostic Sensing}

Diagnostic sensing uses RF measurements to characterize the performance or operating condition of an engineered device or system. The quantity of interest is therefore not the RF field itself, but the device state encoded in its emitted waveform, spectral signature, or other field characteristics.

One application is the characterization of high-frequency electronic devices. Bor'owka et al.~\cite{Sensing-Industrial-Borowka2024} developed a Rydberg-based measurement system at $131$~GHz and used it to characterize the electric field emitted by an on-chip automotive radar for vital-sign sensing. Photonic wave plates controlled the propagation direction and polarization of the millimeter-wave field at the vapor cell, allowing the atomic sensor to measure the device output under a well-defined field configuration.

Fault diagnosis places different demands on the receiver because the relevant information is often carried by broadband transient emissions. Wang et al.~\cite{Sensing-Industrial-Wang2025} applied a three-photon Cs Rydberg sensor to partial-discharge detection and measured the time-domain responses of void, floating, particle, and corona discharges. The atomic readout preserved the characteristic periodic behavior of the four discharge types, although its finite instantaneous bandwidth broadened the fast discharge pulses. Once such discharge fingerprints can be acquired reproducibly, automated classification becomes possible. Yin et al.~\cite{Sensing-Industrial-Yin2026} used a one-dimensional ResNet to identify four partial-discharge categories from the Rydberg measurements and reported an accuracy of approximately $94\%$ under increased source--antenna separation. To measure discharges inside gas-insulated switchgear, Pan et al.~\cite{Sensing-Industrial-Pan2026} placed a nonmetallic vapor-cell probe near the discharge region and detected particle-induced near-field transients in the $0.5$--$3$~GHz band.


\section{Future Directions}

\subsection{Performance Limits}

Rydberg atomic receivers have achieved substantial progress in sensitivity, instantaneous bandwidth and frequency coverage. These advances have largely been obtained through separate optimization of different operating regimes. Future development requires a clearer understanding of the performance that can be sustained within one receiver architecture. This question concerns the sensitivity of the complete receiver, its response to fast signals and its ability to preserve useful performance across frequency.

Receiver sensitivity should ultimately be referenced to the incident RF field. The field experienced by the atoms can differ from the external field because of the RF coupling process. Additional noise is introduced during optical readout before the signal reaches the electrical output. The sensitivity reported at the atomic sensing volume therefore does not fully characterize the receiver noise floor~\cite{Sensing-Backes2024,noise-tang2025}. Future research should establish an input-referred model that connects the external field to the final receiver output. Such a model would identify where sensitivity is lost along the detection chain and provide a consistent basis for comparison with conventional antenna--LNA receivers. It would also help determine which part of the receiver should be improved when the system approaches its practical sensitivity limit.

The achievable sensitivity is closely linked to response speed. Strong atom--field interaction favors weak-signal detection, and fast reception requires the atomic response to follow rapid waveform variations. Existing bandwidth-enhancement experiments indicate that faster response can be accompanied by a weaker optical signal or increased noise~\cite{Sensing-Bandwidth-Hu2023,Sensing-Bandwidth-Manchaiah2026}. Future work should therefore determine the sensitivity--bandwidth region that can be reached under a common operating condition. This performance boundary would provide a more useful design target than separate sensitivity and bandwidth records. Possible progress may come from faster readout mechanisms that preserve the atomic response. Parallel sensing channels may also distribute the required bandwidth without placing the full dynamic burden on one atomic channel.

Useful receiver performance must also be maintained as the carrier frequency changes. Current tuning methods can provide broad continuous coverage, but the operating state often changes together with the target frequency. Magnetic tuning may require a new optical locking condition, and auxiliary-field schemes can introduce additional control variables~\cite{Sensing-ContTune-Simons2021,Sensing-ContTune-Shi2023}. Future research should make frequency tuning faster and more repeatable at receiver level. The required operating point should be recovered automatically after a frequency change, with limited loss of sensitivity. Electrically controlled level tuning and programmable optical addressing may help simplify this process. Practical frequency agility will depend on both the accessible tuning range and the time required to restore a stable high-sensitivity response.

\subsection{Scalable Atomic Arrays}


Rydberg atomic arrays have already demonstrated the basic feasibility of spatial reception with several sensing channels~\cite{Eng-Multi-Otto2021,Eng-Array-Mao2024}. The next challenge is scale. Increasing the number of atomic channels can provide a larger aperture and more spatial degrees of freedom, yet the associated optical and calibration burden can grow at a similar rate. A scalable architecture should therefore increase the available spatial information without requiring a complete receiver chain for every additional degree of freedom.

One route is to enlarge the aperture through distributed vapor cells. This approach provides flexible baselines and allows the array geometry to be adapted to different spatial processing tasks. Its scalability is mainly constrained by the cost of reproducing and maintaining many matched receiver units. Future work should therefore focus on modular atomic elements that can share optical and RF resources across the array. Reproducible cell fabrication and automated channel calibration will also become increasingly important as the array size grows. The objective is to make the incremental cost of adding a sensing element decrease with system scale.

A different scaling mechanism is available inside a shared vapor cell. Several sensing regions can be created through spatially separated optical channels, while recent continuous-aperture concepts further use the spatial atomic response itself as the receiving aperture~\cite{Comm-CA-Cui2026a,Comm-CA-Cui2026b}. This approach reduces repeated hardware and offers a much higher channel density within the same atomic medium. Its future potential depends on how many useful spatial modes can be controlled independently. Optical attenuation and nonuniform RF fields will eventually limit this number. Programmable optical addressing and spatial control of the local-oscillator field may provide a way to increase the usable degrees of freedom while preserving a predictable gain and phase response.

Large-scale atomic arrays may ultimately combine these two forms of spatial scaling. Distributed cells can provide the physical aperture, while each cell supports several controllable spatial modes. Such an architecture would allow the large-scale geometry and the local atomic response to be designed at different levels. The resulting scalability should be evaluated by the number of calibrated spatial degrees of freedom that can be supported for a given hardware and control burden. Establishing this relationship will be essential before atomic arrays can move from small experimental demonstrations toward large-scale wireless spatial processing.

\subsection{Transceiver Co-Design}
Existing research on Rydberg-atom-based communications has largely focused on receiver-side design, including signal detection and receiver algorithms, leaving transmitter-side optimization largely unexplored.

Rydberg atomic receivers generally exhibit frequency-dependent responses and noise levels, resulting in unequal reception quality across frequency channels. Their finite dynamic range further constrains the transmit power that can be effectively accommodated. Transmitter design should therefore account for both the wireless channel and the atomic receiver response. Future research may therefore focus on transmitter design tailored to the characteristics of Rydberg atomic receivers. Key directions include subcarrier selection, power allocation and waveform design, along with corresponding receiver detection.

This co-design framework can also be extended from the frequency domain to the spatial domain. In atomic MIMO systems, transmit precoding can be designed together with the spatial configuration of the atomic receiver and the local-oscillator field, allowing the transmitter to better match the spatial response of the atomic front end. RIS/FRIS can further adjust the propagation channel to better match the spatial response of the atomic receiver. Also, operating-point variations and other system perturbations can make the atomic response time-varying. Therefore, pilots or online measurements could be used to track the wireless channel and atomic response. Future work may investigate how these measurements can support real-time adaptation of the transmitter design, while maintaining robustness to estimation errors and system drift.


\subsection{Standing-Wave Estimation and Processing}
Standing waves are an inherent vapor-cell-induced effect in Rydberg receivers. 
Because Rydberg atoms are confined in dielectric vapor cells, incident RF fields are partially reflected and refracted by the glass walls. This process generates standing waves inside the vapor cell that interfere with the original RF fields.
Experiments have shown that significant standing-wave effects generally emerge when the cell dimension exceeds a quarter wavelength, i.e., $D\gtrsim\lambda_{\mathrm{RF}}/4$. This situation commonly arises because a Rydberg-atom vapor cell may operate across widely separated frequency bands. For example, a 2-cm cell may exhibit relatively weak standing-wave effects at 3 GHz, where \(D/\lambda_{\mathrm{RF}}=0.2\), but pronounced effects at 30 GHz, where \(D/\lambda_{\mathrm{RF}}=2\). 

A major consequence of the standing-wave effect is that the vapor cell itself splits an incident RF wave into multiple propagation directions, thereby creating multiple propagation paths even in a line-of-sight wireless environment. This effect is particularly significant in superheterodyne Rydberg receivers. The superposition of the original LO field and its standing-wave components distorts the otherwise stable amplitude and phase profiles of the LO field along the vapor cell. On the one hand, the spatially varying LO amplitude may reduce the linear dynamic range of RAREs and induce significant nonlinear responses. On the other hand, the reshaped phase profile can substantially distort the beam pattern of the vapor cell, producing strong and undesirable sidelobes.

Therefore, accurately modeling, estimating, and mitigating standing waves represents an important direction for the practical deployment of Rydberg receivers in wireless communication and sensing systems. For instance, more accurate electromagnetic simulations that account for the geometry and dielectric properties of the vapor cell are required to model the standing-wave field. 
Joint pilot designs for the LO and signal channels could be developed to estimate their respective standing-wave responses. 
Furthermore, new vapor-cell geometries, such as irregular or nonparallel cell structures, could be designed to suppress the standing-wave formation at the electromagnetic level. 
Last but not least, dedicated post-processing algorithms are needed to compensate for residual standing-wave distortions that cannot be eliminated through physical cell design.

\subsection{Near-Field Modeling and Beam Management}
Near-field effects are an important consideration for practical Rydberg receivers.
For a single vapor cell, the LO source is typically placed close to the cell and may therefore lie within its Rayleigh distance, approximately given by \(d_{\mathrm{R}}=2D^2/\lambda_{\mathrm{RF}}\), where \(D\) denotes the effective aperture size. 
For example, for \(D=10\) cm and \(\lambda_{\mathrm{RF}}=1\) cm (approximately 30 GHz), the Rayleigh distance reaches \(2\) m, making a nearby LO naturally operate in the near-field region. 
This effect becomes even more pronounced for atomic arrays, whose larger aperture substantially extends the near-field region. 
Then, both the LO and external RF signals can exhibit spherical-wave propagation across the array. 
Therefore, future studies should characterize the amplitude and phase distributions of near-field LO signals across vapor cells and investigate their interaction with external signals under both far-field and near-field propagation. 
The spatial mixing of the LO and signal fields can produce fundamentally different beat-note distributions and consequently reshape the receive beam pattern or beam-focusing behavior. 
Based on such models, dedicated near-field beam training, beam tracking, and other beam management methods will be needed, potentially enabling joint angle-and-range discrimination and near-field spatial focusing for communication and sensing.

\section{Conclusions}\label{sec:7}

In this paper, we have presented a comprehensive survey of Rydberg atomic receivers for wireless communication and sensing, with an emphasis on their physical-layer capabilities and experimental implementations. We first introduced the evolution and operating principles of Rydberg atomic receivers and compared their architectures with conventional electronic receivers. We then reviewed key detection technologies, including sensitivity enhancement, bandwidth extension, continuous frequency tuning, and receiver engineering. Building on these fundamentals, we discussed recent advances in Rydberg atomic communications, covering waveform modulation, array processing, and multiband reception. We also reviewed atomic sensing applications, including field imaging, source localization, and radar sensing, with particular attention to the receiver configurations and capabilities demonstrated in experiments. Throughout these discussions, we examined how atomic responses and hardware constraints affect communication and sensing performance. Finally, we identified open challenges and future research directions concerning receiver performance limits, scalable atomic arrays, transceiver co-design, standing-wave effects, and near-field beam management. Although substantial progress has been made, further advances in receiver integration and signal processing are needed to translate experimental demonstrations into practical wireless systems. We hope that this survey will provide a useful reference for researchers and stimulate further research on Rydberg atomic communication and sensing.


\bibliographystyle{IEEEtran}
\bibliography{Reference}
\end{document}